\documentclass[aps,prb,twocolumn,groupedaddress,citeautoscript,nopacs,10pt,amsmath,amssymb]{revtex4-2}

\usepackage[colorlinks=true]{hyperref}
 \usepackage{lipsum}
 \usepackage[normalem]{ulem}
 \usepackage{graphicx}
 \usepackage{comment}
 \usepackage{textcomp}
\usepackage[dvipsnames]{xcolor}
\usepackage{latexsym}
\usepackage{amsmath}
\usepackage{amsthm}
\usepackage{amssymb}
\usepackage{mathtools}
\usepackage{epstopdf}
\usepackage{enumitem}
\usepackage{soul}
\usepackage{bm}
\usepackage{setspace} % controllable line spacing
\usepackage{dcolumn} % Align table columns on decimal point
\usepackage{bm} % bold math
\usepackage{setspace} % controllable line spacing
\usepackage{ulem} %for strikethroughs
\usepackage{slashed}
\usepackage{color}
\usepackage{xcolor}
\usepackage{youngtab}
\usepackage{tikz}
\usepackage{tikz-3dplot}
\usepackage{braket}
\usepackage{mhchem}
\usepackage{dsfont}

\usetikzlibrary{shapes,snakes,arrows,chains,matrix,positioning,scopes,calc}
\usetikzlibrary{decorations.markings}

\usepackage[tabbotcap]{subfigure}

\allowdisplaybreaks

\usepackage{lipsum}

\newlength{\diamondrulelength}
\newlength{\diamondrulethickness}
\begin{document}
\title{Quantum-interference correction to conductivity in disordered electronic bands with non-uniform quantum geometry}

\author{Yip Chun Wong}
\thanks{These authors contributed equally to this work.}
\affiliation{Department of Physics, University of Toronto, Toronto, Ontario M5S 1A7, Canada}

\author{Asimpunya Mitra}
\thanks{These authors contributed equally to this work.}
\affiliation{Department of Physics, University of Toronto, Toronto, Ontario M5S 1A7, Canada}

\author{Yong Baek Kim}
\affiliation{Department of Physics, University of Toronto, Toronto, Ontario M5S 1A7, Canada}

\date{\today}

\begin{abstract}
Weak localization in two-dimensional disordered electron systems originates from constructive interference between time-reversed scattering paths, giving rise to a logarithmically divergent quantum-interference correction to the dc conductivity. This behavior is generally suppressed when time-reversal symmetry (TRS) is broken. In this study, using a band-projected framework motivated by spin- and valley-polarized rhombohedral $N$-layer graphene (RNG), we show that nonuniform quantum geometry can qualitatively modify quantum-interference corrections to dc transport. In particular, continuously tuning the quantum geometry can change the sign of the conductivity correction, corresponding to a crossover between localization and antilocalization tendencies. Remarkably, at special values of the parameter controlling the quantum geometry, weak-localization (WL) or weak-antilocalization (WAL) can re-emerge despite broken microscopic TRS. At these special points, the orbital structure of the Bloch states gives rise to an emergent antiunitary symmetry that protects a gapless Cooperon mode and restores the logarithmically divergent quantum-interference correction characteristic of WL or WAL. Our results establish nonuniform quantum geometry as a mechanism for controlling quantum interference and electron localization in multiorbital electronic systems.

\end{abstract}
\maketitle

\section{Introduction}

Electron localization induced by impurity scattering in disordered metals is a paradigmatic manifestation of quantum interference phenomena in condensed matter systems. In conventional two-dimensional disordered metals with time-reversal symmetry (TRS), constructive interference between time-reversed scattering paths gives rise to weak localization (WL), characterized by a logarithmically divergent quantum-interference correction to the conductivity \cite{VanDenDries1981, Lin1984}. Breaking TRS, for example by applying a magnetic field, generally suppresses WL as it disrupts this constructive interference \cite{Hikami_1980, VanHaesendonck1982}. In multiorbital systems, however, impurity scattering is governed not only by disorder but also by momentum-dependent overlaps between Bloch states, which encode the underlying quantum geometry of the electronic bands \cite{Ando1998, Oh2024, mccannWeakLocalizationMagnetoresistanceValley2006}. How this quantum geometry modifies quantum-interference corrections, particularly when TRS is absent, lies beyond the conventional theory of weak localization \cite{Abrahams_1979, suzuuraCrossoverSymplecticOrthogonal2002}.

In recent years, the role of quantum geometry in the electronic properties of two-dimensional systems has attracted considerable attention \cite{HaldaneBerryCurvature2004, NagaosaAnomalousHall2010, XiaoBerryPhase2010, royBandGeometryFractional2014, Peotta2015, Sodemann2015, Gao2023, Li2024, Kim2025, LeNir2026,Hirsbrunner_2026,froese2026,desrochersElectronicCrystalPhases2026}, driven in part by remarkable discoveries in rhombohedral $N$-layer graphene (RNG) \cite{zhouHalfQuartermetalsRhombohedral2021, Lu2024FQAH,seo2026family,zhou2021,HanChiralSuperconductivity2025,KumarPervasiveSpinTriplet2026,qin2026stripeordermetallicsuperconducting} and other two-dimensional materials \cite{caoUnconventionalSuperconductivityMagicangle2018b, Park2023FQAH, yuQuantumGeometryQuantum2025, tanakaSuperfluidStiffnessMagicangle2025}. In such systems, the internal orbital or layer structure of Bloch states can give rise to highly nontrivial momentum-space textures and quantum geometry. Importantly, these geometric structures can govern how scattering amplitudes associated with different scattering paths are related and how Bloch states transform under antiunitary symmetries \cite{suzuuraCrossoverSymplecticOrthogonal2002, mccannWeakLocalizationMagnetoresistanceValley2006}. This raises an intriguing question: Can quantum geometry alter quantum interference and hence electron localization? Despite the growing appreciation of quantum geometry in clean electronic systems, its interplay with disorder and quantum-interference corrections to transport remains relatively unexplored.

In this work, we investigate quantum-interference corrections to the dc conductivity in disordered two-dimensional electronic systems with nonuniform quantum geometry and broken microscopic TRS. Motivated by the spin- and valley-polarized quarter-metal state of RNG \cite{zhouHalfQuartermetalsRhombohedral2021}, we employ low-energy models whose active band is polarized onto a single sublattice. Its layer-pseudospin textures produce the nonuniform quantum geometry. In particular, we study two rotationally symmetric $N$-orbital (or $N$-layer) models: the ``ideal model of RNG''\cite{tan2025ideallimitrhombohedralgraphene} and ``$\lambda_N$-model''\cite{desrochersElectronicCrystalPhases2026,Hirsbrunner_2026,LeNir2026,froese2026}. Both retain the multilayer structure and explicitly break microscopic TRS, while exhibiting distinct quantum geometric profiles. For $N=2$, both the models reduce to the so-called ``$\lambda$-jellium'' model \cite{soejima$l$JelliumModelAnomalous2025}.

In conventional two-dimensional disordered electron systems with TRS, repeated impurity scattering in the particle-particle channel supports a gapless Cooperon mode in the absence of dephasing effects. This gapless mode leads to the logarithmically divergent quantum-interference correction to the conductivity \cite{Hikami_1980, vollhardtAndersonLocalization<1980}. By contrast, in the band-projected ``ideal RNG'' and ``$\lambda_N$-models'', microscopic TRS is broken and the Cooperon modes are generally gapped. However, we find that the Cooperon spectrum depends sensitively on the underlying quantum geometric profile. As this geometric profile is continuously tuned, the interference correction can change both in magnitude and sign, signaling a crossover between localization and antilocalization regimes. Therefore, electron localization can be enhanced or suppressed solely by changing the quantum geometry, without modifying the disorder or electronic dispersion.

Remarkably, at special values of the dimensionless quantum geometric parameter $\beta=\mathcal{B}k_F^2/2$, where $k_F$ is the Fermi momentum and $\mathcal{B}$ is the inverse width of the Berry curvature distribution, a Cooperon gap closes, restoring the logarithmically divergent correction characteristic of WL or WAL. We find that this gap closing is protected by an emergent antiunitary symmetry $\mathcal{T}_N$, which acts as an effective TRS at these special values of $\beta$. 
In the ideal RNG model with $N$-orbitals, a Cooperon gap closes at $\beta=1$, where the aniunitary operator squares to $\mathcal{T}_N^2=(-1)^{N-1}$. Therefore, ideal RNG models with odd (even) $N$ exhibit WL (WAL), consistent with the orthogonal (symplectic) symmetry class behavior in the Altland-Zirnbauer (AZ) classification \cite{Altland1997}. By contrast, in the $\lambda_N$-model with $N>3$, no finite value of $\beta$ leads to a gapless Cooperon mode or a logarithmically divergent interference correction. Nevertheless, the nonmonotonic dependence of the Cooperon gaps on $\beta$ gives rise to an oscillatory quantum-interference correction to the dc conductivity as the quantum geometry is tuned. Together, these results demonstrate that nonuniform quantum geometry can control both the sign and magnitude of quantum-interference correction. 

The rest of the paper is organized as follows. In Sec.~\ref{sec:model}, we calculate the quantum-interference correction in the two-orbital $\lambda$-jellium model and identify its emergent antiunitary symmetry. In Sec.~\ref{sec:N>2_models} we introduce the ideal RNG model and the $\lambda_N$-model. 
In Sec.~\ref{sec:antiunitary_symm}, we derive the criterion for Cooperon gap closing in rotationally invariant multiorbital models and relate it to an antiunitary symmetry of the band-projected Hamiltonian at the Fermi surface. 
In Sec.~\ref{sec:results_N_orb_models}, we present the quantum-interference corrections in the two multiorbital models of RNG. Finally, we discuss experimental implications and broader connections in Sec.~\ref{sec:discussion}.

\section{Quantum-interference correction in the two-orbital $\lambda$-jellium model}\label{sec:model}
To elucidate the role of band quantum geometry, we begin with the two-orbital ``$\lambda$-jellium'' model, which possesses an intrinsic quantum geometry. 

\subsection{The $\lambda$-jellium model}
\begin{figure}
    \centering
    
    \includegraphics[scale=1]{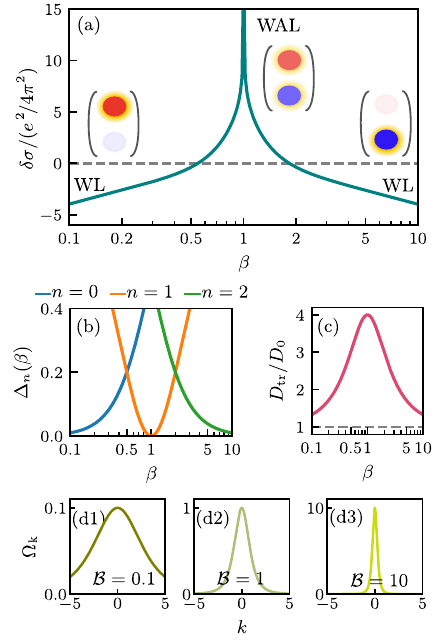}
    \caption{(a) The normalized quantum-interference correction to the dc conductivity, $\delta\sigma/(e^2/4\pi^2)$, in the $\lambda$-jellium model as a function of the dimensionless parameter $\beta=\mathcal{B}k_F^2/2$, that controls the ``active'' band quantum geometry. Weak antilocalization (WAL) due to the non-uniform quantum geometry occurs at $\beta=1$, and weak localization (WL) occurs in the trivial limits of $\beta\to 0$ and $\beta\to \infty$. (b) At $\beta=1$, equal weights of the two orbitals on the Fermi surface is responsible for an emergent antiunitary symmetry $\mathcal{T}$ that protects a gapless Cooperon channel, $\Delta_{n=1}(\beta=1)\to0$, and results in WAL since $\mathcal{T}^2=-1$. The corresponding orbital spinors are shown in the insets of panel (a) for representative values of $\beta$. (c) The normalized transport diffusion coefficient $D_{\text{tr}}/D_{\text{bare}}$, a proxy of the Drude conductivity $\sigma^{\text{Drude}}$, is also enhanced at $\beta=1$. (d1-d3) Berry curvature profiles for different values of $\mathcal{B}$ in the $\lambda$-jellium model.}
    \label{fig:fig_1_summary}
\end{figure}

The $\lambda$-jellium model is constructed from the ordinary spinless two-dimensional (2D) jellium model by engineering a momentum-dependent Bloch wavefunction to have a nonuniform quantum geometry. This geometry is encoded in the lowest eigenstate (or zero-mode) of $L^\lambda(\mathbf{k})$ in the Hamiltonian in Eq.~\eqref{eq:Hamiltonian_lambda_jellium}. The zero mode forms the ``active'' low-energy band of this two-orbital model, while retaining the dispersion of ordinary jellium. Its Hamiltonian is given by \cite{soejima$l$JelliumModelAnomalous2025}
\begin{gather}
\mathcal{H}^{\lambda}_{\mathbf{k}}=\xi_{\mathbf{k}}\mathds{1}_{2\times 2}+J L^{\lambda}(\mathbf{k})_{2\times 2}\label{eq:Hamiltonian_lambda_jellium},
\end{gather}
where $L^\lambda(\mathbf{k})$ is defined as follows
\begin{gather}
L^{\lambda}(\mathbf{k})=
\begin{pmatrix}
\frac{\mathcal{B}k^2}{2} & -\sqrt\frac{\mathcal{B}}{2}(k_x+ik_y)\\-\sqrt\frac{\mathcal{B}}{2}(k_x-ik_y) & 1
\end{pmatrix}\label{eq:Hamiltonian_lambda_jellium_D_matrix}.
\end{gather}
In the following, we take a parabolic dispersion $\xi_{\mathbf{k}}=k^2/(2m)-\mu$, which gives a circular Fermi surface. In Eq.~\eqref{eq:Hamiltonian_lambda_jellium}, $J$ controls the energy separation between the active and higher-energy bands. Within the band-projected formalism adopted here, we formally take $J\rightarrow\infty$. The parameter $\sqrt{\mathcal{B}}$ in Eq.~\eqref{eq:Hamiltonian_lambda_jellium_D_matrix} has dimensions of length and controls the momentum-space profile of the quantum geometry. Throughout, we assume $\mathcal{B}>0$. The Berry curvature is $\Omega_{\mathbf{k}}=\frac{4\mathcal{B}}{(2+\mathcal{B}|\mathbf{k}|^2)^2}$, where $\mathcal{B}$ sets the height of the peak at $\mathbf{k} = 0$, as shown in Fig.~\ref{fig:fig_1_summary}~(d1-d3). The active band, defined over the momentum continuum, carries a Chern number $\mathcal{C}=1$ and additionally satisfies the condition for an ``ideal" quantum geometry \cite{ledwithVortexabilityUnifyingCriterion2023,wangExactLandauLevel2021,royBandGeometryFractional2014}. 

The quantum geometry of the active band is encoded within the cell-periodic part of the Bloch wavefunction of the zero-mode
\begin{gather}
    \ket{s^0_{\mathbf{k}}}=\frac{1}{\sqrt{1+\frac{\mathcal{B}k^2}{2}}}\begin{pmatrix}
        1\\
        \sqrt\frac{\mathcal{B}}{2}(k_x-ik_y)
    \end{pmatrix}.\label{eq:Bloch_state_jellium}
\end{gather}
In the trivial limits of $\mathcal{B}\to0$, $\mathcal{B}\to \infty$, the Bloch state becomes orbital-polarized (see insets in Fig~\ref{fig:fig_1_summary}~(a)). Here the model essentially reduces to the ordinary jellium model and all effects of local quantum geometry vanish. In the transport calculations below, the Bloch states enter through the form factors $\mathcal{F}_{\mathbf{k},\mathbf{k}'}=\langle s_{\mathbf{k}}^0 | s_{\mathbf{k}'}^0\rangle$. After band projection, these Bloch states overlaps modify the disorder vertices, and through the self energy, the Green's functions. 

\subsection{Diffusive transport: diffusons and Cooperons}
Here we provide a brief overview of the concepts underlying diagrammatic transport, aimed at readers with a quantum-geometry background who may be less familiar with these techniques. 

\begin{figure}
    \centering
    \includegraphics[scale=0.25]{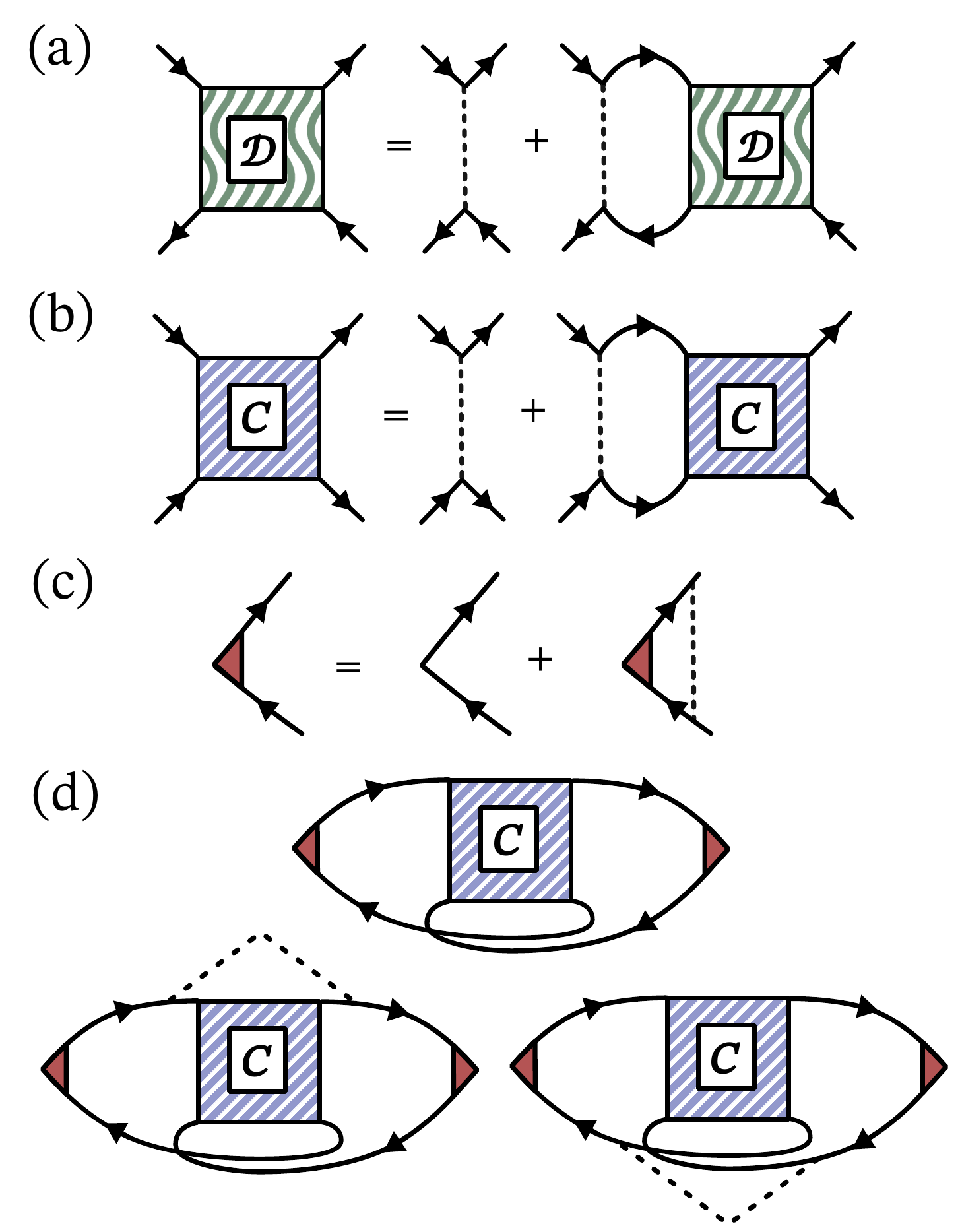}
    \caption{Diagrammatic representation of the Bethe-Salpeter equation for: (a) diffuson, (b) Cooperon. Dashed lines denote impurity vertices, while arrows denote the disordered electronic Green's functions. (c) Vertex correction to the velocity from ladder diagrams. (d) Diagrams
    contributing to the quantum-interference correction of the conductivity: the bare and the two dressed Hikami boxes (top and bottom respectively).}
    \label{fig:fig_2_diagrams}
\end{figure}

In a weakly disordered metal, quasiparticle transport is governed by repeated elastic scattering from impurities. In this regime, the quasiparticle linewidth $1/\tau$ is much smaller than the Fermi energy $\mu$, i.e. $\mu\gg1/\tau\Rightarrow k_F l_0\gg1$, where $l_0=v_F\tau$ is the mean free path and $v_F$ is the Fermi velocity \cite{Akkermans_Montambaux_2007}. Therefore, quasiparticles remain well defined between successive elastic scattering events, which conserve their energy but progressively randomize their momenta. This causes the motion to become diffusive on length scales much larger than the mean-free path $l_0$. At these scales, the microscopic details of the scattering events become irrelevant, and the long-wavelength dynamics is governed by the diffusive relaxation of density fluctuations. This behavior is described by the diffuson, a disorder-averaged two-particle propagator which is obtained by resumming impurity-ladder diagrams in the particle-hole channel (see Fig.~\ref{fig:fig_2_diagrams}~(a)). It features the characteristic gapless diffusion pole required by particle-number conservation \cite{vollhardtAndersonLocalization<1980}. The particle-hole impurity ladders also give rise to the transport diffusion coefficient $D_{\mathrm{tr}}$, that enters the Drude conductivity $\sigma^{\text{Drude}}$.

Beyond the Drude description of diffusive transport, quantum interference between multiple scattering paths gives rise to quantum corrections to the dc conductivity. On length scales between mean free path $l_0$ and coherence length $L_{\phi}$, 
the quasiparticles can undergo multiple scattering events while remaining phase coherent. In systems with microscopic TRS, this enables coherent interference between pairs of scattering paths related by time reversal. These interference processes are described by the Cooperon propagator, obtained by resumming impurity-ladder diagrams in the particle-particle (see Fig.~\ref{fig:fig_2_diagrams}~(b)). The resulting constructive or destructive interference between the paths enhances or suppresses backscattering, generating the logarithmically divergent quantum-interference correction to the dc conductivity associated with WL or WAL, respectively \cite{Hikami_1980, Lee_1985}. Unlike the diffuson, the Cooperon can acquire a finite gap depending on the symmetries of the system, which can cut off these divergent quantum-interference corrections.

In the band-projected formalism, the Bloch-state form factors, encoding the quantum geometry, make the matrix elements of a scalar impurity scattering momentum dependent. As we shall see, this modifies both the diffuson and Cooperon propagators, and the resulting quantum interference corrections.

\subsection{Disorder model and impurity averaging}\label{sec:disordered_GF_impurity}
We model disorder as elastic scattering of electrons from point-like impurities described by an orbital-independent scalar potential
\begin{gather}
    \Tilde U(\mathbf r)
    =u_0
    \sum_{i=1}^{N_{\rm imp}}
    \delta(\mathbf r-\mathbf r_i)\,
    \mathds{1}_{\rm orb},\label{eq:disorder_structure}
\end{gather}
where $u_0$ is the impurity potential and $\mathbf r_i$ is the position of the $i$-th impurity. The impurities are independently and uniformly distributed with a density $n_{\rm imp}=N_{\rm imp}/\mathcal V$. The disorder-averaged potential is spatially uniform, and is absorbed into the chemical potential. The remaining disorder fluctuations are modeled as a Gaussian white noise which is fully characterized by the connected two-point correlator,
$
   \langle
    \Tilde U(\mathbf r)
    \Tilde U(\mathbf r')
    \rangle_c
    =
    n_{\rm imp}u_0^2
    \delta(\mathbf r-\mathbf r')
$, with all higher connected cumulants taken to zero \cite{Akkermans_Montambaux_2007}. Upon projection onto the ``active'' band, the impurity vertex in momentum space acquires the Bloch-state form factor,
$
    U_{\mathbf k,\mathbf p}=\Tilde U_{\mathbf k-\mathbf p}\,
    \mathcal F_{\mathbf k,\mathbf p}
$. The impurity average then gives the disorder vertex
$
    \frac{1}{\mathcal V}
    \left\langle
    U_{\mathbf k,\mathbf p}
    U_{\mathbf p,\mathbf k}
    \right\rangle_{\rm imp,c}
    =
    n_{\rm imp}u_0^2
    \left|
    \mathcal F_{\mathbf k,\mathbf p}
    \right|^2
$, which enters the band-projected self-energy.

We evaluate the disorder self-energy within the first Born approximation, assuming the Fermi level is sufficiently far from the band edge \cite{Akkermans_Montambaux_2007}.
Evaluating the resulting retarded self-energy at the Fermi energy, we find (details in Appendices.~\ref{sec_app:G_F_band_proj} and \ref{sec_supp:self_energy})
\begin{align}
\Sigma^R_{\mathbf{k}}(\omega)=&\frac{n_{\text{imp}}u_0^2}{\mathcal{V}}\sum_{\mathbf{p}}\frac{|\mathcal{F}_{\mathbf{k},\mathbf{p}}|^2}{\omega+i0^+-\xi_{\mathbf{p}}}\nonumber\\
\Rightarrow \Sigma^R_{\mathbf{k}_F}(0)\simeq&-i\pi n_{\text{imp}}u_0^2 \rho(0)\frac{1}{\gamma} =-\frac{i}{2\tau},\label{eq:self_energy_final}
\end{align}
where $\rho(0)$ is the density of states at the Fermi energy. The self-energy contains no interband components as expected within a band-projected description. In Eq.~\eqref{eq:self_energy_final}, the effect of form factors on the quasiparticle lifetime $\tau$ is contained within $\gamma$, where
\begin{gather}
    \gamma=\frac{\left(1+\beta\right)^2}{1+\beta^2},\quad \beta=\frac{\mathcal{B}k_F^2}{2},
\end{gather}
and $k_F$ the Fermi momentum. Using the quasiparticle lifetime of ordinary jellium $\tau_0=[2\pi n_{\mathrm{imp}}u_0^2\rho(0)]^{-1}$, we find that band geometry modifies the quasiparticle lifetime to
$\tau=\gamma\tau_0$. Since $\gamma\geq1$, the band quantum geometry suppresses elastic impurity scattering, which enhances the quasiparticle lifetime relative to ordinary jellium. Using Eq.~\eqref{eq:self_energy_final}, the disorder-averaged retarded and advanced Green's functions of the active band can be written as
\begin{gather}
    G^{R,A}_{\mathbf{k}}(\omega)=\frac{1}{\omega-\xi_{\mathbf{k}}\pm i/2\tau},
\end{gather}
which we use in all our subsequent calculations.

\subsection{Diffuson and Drude conductivity}
Before turning to the quantum correction to the conductivity, we first examine the diffuson and the Drude conductivity. The ``diffuson'' describes the long-wavelength density diffusion in the particle-hole channel and satisfies the Bethe-Salpeter equation shown in Fig.~\ref{fig:fig_2_diagrams}~(a), which is written as \cite{vollhardtAndersonLocalization<1980}
\begin{gather}
\mathcal{D}_{\mathbf{k,k'}}(\mathbf{q},\omega)=\mathcal{D}^0_{\mathbf{k,k'}}+\frac{1}{\mathcal{V}}\sum_{\mathbf{p}}\mathcal{D}_{\mathbf{k,p}}G^R_{\mathbf{p}}G^A_{\mathbf{p+q}} \mathcal{D}^0_{\mathbf{p,k'}} \label{eq:BS_diffuson}.
\end{gather}
Here $\mathcal{D}^0_{\mathbf{k,k'}}\simeq n_{\text{imp}}u_0^2|\mathcal{F}_{\mathbf{k,k'}}|^2$ is the bare diffuson associated with a single scattering event, and $\mathbf{q}$ and $\omega$ are the external momentum and frequency transferred in the particle-hole channel.

In the weak-disorder regime, $\mu\tau\gg1$, the product
$G^R_{\mathbf{p}}G^A_{\mathbf{p+q}}$ is sharply peaked near the Fermi surface. The momentum sum in Eq.~\eqref{eq:BS_diffuson} is therefore dominated by momenta at the Fermi surface $|\mathbf{p}| = k_F$. In the diffusive limit, $\omega\tau \ll 1$ and $qv_F\tau \ll 1$, this simplifies the second term in Eq.~\eqref{eq:BS_diffuson} to
$
    \rho(0) \int \frac{d\phi_\mathbf{p}}{2\pi}\mathcal{D}_{\mathbf{k,p}}\mathcal{D}^0_{\mathbf{p,k'}}\frac{2\pi\tau}{1-i\omega\tau-i\tau \mathbf{v_p\cdot q}}.\label{eq:kernel_diffuson}
$
Restricting remaining momenta to the circular Fermi surface (on-shell) enables the diffuson to be expanded in angular harmonics as
\begin{gather}
\mathcal{D}_{\mathbf{k,k'}}(\mathbf{q})=\sum_{mn} \mathcal{D}_{m,n}(\mathbf{q}) e^{im\phi_{\mathbf{k}}} e^{-in\phi_{\mathbf{k'}}}.\label{eq:angular_harmonics_diffuson}
\end{gather}
This decomposition decouples the angular dependence in Eq.~\eqref{eq:BS_diffuson}, reducing the Bethe-Salpeter equation to a matrix equation in the angular channels. Solving this equation in the $\mathbf{q}\to 0$ limit, we obtain
(details in Appendix.~\ref{sec_supp:diffuson})
\begin{gather}
    \mathcal{D}_{\mathbf{k,k'}}(\mathbf{q},\omega)|_{\mathbf{k,k'}\in\mathrm{FS}} \sim \frac{\tau^{-1}}{-i\omega + D_\mathrm{tr}|\mathbf{q}|^2}. \label{eq:diffuson_final}
\end{gather}
Here, $D_{\mathrm{tr}}$ is the transport diffusion coefficient, which accounts for the current relaxation from scattering events. In two dimensions, $D_{\mathrm{tr}}=v_F^2\tau_{\mathrm{tr}}/2$, where $\tau_{\mathrm{tr}}$ is the transport relaxation time. For the $\lambda$-jellium model, we obtain (see Appendix.~\ref{sec_supp:diffuson})
\begin{gather}
D_{\mathrm{tr}}=D_{\mathrm{bare}}\eta\gamma,\label{eq:Dtr_diffuson}
\end{gather}
where $D_{\mathrm{bare}} = \frac{v_F^2\tau_0}{2}$ is the bare diffusion coefficient of ordinary jellium in the absence of form factors  \cite{Akkermans_Montambaux_2007}. Here, $\eta=\tau_{\mathrm{tr}}/\tau$ relates the transport relaxation time to the quasiparticle lifetime. For the $\lambda$-jellium model we find that this ratio is (details in Appendix.~\ref{sec_supp:vertex_correction}),
\begin{align}
\eta =& \left[1 - \gamma \int\frac{d\phi_{\mathbf{k,k'}}}{2\pi}|\mathcal{F}_{\mathbf{k,k'}}|^2\cos\phi_{\mathbf{k,k'}}\right]^{-1}\label{eq:eta_general}\\
=&\frac{1+\beta^2}{1-\beta+\beta^2}.\label{eq:eta_jellium}
\end{align}
In Eq.~\eqref{eq:eta_general}, the form factor $|\mathcal{F}_{\mathbf{k},\mathbf{k}'}|^2$ determines the angular dependence of the scattering between momenta $\mathbf{k}$ and $\mathbf{k}'$, separated by an angle $\phi_{\mathbf{k,k'}}=\phi_{\mathbf{k}}-\phi_{\mathbf{k}'}$. Since forward scattering events are less effective at relaxing the current, $\tau_{\mathrm{tr}}$ is enhanced relative to $\tau$, i.e. we find $\eta\geq1$. Therefore, the band quantum geometry modifies $D_{\mathrm{tr}}$ through both the renormalized quasiparticle lifetime $\tau=\gamma\tau_0$ and the angular dependence of scattering encoded through $\eta$.

Next, we evaluate the longitudinal conductivity by including impurity-ladder corrections to the current vertex, as shown in Fig.~\ref{fig:fig_2_diagrams}~(c),
\begin{align}
\sigma_i(\mathbf{q},\omega) = &\frac{e^2}{2\pi\mathcal{V}}\sum_\mathbf{k}\overline{v}_\mathbf{k}^{i}v_\mathbf{k}^{i} G_\mathbf{k}^RG_\mathbf{k+q}^A,\end{align}
where $\overline{v}^i_{\mathbf{k}}$ is the renormalized velocity vertex \cite{Akkermans_Montambaux_2007} (details in Appendices.~\ref{sec_supp:vertex_correction}, \ref{sec_supp:drude_conductivity}). In the dc limit $(\mathbf{q}=0$, $\omega\rightarrow0)$, we recover the standard Drude form
\begin{gather}
    \sigma_i^\mathrm{Drude} = e^2\rho(0)D_\mathrm{tr},\label{eq:Drude_dc}
\end{gather}
with the same transport diffusion coefficient $D_{\mathrm{tr}}$. 
Note here, we obtain the same  $D_{\mathrm{tr}}$ in both the diffuson and Drude conductivity. Since $\rho(0)$ is constant, in Fig.~\ref{fig:fig_1_summary}(c) we show the behavior of the Drude conductivity through $D_{\mathrm{tr}}/D_{\mathrm{bare}}$. Using Eq.~\eqref{eq:Dtr_diffuson}, we find $D_{\mathrm{tr}}/D_{\mathrm{bare}} =\eta\gamma = (1+\beta)^2/(1-\beta+\beta^2)\geq1$. Therefore, it reaches its maximum value of $D_{\mathrm{tr}}/D_{\mathrm{bare}}=4$ at $\beta=1$. In the limiting cases $\beta\to 0$ and $\beta\to\infty$, it approaches unity, as the Bloch states on the Fermi surface become fully orbital-polarized (see the insets of Fig.~\ref{fig:fig_1_summary}). Therefore, the system is effectively reduced to the limit of ordinary jellium with no quantum geometric effects. Away from these extreme limits however, we observe that the orbital structure of the Bloch state modifies the impurity scattering rate and enhances the Drude conductivity.

\begin{figure}
    \centering
    \includegraphics[scale=1]{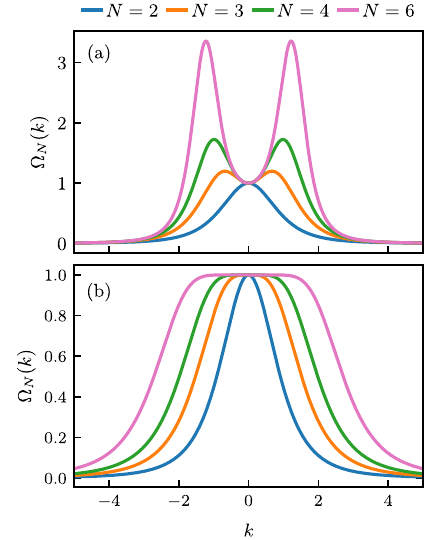}
    \caption{Berry curvature $\Omega_N(k)$: for (a) ideal RNG model, and (b) $\lambda_N$-model at $\mathcal{B}=1$, shown for different values of $N$. For $N=2$, both models reduce to the $\lambda$-jellium model.}
    \label{fig:omega_profile}
\end{figure}

\subsection{Cooperon} \label{sec:Cooperon}
In the diffusive regime, pairs of scattering paths traversed in opposite directions can interfere coherently if their lengths are shorter than the phase coherence length $L_{\phi}$. These processes are encoded in the Cooperon, obtained by summing the particle-particle impurity ladder, and give rise to a quantum-interference correction to the dc conductivity \cite{Akkermans_Montambaux_2007}. The Bethe-Salpeter equation for the Cooperon following Fig.~\ref{fig:fig_2_diagrams}~(b) is 
\begin{gather}
    C_{\mathbf{k,k'}}(\mathbf{Q,\omega)}=C^0_{\mathbf{k,k'}}+\frac{1}{\mathcal{V}}\sum_{\mathbf{p}}C_{\mathbf{k,p}} G^R_{\mathbf{p}}G^A_{\mathbf{Q-p}} C^0_{\mathbf{p,k'}},\label{eq:BS_Cooperon}
\end{gather}
where $C^0_{\mathbf{k},\mathbf{k'}}\simeq n_{\text{imp}}u_0^2\mathcal{F}_{\mathbf{k,k'}}^2$ is the bare Cooperon and $\mathbf{Q}$ is the total momentum of the particle-particle pair. As in the diffuson calculation, we restrict all momenta to the Fermi surface. The second term in Eq.~\eqref{eq:BS_Cooperon} then becomes %$\frac{1}{\mathcal{V}}\sum_{\mathbf{p}}C_{\mathbf{k,p}}G^R_{\mathbf{p}}G^A_{\mathbf{Q-p}} C^0_{\mathbf{p,k'}} =
$\rho(0) \int \frac{d\phi_\mathbf{p}}{2\pi}C_{\mathbf{k,p}}C^0_{\mathbf{p,k'}}\frac{2\pi\tau}{1-i\omega\tau+i\tau \mathbf{v_p\cdot Q}}$. Similar to Eq.~\eqref{eq:angular_harmonics_diffuson}, expanding the bare and dressed Cooperons in angular harmonics reduces Eq.~\eqref{eq:BS_Cooperon} to a matrix equation. Solving this equation at $\omega=0$, we obtain the Cooperon matrix element in the angular-harmonic basis as (details in Appendix.~\ref{sec_supp:Cooperon})
\begin{gather}
C_{m,n}=\frac{C^{0}_{m,n}\delta_{m,n}}{\Delta_n+D_{n}Q^2\tau},\label{eq:Cooperon_comps}
\end{gather}
where $Q=|\mathbf{Q}|$, and $m$ and $n$ labels its angular harmonics. The bare Cooperon $C^0_{n,n}$ determines which channels are present and sets their respective weights, while the Cooperon gap $\Delta_n$ and diffusion stiffness $D_n$ also depend on the channel index $n$. Summing over the angular channels gives the complete Cooperon propagator
\begin{gather}
    C_{\mathbf{Q-k},\mathbf{k'}}(\mathbf{Q})|_{\mathbf{k,k'}\in\mathrm{FS}}=\sum_n\frac{C_{n,n}^{0}(-1)^ne^{in\phi_{\mathbf{k,k}'}}}{\Delta_n+D_nQ^2\tau}\label{eq:dressed_C_expansion}.
\end{gather}
Note that the odd and even $n$ channels contribute with alternating signs. This behavior becomes evident in their respective contributions to the quantum-interference correction. For the $\lambda$-jellium model, $n=0,1,2$ channels contribute, since the corresponding bare Cooperon $C_{n,n}^{0}$ are nonzero. The respective Cooperon gaps are,
\begin{gather}
    \Delta_n \equiv
    \begin{dcases}
        \frac{\beta^2}{1+\beta^2} & :n=0,\\
        \frac{(1-\beta)^2}{1+\beta^2} & :n=1,\\
        \frac{1}{1+\beta^2} & :n=2.
\end{dcases}\label{eq:Cooperon}
\end{gather}
and are shown in Fig.~\ref{fig:fig_1_summary}~(b). These gaps close at distinct values of $\beta$: $\Delta_0=0$ at $\beta=0$, $\Delta_1=0$ at $\beta=1$, and $\Delta_2\to0$ as $\beta\to\infty$. In the first and last limits, the Bloch states are orbital-polarized. By contrast, the gap closing at $\beta=1$ occurs when the two-orbitals have equal weights. 

\begin{figure}
    \centering
    \includegraphics[scale=1]{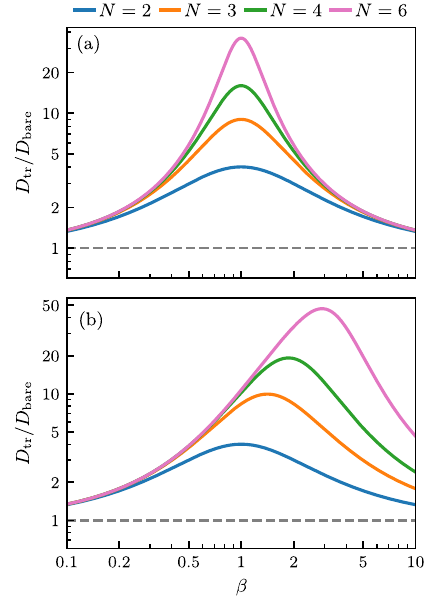}
    \caption{Transport diffusion coefficient $D_{\mathrm{tr}}/D_{\text{bare}}$ for: (a) ideal RNG model, and (b) $\lambda_N$-model for different values of $N$. For $N=2$, both models reduce to the $\lambda$-jellium model.}
    \label{fig:fig_5_diffsion_const}
\end{figure}
\subsection{Quantum interference correction to the conductivity} \label{sec:quantum_correction}
We now use the Cooperon propagator to evaluate the quantum-interference correction to the dc conductivity. 
Diagrammatically, the Cooperon enters the conductivity through the Hikami-boxes, and has two contributions 
\begin{gather}
\delta\sigma_i = \delta\sigma_{1,i} + 2\delta\sigma_{2,i},   
\end{gather}
where $\delta\sigma_{1,i}$ and $\delta\sigma_{2,i}$, respectively, come from the bare and dressed Hikami boxes \cite{Akkermans_Montambaux_2007, Hikami_1980}. The corresponding diagrams are shown in Fig.~\ref{fig:fig_2_diagrams}~(d), and are expressed as
\begin{align}
    \delta\sigma_{1,i}&=\frac{e^2\eta^2}{2\pi\mathcal{V}^2}\sum_{\mathbf{k,Q}}v_{\mathbf{k}}^iv_{\mathbf{-k}}^i G^R_{\mathbf{k}} G^A_{\mathbf{k}} G^R_{\mathbf{-k}} G^A_{\mathbf{-k}} C_{\mathbf{Q-k,k}}, \\
    \delta\sigma_{2,i}&=\frac{e^2\eta^2}{2\pi \mathcal{V}^3}
    \sum_{\mathbf{k,p,Q}}  
    {v}^i_{\mathbf{k}}{v}^i_{\mathbf{-p}}
    G^A_{\mathbf{k}} G^A_{\mathbf{-p}}  G^R_{\mathbf{k}} G^R_{\mathbf{p}}\nonumber\\ &\qquad\qquad\qquad \times G^R_{\mathbf{-k}} G^R_{\mathbf{-p}} C^0_{\mathbf{p,k}} C_{\mathbf{Q-k,p}}.\label{eq:conductivity_correction}
\end{align}
The bare Hikami box contains the Cooperon $C_{\mathbf{Q-k},\mathbf{k}}$, which connects nearly opposite momenta and therefore contributes to the exact-backscattering contribution to the interference correction. By contrast, the dressed Hikami box contains $C_\mathbf{Q-k,p}$ with independent momenta $\mathbf{k}$ and $\mathbf{p}$, and accounts for non-backscattering contributions. Although the dressed Hikami box contains an additional impurity line, the additional Green functions compensate for the extra disorder vertex, so that it contributes at the same order as the bare Hikami box. Because the Cooperon $C_{\mathbf{Q-k},\mathbf{p}}$ depends both on $\beta$ and the angular channel $n$, the relative weights of the dressed and bare Hikami boxes vary between channels (details in Appendix. \ref{sec_supp:conductivity_correction}). This contrasts with analogous calculations in graphene \cite{mccannWeakLocalizationMagnetoresistanceValley2006} and nodal-line semimetals \cite{chenWeakLocalizationAntilocalization2019}, where the dressed Hikami-box differs from the bare by a constant factor of $-1/4$. Substituting the Cooperon expansion from Eq.~\eqref{eq:dressed_C_expansion}, we obtain (details in Appendix \ref{sec_supp:conductivity_correction})
\begin{align}
    \delta \sigma_i&=-\frac{e^2\eta^2\gamma^2}{4\pi^2 }\sum_{n} \left\{\frac{(-1)^n[\mathcal{F}^2]_{n,n}}{D_n/D_{\mathrm{bare}}}\right.\nonumber\\
    &\quad\times\left[1-\frac{\gamma}{2} ([\mathcal{F}^2]_{n+1,n+1}+[\mathcal{F}^2]_{n-1,n-1})\right]\nonumber\\
    &\quad\times \left. \log \left[\frac{\Delta_n+D_n/(2\gamma\, D_{\mathrm{bare}})}{\Delta_n+D_n/(2\Lambda^2\gamma\, D_{\mathrm{bare}})} \right] \right\}\label{eq:delta_sigma_main_text},
\end{align}
where 
$[\mathcal{F}^2]_{n,n} = C_{n,n}^{0}/(n_\mathrm{imp}u_0^2)$ is proportional to the bare Cooperon in channel $n$. $[\mathcal{F}^2]_{n,n}$ denotes the $n$-th angular harmonic of the Fermi surface form factor,
\begin{gather}
    [\mathcal{F}^2]_{n,n}=\int \frac{d\phi_{\mathbf{k},\mathbf{k'}}}{2\pi} \mathcal{F}(k_F,\phi_{\mathbf{k},\mathbf{k'}})^2e^{-in\phi_{\mathbf{k},\mathbf{k'}}}.\label{eq_supp:comp_F2_nn}
\end{gather}
Here the $\mathbf{Q}$-summation is bounded above by the inverse mean free path $l_0^{-1}$, where $l_0=v_F\tau$, and from below by $L_{\max}^{-1}$, where $L_{\text{max}}=\min\{L,L_{\phi} \}$ is the smaller of the sample size $L$ or the coherence length $L_\phi$. For the plots of $\delta\sigma_i$, we use $\Lambda=L_{\max}/l_0\sim10^4$ as the infrared cutoff regulating the logarithmic divergence.

In Fig.~\ref{fig:fig_1_summary}~(a), we show the quantum-interference correction to the conductivity for the $\lambda$-jellium model, with the corresponding Cooperon gaps shown in Fig.~\ref{fig:fig_1_summary}~(b). At each of the three gap closing points, $\delta\sigma$ recovers the logarithmic scaling characteristics of two-dimensional quantum interference such that, $\delta\sigma \propto(-1)^{n+1}\log\Lambda$, where $n$ labels the Cooperon channel which becomes gapless. For the even channels $n=0,2$, whose gaps vanish in the limits of $\beta\to0$ and $\beta\to\infty$, respectively, the correction is negative which indicates WL. This is expected, as in these orbital-polarized limits, the quantum geometric effects vanish and the system effectively reduces to ordinary jellium. Between these two limits, we observe that tuning the quantum geometry using $\beta$ can reverse the sign of the correction, making it positive near $\beta = 1$. At $\beta=1$, the $n=1$ Cooperon gap $\Delta_1$ vanishes, giving rise to WAL. At this point, the Bloch state carries equal weight on the two orbitals. In Sec.~\ref{sec:antiunitary_symm}, we show that this reflection symmetry of the orbital weights provides the exact criterion for a Cooperon channel to become gapless.

\subsection{An emergent antiunitary symmetry}\label{subsec:antiunitary_symm}

By construction, the Hamiltonian in Eq.~\eqref{eq:Hamiltonian_lambda_jellium} breaks physical TRS, $T\mathcal{H}_{\mathbf{k}}^{\lambda}T^{-1}\neq \mathcal{H}_{\mathbf{-k}}^{\lambda}$, where $T=\mathcal{K}$ is the spinless time-reversal operator and $\mathcal{K}$ denotes complex conjugation. However, at $\beta=1$, the Hamiltonian projected onto the active band possesses an emergent antiunitary symmetry. To make this explicit, we define
\begin{gather}
    \mathcal{T} = -i\sigma^y_{\text{orb}} \mathcal{K},\label{eq:2x2antiunitaryoperator}
\end{gather}
where $\sigma^y_{\text{orb}}$ acts within the orbital subspace. Using the the active-band projector $P^{\lambda}_{\mathbf{k}} = \ket{s^0_\mathbf{k}}\bra{s^0_\mathbf{k}}$ constructed from the Bloch states in Eq.~\eqref{eq:Bloch_state_jellium}, the projected Hamiltonian is $\widetilde H_\mathbf{k} = P_\mathbf{k}^\lambda H_\mathbf{k} P_\mathbf{k}^\lambda = \xi_{\mathbf{k}} P_\mathbf{k}^\lambda$. Since the dispersion is inversion symmetric ($\xi_\mathbf{k} = \xi_\mathbf{-k}$), a symmetry of the projector $\mathcal{T}P_\mathbf{k}^\lambda\mathcal{T}^{-1} = P_\mathbf{-k}^\lambda$ implies the corresponding symmetry in the projected Hamiltonian $\mathcal{T} \widetilde{H}_{\mathbf{k}}\mathcal{T}^{-1} = \widetilde{H}_\mathbf{-k}$. Explicit evaluation gives (details in Appendix~\ref{sec_supp:effective_TRS})
\begin{align}
    &\quad \mathcal{T}P_\mathbf{k}^\lambda\mathcal{T}^{-1} 
    = \begin{cases}
        P_{-\mathbf{k}}^{\lambda},
        & \displaystyle \frac{\mathcal{B}k^2}{2}=1,\\[2mm]
        \neq P_{-\mathbf{k}}^{\lambda},
        & \displaystyle  \frac{\mathcal{B}k^2}{2}\neq 1.
        \end{cases}
\end{align}
Within the on-shell approximation ($|\mathbf{k}|=k_F$) used in the Cooperon calculation, this condition becomes
$\mathcal{B}k_F^2/2=\beta=1$. Thus, at this point, the on-shell active band states have an emergent antiunitary symmetry $\mathcal{T}$, even though physical TRS remains broken. Further, from Eq.~\eqref{eq:2x2antiunitaryoperator} the operator satisfies 
\begin{gather}
    \mathcal{T}^2=-\mathds{1}_{2\times 2}.
\end{gather}
Since particle-hole and chiral symmetries are absent, the projected on-shell Hamiltonian has the symplectic structure of AZ class AII \cite{Altland1997}. Note here that the AZ classification is based on the emergent antiunitary symmetry $\mathcal{T}$, and not by physical time-reversal $T$, which remains broken. This emergent symmetry protects the gapless $n=1$ Cooperon channel at $\beta=1$ and gives rise to WAL \cite{hikamiAndersonLocalizationNonlinearsigmamodel1981}. In Sec.~\ref{sec:antiunitary_symm}, we demonstrate how the occurrence of a gapless Cooperon mode is tied to the presence of such an emergent antiunitary symmetry at the Fermi surface in models with more than two orbitals.

\section{Models with more than two orbitals}\label{sec:N>2_models}
We now extend our analysis to multiorbital models while retaining the parabolic active-band dispersion and scalar disorder potential introduced in Sec.~\ref{sec:model}. We introduce the two band-projected models with nonuniform quantum geometry: the ``ideal model of RNG'' \cite{tan2025ideallimitrhombohedralgraphene} and the ``$\lambda_N$-model''  \cite{desrochersElectronicCrystalPhases2026}, both motivated by the quarter-metal state of $N$-layer rhombohedral graphene (RNG) \cite{zhouHalfQuartermetalsRhombohedral2021}. These spinless models contain $N$ internal degrees of freedom, which microscopically correspond to the different layers of RNG. For simplicity, we refer to these degrees of freedom as ``orbitals'' throughout the following discussion. For $N=2$, both models reduce to the $\lambda$-jellium model which was introduced in Sec.~\ref{sec:model}.

\subsection{Ideal model of RNG}
\label{sec:ideal_model}
This $N$-orbital model is derived from a minimal model of RNG in an experimentally relevant displacement-field regime, retaining the nearest-neighbor intralayer hopping and the dominant interlayer hopping between adjacent layers. It further assumes that the $A$ and $B$ sublattices on the bottom layer are decoupled. This idealization generates a zero mode for the model that only has support on the $A$ sublattices \cite{tan2025ideallimitrhombohedralgraphene}. The corresponding Bloch-state components of this zero mode acquire successive momentum-space windings $(k_x-ik_y)^n$ across the $N$ layers. The result is a model with an ideal quantum geometry, and a nonuniform Berry curvature $\Omega_{\mathbf{k}}$, with a characteristic ring-like profile centered around $\mathbf{k}=0$, as shown in Fig.~\ref{fig:omega_profile}~(a). The zero mode, which forms the active band, has the following Bloch state
\begin{gather}
\ket{s^0_{\mathbf{k}}}^{\mathrm{ideal\ RNG}}
=
\frac{1}{
\sqrt{\displaystyle\sum_{n=0}^{N-1}\beta_k^n}
}
\begin{pmatrix}
1\\
\sqrt{\beta_k}\,e^{-i\phi_{\mathbf k}}\\
\beta_k e^{-2i\phi_{\mathbf k}}\\
\vdots\\
\beta_k^{(N-1)/2}e^{-i(N-1)\phi_{\mathbf k}}
\end{pmatrix},\label{eq:Bloch_ideal_RNG}
\end{gather}
where $\beta_k=\mathcal{B}|\mathbf{k}|^2/2$. As seen in Fig.~\ref{fig:omega_profile}~(a), the parameter $\mathcal{B}$ controls the value of $\Omega_N(\mathbf{k})$ at $\mathbf{k}=0$, while increasing $N$ enhances the prominence of the ring of Berry curvature. The active band has a Chern number $\mathcal{C}=N-1$. The conductivity correction computed using the form factors of this model is discussed in Sec.~\ref{sec:results_N_orb_models}.

\subsection{$\lambda_N$-model}
The $\lambda_N$-model is an $N$-orbital generalization of the $\lambda$-jellium model, constructed such that its form factor continuously approach that of the lowest Landau level (LLL) as $N\to\infty$. The model can be related to a modified minimal model of RNG in a displacement field. In the low-energy regime, the lowest electronic band is approximately polarized onto the $A$ sublattice, allowing the $B$ sublattice to be integrated out. Further by taking the intralayer and interlayer hoppings to be layer dependent, the resulting effective single-sublattice model supports a zero mode whose quantum geometry continuously approaches the uniform LLL limit as $N$ increases \cite{desrochersElectronicCrystalPhases2026}.
In this case, the form factor can be expressed as the product of the LLL form factor  ${F}^{LLL}_{\mathbf{k},\mathbf{k}'}=\exp{\left(-\frac{\mathcal{B}}{4}\left(|\mathbf{k}-\mathbf{k}'|^2+2i\mathbf{k}\wedge \mathbf{k'} \right) \right)}$, where $\mathbf{k}\wedge \mathbf{k'}=k_xk_y'-k_yk_x'$, and an $N$-dependent prefactor $\mathcal{M}^N_{\mathbf{k},\mathbf{k}'}$,
\begin{gather}
\mathcal{F}^{\lambda_N}_{\mathbf{k},\mathbf{k}'}=\mathcal{M}^N_{\mathbf{k},\mathbf{k}'}\mathcal{F}^{LLL}_{\mathbf{k},\mathbf{k}'}\label{eq:L_N_ff}.
\end{gather}
The prefactor is expressed in terms of the upper incomplete gamma function $\Gamma(N,z)$ as follows
\begin{gather}
\mathcal{M}^N_{\mathbf{k},\mathbf{k}'}=\frac{\Gamma(N,\mathcal{B}k_zk_z'^*)}{\sqrt{\Gamma(N,\mathcal{B}|\mathbf{k}|^2/2)\Gamma(N,\mathcal{B}|\mathbf{k}'|^2/2)}},\label{eq:M_N_factor}
\end{gather}
where $k_z=(k_x+ik_y)/\sqrt{2}$. In the limit $N\to \infty$, we find $\mathcal{M}^N_{\mathbf{k},\mathbf{k}'}\to 1$, recovering the uniform quantum geometry of the LLL. The structure of Eq.~\eqref{eq:M_N_factor} produces a characteristic factorial dependence of the orbital weights in the Bloch state for the zero mode,
\begin{gather}
\ket{s^0_{\mathbf{k}}}^{\lambda_N}
=
\frac{1}{
\sqrt{\displaystyle\sum_{n=0}^{N-1}\frac{\beta_k^n}{n!}}
}
\begin{pmatrix}
1\\
\sqrt{\beta_k}\,e^{-i\phi_{\mathbf{k}}}\\[2pt]
\dfrac{\beta_k e^{-2i\phi_{\mathbf{k}}}}{\sqrt{2!}}\\
\vdots\\
\dfrac{\beta_k^{(N-1)/2}e^{-i(N-1)\phi_{\mathbf{k}}}}{\sqrt{(N-1)!}}
\end{pmatrix}.\label{eq:Bloch_L_N}
\end{gather}
Note that the difference from the ideal RNG model is only through the additional factorial factors. The Berry-curvature profile $\Omega_N(\mathbf{k})$ for this model is shown in Fig.~\ref{fig:omega_profile}~(b). It exhibits a plateau centered at $\mathbf{k}=0$, where the parameter $\mathcal{B}$ controls its height. The plateau broadens with increasing $N$, reflecting the gradual approach towards uniform quantum geometry. Similar to the ideal RNG model, the active band also has a Chern number $\mathcal{C}=N-1$. The resulting conductivity correction is discussed in Sec.~\ref{sec:results_N_orb_models}. 

\section{Gapless Cooperons and Emergent Antiunitary Symmetry}\label{sec:antiunitary_symm}

\begin{figure*}
    \centering
    \includegraphics[scale=1]{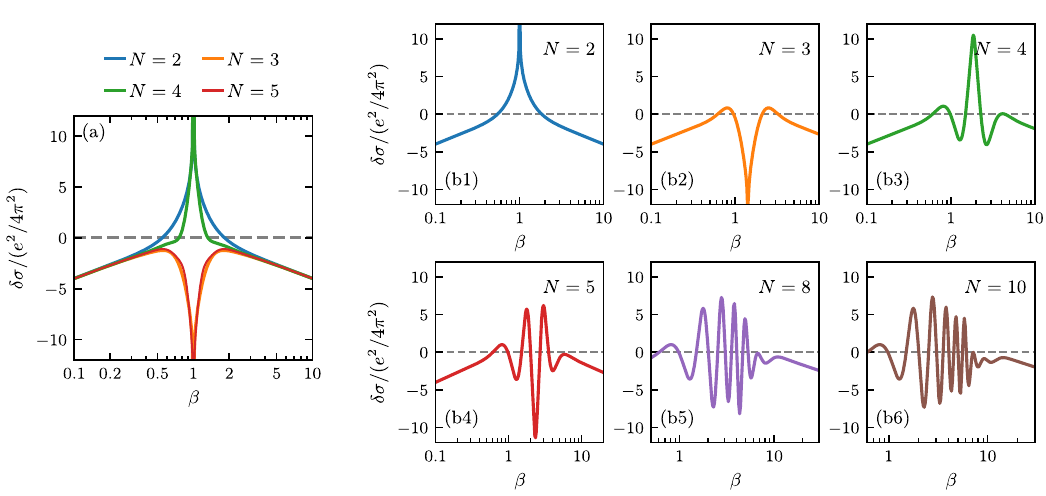}
    \caption{The normalized quantum-interference correction to the dc conductivity, $\delta\sigma/(e^2/4\pi^2)$, in: (a) the ideal model of RNG and (b) the $\lambda_N$-model. At $\beta=1$, the ideal RNG model exhibits weak localization (WL) for odd $N$ and weak anti-localization (WAL) for even $N$. The $\lambda_N$-model exhibits WAL for $N=2$ and WL for $N=3$. For $N>3$, no logarithmically divergent corrections are present.}
    \label{fig:fig_3_quantum correction_N>2_models}
\end{figure*}

Now we derive a general condition for the closing of a Cooperon gap in rotationally symmetric $N$-orbital models of quarter-metal RNG, expressed in terms of the orbital weights of the Bloch states.
We show that the same condition identifies an emergent antiunitary symmetry that protects the gapless Cooperon mode. 

\subsection{Condition for a gapless Cooperon}
We consider an active band with a circular Fermi surface. Its Bloch states have $N$-orbital components and takes the following form when the momentum is restricted to the Fermi surface $|\mathbf{k}|=k_F$,
\begin{gather}
    \ket{s^0_{\mathbf{k}}} = \ket{s^0_{k_F}(\phi_{\mathbf{k}})} = \sum_{m=0}^{N-1}\sqrt{f_m}e^{-im\phi_{\mathbf{k}}}\ket{m}.
    \label{eq:bloch_states_fourier_expansion}
\end{gather}
Here, $\{\ket{m}\}$ is an orthonormal orbital basis, and $\sqrt{f_m}$ denotes the weight of orbital $m$ on the Fermi surface, with the corresponding Bloch state component carrying the angular dependence $e^{-im\phi_\mathbf{k}}$. This is the orbital basis in which the ideal RNG and $\lambda_N$-models are naturally defined.

The Cooperon gap of the $n$-th angular channel can then be written as follows (see Appendix~\ref{sec_supp:Cooperon})
\begin{gather}
    \Delta_n = 1 - \gamma[\mathcal{F}^2]_{n,n},
\end{gather}
where we used the definitions $\gamma^{-1} = \sum_{m=0}^{N-1} f_m^2$ and $ [\mathcal{F}^2]_{n,n} = \sum_{m=0}^{N-1} f_m f_{n-m}$. Since the model has $N$-orbitals, $f_m=0$ if  $m$ is outside of $\{0,\cdots, N-1\}$. Further, the Cauchy-Schwarz inequality gives $\gamma[F^2]_{n,n}\leq1$, and ensures that the Cooperon gaps are non-negative. The gaps can close if and only if the following condition is satisfied for all $m$,
\begin{gather}
    \Delta_n = 0 \Longleftrightarrow f_{n-m} = f_m, \quad \forall m\in\{0,\cdots,N-1\}.\label{eq:cooperon_gap_condition}
\end{gather}
For nonzero orbital weights, this condition requires $n=N-1$. Therefore, the gap $\Delta_{N-1}$ closes when the orbital weights of the Bloch-states on the Fermi surface, $\{f_0,\ldots,f_{N-1}\}$, are reflection symmetric about $m=(N-1)/2$. However, all Cooperon channels with $n\neq N-1$ always remain gapped from above condition.

\subsection{Constructing an antiunitary operator}\label{sec:const_antiunitay_op}
The reflection symmetry of the orbital weights at the gapless point can be associated with an antiunitary operator $\mathcal{T}_{N}$ acting within the Hilbert space spanned by the orbital basis states restricted to the Fermi surface. The antiunitary operator is defined as
\begin{gather}
\mathcal{T}_{N}=\mathcal{R}_{N}\mathcal{K}, \label{eq:emergent_TRS_def}
\end{gather}
where $\mathcal{K}$ denotes complex conjugation. $\mathcal{R}_N$ is a unitary operator (see Appendix~\ref{sec_supp:effective_TRS}) that reflects the orbital index about $m=(N-1)/2$, and is defined as follows
\begin{gather}
\mathcal{R}_N\ket{m} = (-1)^m\ket{N-1-m}.\end{gather}
When the orbital weights satisfy the reflection condition in Eq.~\eqref{eq:cooperon_gap_condition}, this operator $\mathcal{T}_N$ maps the Bloch state on the Fermi surface at $\mathbf{k}$ to its ``effective'' time-reversed partner at $-\mathbf{k}$,
\begin{align} 
    \mathcal{T}_N\ket{s^0_{k_F}(\phi_{\mathbf{k}})} = (-1)^{N-1} e^{i(N-1)\phi_{\mathbf{k}}}
    \ket{s^0_{k_F}(\phi_{\mathbf{-k}})}. \label{eq:symmetry_sewing_relation}
\end{align}
However, microscopic time-reversal $T$ would map $\ket{s^0_{k_F}(\phi_{\mathbf{k}})}$ to its complex conjugate $\ket{s^0_{k_F}(\phi_{\mathbf{k}})}^*$, which is distinct from $\ket{s^0_{k_F}(\phi_{-\mathbf{k}})}$ in these TRS-broken models. Evaluating the active-band projector $P_{\mathbf{k}}$ using Eq.~\eqref{eq:symmetry_sewing_relation}, we find that the phase factors cancel out in 
$P_{\mathbf{k}}=\ket{s^0_{k}(\phi_{\mathbf{k}})}\bra{s^0_{k}(\phi_{\mathbf{k}})}$, leading to the following symmetry
\begin{gather}
    \mathcal{T}_N P_{\mathbf{k}}\mathcal{T}_N^{-1} = P_{\mathbf{-k}}, \quad\mathbf{k}\in\mathrm{FS}.
\end{gather}
Similar to Sec.~\ref{subsec:antiunitary_symm}, here $\mathcal{T}_N$ describes an emergent antiunitary symmetry of projected the on-shell Hamiltonian. Additionally, the definition in Eq.~\eqref{eq:symmetry_sewing_relation} implies that
\begin{gather}
    \mathcal{T}_N^2=(-1)^{N-1}\label{eq:T_n^2_action},
\end{gather}
provided that a Cooperon is gapless. For $\mathcal{T}_N^2=+1$, the system therefore belongs to the orthogonal class (AI) associated with WL. Whereas $\mathcal{T}_N^2=-1$ corresponds to the symplectic class (AII) associated with WAL \cite{hikamiAndersonLocalizationNonlinearsigmamodel1981}. Hence, when a Cooperon is gapless as implied by Eq.~\eqref{eq:cooperon_gap_condition}, an emergent antiunitary symmetry $\mathcal{T}_N$ arises that protects the gapless Cooperon mode. At this gapless point, Eq.~\eqref{eq:T_n^2_action} predicts WAL for models with an even number of orbitals $N$, and WL for models with odd $N$. In Sec.~\ref{sec:Berry_phase}, we provide a complementary physical interpretation of WL and WAL in terms of the Berry phases accumulated along ``effective'' time-reversed scattering trajectories.

More broadly, antiunitary symmetries control the sign of quantum-interference corrections in a variety of systems. On the surface of a three-dimensional topological insulator, spin-momentum locking together with $\mathcal{T}^2=-1$ gives rise to WAL \cite{Tkachov_2011,Lu_2011,Chen_2010,He_2011}. By contrast, in magnetic Euler bands, an emergent effective TRS allows WL to persist despite broken microscopic TRS \cite{kim2026weaklocalizationmagneticeuler}, while in nodal-line semimetals, distinct antiunitary symmetries can give rise to WL or WAL in different interference channels \cite{chenWeakLocalizationAntilocalization2019}. 

\begin{figure*}
    \centering
    \includegraphics[scale=1]{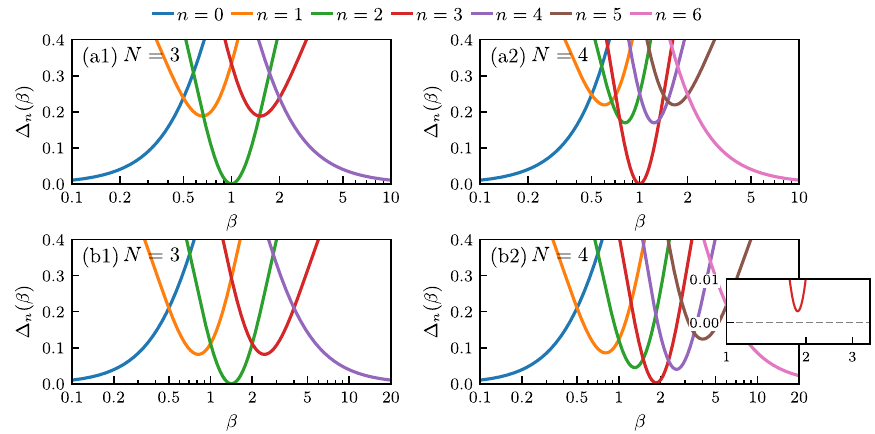}
    \caption{The Cooperon gaps $\Delta_n(\beta)$ in the angular channels for: (a1-a2) the ideal RNG model, and (b1-b2) the $\lambda_N$-model. In the ideal RNG model, the Cooperon channel $\Delta_{n=N-1}$ becomes gapless at $\beta=1$ for all $N$. In the $\lambda_N$-model with $N>3$, all Cooperon channels remain gapped (see inset) at finite $\beta$. Both models have gapless Cooperon in the trivial limits of $\beta\to0$ and $\beta\to\infty$, where the Bloch states become orbital polarized.}
    \label{fig:fig_4_Cooperon_channels}
\end{figure*}

\section{Quantum-interference correction in models with more than two orbitals}\label{sec:results_N_orb_models}
We now analyze the conductivity correction in the two multiorbital models introduced in Sec.~\ref{sec:N>2_models}, and interpret the results using the Cooperon gap-closing criterion and emergent antiunitary symmetries discussed in Sec.~\ref{sec:antiunitary_symm}.

\subsection{Ideal RNG model}
To begin, we first examining the behavior of the Drude conductivity in the ideal RNG model. From the Bloch state of this model in Eq.~\eqref{eq:Bloch_ideal_RNG}, we find that at $\beta=\mathcal{B}k_F^2/2=1$, all orbitals have equal weights. The form factor can then be expressed as follows
\begin{gather}
    |\mathcal{F}_\mathbf{k,k'}|^2 = \frac{1}{N^2}\frac{\sin^2(N\phi_\mathbf{k,k'}/2)}{\sin^2(\phi_\mathbf{k,k'}/2)}.
\end{gather}
As $N$ increases, $|\mathcal{F}_\mathbf{k,k'}|^2$ becomes increasingly peaked near $\phi_{\mathbf{k},\mathbf{k}'}=0$. From Eq.~\eqref{eq:eta_general}, we see that this favors forward-scattering processes, and therefore enhances the transport diffusion coefficient $D_{\text{tr}}$, and also the Drude conductivity $\sigma^{\text{Drude}}$. At $\beta=1$, this enhancement scales as $D_\mathrm{tr}/D_\mathrm{bare}=N^2$. As shown in Fig.~\ref{fig:fig_5_diffsion_const}~(a), $D_{\mathrm{tr}}$ reaches its maximum at this point and decreases on either side as $\beta$ is tuned away from unity.

We next analyze the quantum-interference correction to the conductivity $\delta\sigma$ which is shown in Fig.~\ref{fig:fig_3_quantum correction_N>2_models}~(a). For even $N$, $\delta\sigma$ changes sign as a function of $\beta$, changing from negative to positive values near $\beta=1$. This reveals a crossover from electron-localization to anti-localization tendencies. In contrast, for odd $N$, electron-localization tendencies persist throughout the range of $\beta$. At the point $\beta=1$, a clear even–odd trend emerges as $N$ is varied: it exhibits WAL for even $N$, and WL for odd $N$. This even-odd effect can be explained based on the discussions in Sec.~\ref{sec:antiunitary_symm}. From Eq.~\eqref{eq:Bloch_ideal_RNG}, the orbital weights on the Fermi surface for this ideal RNG model are
\begin{gather}
    f_n^{\text{ideal RNG}}=\frac{\beta^n}{\sum_{m=0}^{N-1}\beta^m}.
\end{gather}
At $\beta=1$, all orbital weights are equal and satisfy the condition in Eq.~\eqref{eq:cooperon_gap_condition}, which ensures a gapless Cooperon for all values of $N$ at this point. The resulting emergent antiunitary symmetry $\mathcal{T}_N$ protects the logarithmically divergent quantum interference correction. From Eq.~\eqref{eq:T_n^2_action}, we have $\mathcal{T}_N^2=(-1)^{N-1}$, and this naturally explains the even-odd effect. Further, consistent with our earlier prediction we find that the $n=N-1$ Cooperon channel becomes gapless precisely at $\beta=1$, or
$\Delta_{n=N-1}(\beta=1)=0$. This is shown in Figs.~\ref{fig:fig_4_Cooperon_channels}~(a1,a2) for $N=3$ and $N=4$ respectively. Additionally, we find that a Cooperon gap also closes in the trivial limits: $\beta\rightarrow 0$ and $\beta\rightarrow\infty$, where the conventional WL of ordinary jellium is recovered, with the $n=0$ and $n=2N-2$ channels becoming gapless, respectively. 

\subsection{$\lambda_N$-model}
In the $\lambda_N$-model, the transport diffusion coefficient $D_{\mathrm{tr}}$, and hence the Drude conductivity, exhibits a qualitatively different behavior from that of the ideal RNG model and is shown in Fig.~\ref{fig:fig_5_diffsion_const}~(b). The additional factorials in the Bloch-states $\ket{s^0_{\mathbf{k}}}^{\lambda_N}$, and the form factors, suppress the weight of higher orbitals for small values of $\beta$. With increasing $N$ increases, a larger $\beta$ is required for multiple orbitals to acquire appreciable weights. As a result, the peak in $D_{\mathrm{tr}}$ associated with strongly enhanced forward scattering shifts to larger values of $\beta$.

\begin{figure}
    \centering
    \includegraphics[scale=1]{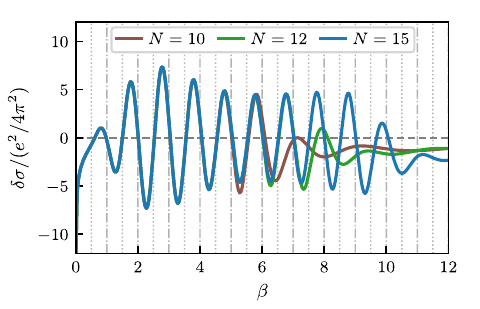}
    \caption{Oscillations in the conductivity correction $\delta\sigma$ with increasing $\beta$. In the limit of uniform quantum-geometry ($N\gg2\beta$) the oscillations exhibit a uniform period of $\Delta\beta_{\text{osc}}=1$. Vertical dotted lines mark odd integer values of $\beta$, and dash-dotted lines mark even integer values. The uniform oscillations persist to larger values of $\beta$ with increasing $N$.}
    \label{fig:oscillations}
\end{figure}

A similar trend appears for the Cooperon gaps, whose minima shift to larger values of $\beta$ with increasing $N$, as shown in Fig.~\ref{fig:fig_4_Cooperon_channels}~(b1,b2). We find that the $N=3$ model is special as a Cooperon gap closes at $\beta=\sqrt{2}$, producing the WL behavior seen in Fig.~\ref{fig:fig_3_quantum correction_N>2_models}~(b2). For $N\geq4$, by contrast, all Cooperon channels remain gapped as shown in Fig.~\ref{fig:fig_4_Cooperon_channels}~(b2). Nevertheless, the Cooperon gaps can become soft as shown in the inset. These give rise to the sharp but finite oscillatory features in $\delta\sigma$ seen in Figs.~\ref{fig:fig_4_Cooperon_channels}~(b3-b6) and Fig.~\ref{fig:oscillations}, rather than a true divergence.

All of the above features can be explained using the Cooperon gap-closing condition in Eq.~\eqref{eq:cooperon_gap_condition}. The $\lambda_N$-model has different orbital weights on the Fermi surface, and from Eq.~\eqref{eq:Bloch_L_N} we find
\begin{gather}
    f^{\lambda_N}_n=\frac{\beta^n/n!}{\sum_{m=0}^{N-1}\beta^m/m!}.
\end{gather}
The gap closing of $\Delta_{N-1}$, requires 
$
    \beta^{N-1-2n}=(N-1-n)!/n!,
$
for all $n\in \{0,\cdots,N-1\}$. At finite $\beta$, these conditions can be satisfied only for $N=2$ (equivalent to $\lambda$-jellium) and $N=3$. For $N=3$, the symmetry $\mathcal{T}_3$ is associated with the gapless Cooperon mode at $\beta=\sqrt{2}$, where $\Delta_2(\beta=\sqrt{2})\to0$. Since $\mathcal{T}_3^2=+1$, this gap closing gives rise to WL as shown in Fig.~\ref{fig:fig_3_quantum correction_N>2_models}~(b2). For $N>3$, all Cooperon channels remain gapped, consistent with Fig.~\ref{fig:fig_4_Cooperon_channels}~(b1-b2). This also explains the absence of a WL or WAL correction in Fig.~\ref{fig:fig_3_quantum correction_N>2_models}~(b3-b6). 

In general, for fixed $N$, increasing $\beta$ produces oscillations in $\delta\sigma$, characterized by $N-1$ peaks and $N$ troughs, and $N$ sign changes. These oscillations, shown in Figs.~\ref{fig:fig_3_quantum correction_N>2_models}~(b), arise due to the redistribution of weights among Cooperon channels that respectively suppress and enhance the conductivity. Owing to the $(-1)^{n+1}$ factor in Eq.~\eqref{eq:delta_sigma_main_text}, even-$n$ channels contribute negatively to $\delta\sigma$, while odd-$n$ channels contribute positively. The total correction depends on the relative contribution between these two groups at a given value of $\beta$.

When the Fermi surface samples regions of approximately uniform quantum geometry, corresponding to the limit $N\gg2\beta$ in the $\lambda_N$-model, the oscillations acquire a nearly constant period, $\Delta\beta_{\mathrm{osc}}\simeq1$, as seen in Fig.~\ref{fig:oscillations}. Here, the bare Cooperon weights approach a Poisson distribution centered at the index $n=2\beta$ (details in Appendix~\ref{sec_app:oscillations}). At $\beta=n/2$, the bare Cooperon weights are approximately reflection symmetric about $n-1/2$, with channels $n-1-m$ and $n+m$ carrying equal weights. A similar behavior is also observed in the gaps $\Delta_n$ and the diffusion stiffness $D_n$ (details in Appendix~\ref{sec_app:oscillations}). Therefore, these paired channels contribute with comparable magnitudes but opposite signs to the dc conductivity, and therefore largely cancel. Increasing $\beta$ away from $n/2$ gradually transfers orbital weight to higher orbital indices, while also shifting the dominant conductivity contributions towards higher $n$. The resulting imbalance between the even and odd $n$ channels produces a net positive or negative correction to the conductivity. However, again at $\beta=(n+1)/2$, the contributions from the even and odd $n$ channels approximately cancel. Moving away from this value once more creates an imbalance between the two parities, but now with the opposite sign to the correction. Repetition of this process generates the oscillatory behavior, with successive sign reversals separated by approximately $\Delta\beta\simeq1/2$, as illustrated in Fig.~\ref{fig:oscillations}.

\subsection{Berry phase picture of WL/WAL}\label{sec:Berry_phase}
In the conventional jellium model, the occurrence of WL or WAL can be understood in terms of the relative phases accumulated along pairs of time-reversed scattering trajectories. In our model however, where TRS is broken, these trajectories are related by an ``effective'' TRS associated with an emergent antiunitary symmetry $\mathcal{T}_N$.

When a Cooperon channel becomes gapless, for instance at $\beta=1$ in the ideal RNG model, Eq.~\eqref{eq:cooperon_gap_condition} implies that the Fermi surface form factors can be written as follows
\begin{gather}
\mathcal{F}_{\mathbf{k},\mathbf{k}'}=\sum_{m=0}^{N-1}f_m e^{im\phi_{\mathbf{k},\mathbf{k'}}}=e^{i\frac{(N-1)}{2}\phi_{\mathbf{k},\mathbf{k'}}}S_{N}(\phi_{\mathbf{k},\mathbf{k'}}),\label{eq_supp:Berryphase_FF}
\end{gather}
where
$S_{N}(\phi_{\mathbf{k},\mathbf{k'}})=\sum_{m=0}^{N-1}f_m e^{i(m-\frac{(N-1)}{2})\phi_{\mathbf{k},\mathbf{k'}}}$, 
and $\phi_{\mathbf{k},\mathbf{k'}}$ is the angle between $\mathbf{k}$ and $\mathbf{k}'$. Note here that Eq.~\eqref{eq:cooperon_gap_condition} guarantees that $S_N(\phi)$ a real and even function, i.e. $S_{N}(\phi)=S_{N}(-\phi)$.
Now consider a scattering trajectory $P$ from $\mathbf{k}$ to $-\mathbf{k}$ through intermediate momenta $\mathbf{k}_1,\cdots,\mathbf{k}_n$. From Eq.~\eqref{eq:symmetry_sewing_relation}, the action of the aniunitary operator $\mathcal{T}_N$ relates trajectory $P$ with a partner trajectory $\overline{P}$, 
%which traverses the same sequence in reverse and with all momenta inverted,
obtained by reversing the sequence of scattering events and inverting all momenta,
\begin{align}
    P:&\,\,\mathbf{k}\rightarrow \mathbf{k}_1 \rightarrow \mathbf{k}_2\rightarrow\cdots \rightarrow\mathbf{k}_n\rightarrow \mathbf{-k},\nonumber\\
    \overline{P}:&\,\, \mathbf{k}\rightarrow -\mathbf{k}_{n} \rightarrow -\mathbf{k}_{n-1}\rightarrow\cdots \rightarrow-\mathbf{k}_1\rightarrow \mathbf{-k}.
\end{align} 
$\mathcal{T}_N$ therefore acts as an ``effective'' TRS for the Bloch states at the Fermi surface (on-shell) that contribute to transport. For a scalar disorder projected onto the active band, the scattering amplitude along a path contains products of the form factors. Using Eq.~\eqref{eq_supp:Berryphase_FF} we can express the scattering amplitude for path $P$ and  $\overline{P}$ as follows
\begin{align}
A_P&\propto
\prod_{j=0}^n
\mathcal{F}_{\mathbf{k}_{j+1},\mathbf{k}_{j}}= e^{i(N-1)\Theta_P/2}
    \prod_{j=0}^{n}
    S_{N}\left(\Delta\phi_j\right),\\
        A_{\overline{P}}&\propto
    e^{-i(N-1)\Theta_P/2}
    \prod_{j=0}^{n}
    S_{N}\left(-\Delta\phi_j\right).
\end{align}
where we defined $\mathbf{k}_0=\mathbf{k}$,
$\mathbf{k}_{n+1}=-\mathbf{k}$, $\Delta\phi_j=\phi_{\mathbf{k}_{j+1}}-\phi_{\mathbf{k}_j}$ and $\Theta_P=\sum_{j=0}^{n} \Delta\phi_j$. Since the trajectories connect $\mathbf{k}$ and $-\mathbf{k}$,
the accumulated angle is $\Theta_P=\pi$. Their amplitudes are related by a phase factor,
\begin{gather}
    A_{\overline{P}}=e^{-i(N-1)\Theta_P}A_P=(-1)^{N-1}A_{P}\label{eq:amp_ratios},
\end{gather}
where the total Berry phase enclosed by these ``effective'' time-reversed paths is $\Phi_B=(N-1)\pi$. Therefore, the backscattering probability resulting from the interference between trajectories $P$ and $\overline{P}$ is
\begin{gather}
I_{\text{backscatter}}\propto|A_P+A_{\overline{P}}|^2=\begin{cases}
    4|A_P|^2 \quad &N\, \text{is odd}\\
    0 \quad &N\, \text{is even.}
\end{cases}\label{eq:intensity}
\end{gather}
Compared with the incoherent sum $2|A_P|^2$, interference enhances the exact backscattering probability for odd $N$ and suppresses it for even $N$, giving rise to WL and WAL, respectively. This provides a more physical interpretation of the even-odd effect obtained from the diagrammatic calculation above.

\section{Discussion}\label{sec:discussion}
In this work, we showed that non-uniform quantum geometry can significantly modify electron localization in two-dimensional multiorbital disordered electron systems, by altering the quantum-interference correction to the conductivity. Even when microscopic TRS is broken, certain configurations of the band quantum geometry can give rise to an emergent antiunitary symmetry (or an ``effective'' TRS) that protects a gapless Cooperon mode. This restores the logarithmically divergent quantum-interference correction, leading to either WL or WAL. The emergent antiunitary symmetry $\mathcal{T}_N$ protecting the gapless Cooperon arises from the orbital-weight structure of the Bloch states, and is a symmetry of the band-projected Hamiltonian restricted to the Fermi surface. The sign of $\mathcal{T}_N^2$ then determines the nature of the correction: $\mathcal{T}_N^2=+1$ gives rise to WL, and $\mathcal{T}_N^2=-1$ to WAL. We demonstrate this mechanism in two band-projected models with non-uniform quantum geometry that are motivated by the quarter-metal state of RNG: the ``ideal model of RNG'' \cite{tan2025ideallimitrhombohedralgraphene} and the ``$\lambda_N$-model'' \cite{desrochersElectronicCrystalPhases2026}. 

Our results point towards experimentally accessible signatures of band quantum geometry in the diffusive transport regime of multilayer RNG. Based on the more microscopically realistic ``ideal model of RNG'' with a circular Fermi surface, we predict an even-odd alternation between antilocalization and localization tendencies with the number of layers (or orbitals), as shown in Fig.~\ref{fig:fig_3_quantum correction_N>2_models}~(a). This follows from $\mathcal{T}_N^2=(-1)^{N-1}$, with the quantum correction evolving toward WL for odd $N$, while for even $N$ it changes sign and evolves toward WAL as the quantum geometry is tuned. Experimentally, this behavior could be qualitatively probed by tuning the quarter-metal filling, which changes the Fermi momentum $k_F$ and the geometric parameter $\beta=\mathcal{B}k_F^2/2$, while $\mathcal{B}$ remains fixed by the microscopic band structure. Measurements of dc conductivity as a function of filling could then probe the sign of the quantum-interference correction, giving potentially distinct responses in the WL and WAL regimes.
%Measurements of low-field magnetoresistivity as a function of filling could then probe the sign of the quantum-interference correction. Since a magnetic field is expected to gap the Cooperon modes and suppress the interference correction, giving potentially distinct responses in the WL and WAL regimes. 
We also predict a nonmonotonic dependence of the Drude conductivity on filling, as shown in Fig.~\ref{fig:fig_5_diffsion_const}~(a), providing an additional transport signature of the underlying band quantum geometry. 

Note that our predicted even-odd dependence of the quantum correction on the number of layers $N$ is opposite to that discussed in an earlier work \cite{zhangBandStructureStacked2010}, which used a TRS preserving low-energy $k\cdot p$ model of RNG \cite{koshinoTrigonalWarpingBerrys2009}. As an extension of the conventional picture of WAL in monolayer graphene, it postulated that the chiral $(A_1,B_N)$ sublattice-pseudospin texture leads to WAL for odd $N$ and WL for even $N$. By contrast, our prediction applies to the interaction-driven quarter-metal state of RNG, where microscopic TRS is broken. Further, we assume an out-of-plane displacement-field regime where the active-band Bloch states become strongly sublattice polarized and generate a momentum-dependent layer-pseudospin structure. The resulting quantum geometry of the active band gives rise to the even-odd effect that we predict. Our mechanism is therefore distinct from the earlier result, which originates from the chiral sublattice-pseudospin of the TRS preserving low-energy bands.

Accessing the predicted signatures will require samples with moderate disorder, such that $l_0$ remains well below both the device size and the phase coherence length $L_{\phi}$. By contrast, in recent ultraclean RNG devices \cite{seo2026family,zhou2021,holleis2026cryogenicshockexfoliationultrahigh} the mean free path $l_0$ is comparable to the device dimensions. A further effect neglected in our RNG models is trigonal warping of the Fermi surface \cite{koshinoTrigonalWarpingBerrys2009}, which breaks the $\mathbf{k}\to-\mathbf{k}$ symmetry within a given valley and can gap the corresponding Cooperon modes \cite{mccannWeakLocalizationMagnetoresistanceValley2006,Kechedzhi_2007_PRL}. This introduces a length scale $L_{\text{TW}}$ dependent inversely on the energy mismatch $\xi_{\mathbf{k}}-\xi_{-\mathbf{k}}$. If this warping length scale is larger than the phase coherence length, $L_{\mathrm{TW}}\gtrsim L_\phi$, the Cooperon mode can remain effectively soft over the phase-coherent regime. So signatures of the even–odd layer dependence, as predicted by the ideal RNG model, may still persist for sufficiently weak trigonal warping. A more quantitative assessment requires a microscopic treatment, which we leave for future work.

Recent works have highlighted that quantum geometry can influence disordered transport and localization well beyond the weak-disorder limit. In nearly flat bands, quantum geometry can introduce a characteristic length scale that controls the localization length in the presence of strong disorder \cite{Dai_2026,Chau_2026,Wang_2026}. In flat-band systems, it provides a finite geometric contribution to the diffusion coefficient even when conventional band dispersion is strongly quenched \cite{Chau_2026}, while in metals and semimetals with linear Dirac dispersion it can substantially modify the diffusion coefficient \cite{Burkov_2026}. In this study, we identified a complementary effect in the weak-disorder regime, where quantum geometry reshapes the Cooperon spectrum and controls both the sign and magnitude of the quantum-interference correction to the dc conductivity.

\section{Acknowledgements}
The authors thank Mark R. Hirsbrunner for providing useful comments on an initial draft of the manuscript. This work was supported by the Natural Science and Engineering Council of Canada (NSERC) Grant No. RGPIN-2023-03296 and the Center for Quantum Materials at the University of Toronto.

\twocolumngrid
\appendix

\onecolumngrid
\section{Band-projected Green's function}\label{sec_app:G_F_band_proj}
In a multiorbital system the bare retarded Green's function in the orbital basis $\Tilde{G}^R_{0,\mathbf{k}}$ can be written as
\begin{gather}
\Tilde{G}^R_{0,\mathbf{k}}=\sum_n \ket{s^n_{\mathbf{k}}}\bra{s^n_{\mathbf{k}}}G^R_{\mathbf{k},n}=\sum_n\frac{\ket{s^n_{\mathbf{k}}}\bra{s^n_{\mathbf{k}}}}{\omega+i0^+-\xi^n_{\mathbf{k}}}.
\end{gather}
Within the band-projected formalism, we retain only the lowest ``active'' band with dispersion $\xi_{\mathbf{k}}$, as the remaining bands are separated by an energy scale $J$, where we take $J\rightarrow\infty$. The bare retarded Green's function in the orbital basis, projected onto the active band is
\begin{gather}
\Tilde{G}^R_{0,\mathbf{k}}=P^0_{\mathbf{k}}G^R_{0,\mathbf{k}}P^0_{\mathbf{k}}=P^0_{\mathbf{k}}\frac{1}{(\omega+i0^+-\xi_{\mathbf{k}})}P^0_{\mathbf{k}},
\end{gather}
where $P^0_{\mathbf{k}}=\ket{s^0_{\mathbf{k}}}\bra{s^0_{\mathbf{k}}}$ projects onto the ``active'' band, and $G^R_{0,\mathbf{k}}$ denotes the corresponding bare retarded Green's function in the active band. Throughout this work, we adopt this band-projected description.

\section{Self-energy in the first Born approximation}\label{sec_supp:self_energy}
The self-energy in the active band of the $\lambda_N$-model, which generalizes Eq.~\eqref{eq:self_energy_final}, is given by
\begin{align}
    \Sigma^R_{\mathbf{k}}(\omega)=&\frac{n_{\text{imp}}u_0^2}{\mathcal{V}}\sum_{\mathbf{p}}\frac{|\mathcal{F}_{\mathbf{k},\mathbf{p}}|^2}{\omega+i0^+-\xi_{\mathbf{p}}}\\
    =&n_{\text{imp}}u_0^2 \int \frac{d\phi_{\mathbf{p}}}{2\pi}\int d\xi_{\mathbf{p}} \rho(\xi_{\mathbf{p}})\frac{|\mathcal{F}_{\mathbf{k},\mathbf{p}}|^2}{\omega+i0^+-\xi_{\mathbf{p}}}\\
= &n_{\text{imp}}u_0^2 \int d\xi_{\mathbf{p}} \rho(\xi_{\mathbf{p}})\frac{1}{\omega+i0^+-\xi_{\mathbf{p}}}\frac{1}{\left[\sum_{n=0}^{N-1} \frac{\left(\mathcal{B}|\mathbf{p}|^2/2 \right)^n}{n!}\right]\left[\sum_{n=0}^{N-1} \frac{\left(\mathcal{B}|\mathbf{k}|^2/2 \right)^n}{n!}\right]}\nonumber\\&\int \frac{d\phi_{\mathbf{p}}}{2\pi} \sum_{n,m=0}^{N-1}
\frac{(\mathcal{B}|\mathbf{k}||\mathbf{p}|/2)^{n+m}}{n!m!}
\cos((n-m)(\phi_{\mathbf{k}}-\phi_{\mathbf{p}}))\\
=&n_{\text{imp}}u_0^2 \int d\xi_{\mathbf{p}} \rho(\xi_{\mathbf{p}})\frac{1}{\omega+i0^+-\xi_{\mathbf{p}}}\frac{1}{\left[\sum_{n=0}^{N-1} \frac{\left(\mathcal{B}|\mathbf{p}|^2/2 \right)^n}{n!}\right]\left[\sum_{n=0}^{N-1} \frac{\left(\mathcal{B}|\mathbf{k}|^2/2 \right)^n}{n!}\right]}\left[\sum_{n=0}^{N-1}\frac{(\mathcal{B}^2|\mathbf{k}|^2|\mathbf{p}|^2/4)^n}{(n!)^2}\right]\\
\underset{(|\mathbf{k}|\approx|\mathbf{p}|\approx k_F)}{\simeq}&n_{\text{imp}}u_0^2  \rho(0)\frac{1}{\left[\sum_{n=0}^{N-1} \frac{\left(\mathcal{B}k_F^2/2 \right)^n}{n!}\right]^2}\left[ \sum_{n=0}^{N-1} \frac{(\mathcal{B}k_F^2/2)^{2n}}{(n!)^2} \right]
\int d\xi_{\mathbf{p}}\frac{1}{\omega+i0^+-\xi_{\mathbf{p}}}
\\
\Rightarrow \Sigma^R_{\mathbf{k}_F}(0)\simeq&-i\pi n_{\text{imp}}u_0^2 \rho(0) \frac{1}{\left[\sum_{n=0}^{N-1} \frac{\left(\mathcal{B}k_F^2/2 \right)^n}{n!}\right]^2}\left[ \sum_{n=0}^{N-1} \frac{(\mathcal{B}k_F^2/2)^{2n}}{(n!)^2} \right]=-\frac{i}{2\gamma\tau_0}=-\frac{i}{2\tau}.\label{eq_supp:self_energy_final}
\end{align}
In the above, we have restricted both momenta to lie on the Fermi surface (i.e., the momenta are on-shell), and neglected the real part of the self-energy, which can be absorbed into the chemical potential. This renormalizes the self-energy compared to ordinary jellium by $\gamma$, which is defined below
\begin{gather}
    \gamma=\left[\sum_{n=0}^{N-1}\frac{\beta^{n}}{n!} \right]^2/\left[\sum_{n=0}^{N-1}\frac{\beta^{2n}}{(n!)^2} \right].\label{eq_supp:def_delta}
\end{gather} 
where $\beta=\mathcal{B}k_F^2/2$.

\section{Diffuson}\label{sec_supp:diffuson}
In Fig.~\ref{fig:fig_2_diagrams}~(a), the bare diffuson $\mathcal{D}^0_{\mathbf{k,k'}}(\mathbf{q})$ is defined as follows
\begin{gather}
    \mathcal{D}^0_{\mathbf{k,k'}}=\frac{1}{\mathcal{V}}\left\langle U_{\mathbf{k,k'}} U_{\mathbf{k'+q,k+q}} \right\rangle_{\text{imp}}=n_{\text{imp}}u_0^2\mathcal{F}_{\mathbf{k,k'}}\mathcal{F}_{\mathbf{k'+q,k+q}}.
\end{gather}
$\mathcal{D}^0_{\mathbf{k,k'}}(\mathbf{q})$ is nonsingular in $\mathbf{q}$ and its momentum dependence can be evaluated at $\mathbf{q}=0$ when extracting the leading contribution near the diffuson pole.
\begin{gather}
\mathcal{D}^0_{\mathbf{k,k'}}\simeq n_{\text{imp}}u_0^2|\mathcal{F}_{\mathbf{k,k'}}|^2.
\end{gather}
The Bethe-Salpeter equation for the diffuson from Fig.~\ref{fig:fig_2_diagrams}~(a) is
\begin{align}
\mathcal{D}_{\mathbf{k,k'}}=&\mathcal{D}^0_{\mathbf{k,k'}}+\frac{1}{\mathcal{V}}\sum_{\mathbf{p}}\mathcal{D}_{\mathbf{k,p}} G^R_{\mathbf{p}}G^A_{\mathbf{p+q}} \mathcal{D}^0_{\mathbf{p,k'}}\\
=&n_{\text{imp}}u_0^2 |\mathcal{F}_{\mathbf{k,k'}}|^2+n_{\text{imp}}u_0^2\rho(0) \int \frac{d\phi_{\mathbf{p}}}{2\pi}D_{\mathbf{k,p}}\mathcal{F}_{\mathbf{p,k'}}\mathcal{F}_{\mathbf{k'+q,p+q}}\frac{2\pi \tau}{1-i\omega\tau-i\tau \mathbf{v}_{\mathbf{p}}\cdot \mathbf{q}}\label{eq_supp:BS_diffusion_final}.
\end{align}
Constraining the momenta to the Fermi surface, i.e. $\mathbf{|k|=|\mathbf{p}|=|k'|}=k_F$, the diffusons can be expanded in angular harmonics on the Fermi surface as follows
\begin{gather}
    \mathcal{D}_{\mathbf{k,k'}}=\sum_{mn} \mathcal{D}_{m,n}(\mathbf{q}) e^{im\phi_{\mathbf{k}}} e^{-in\phi_{\mathbf{k'}}},\\
    \mathcal{D}^0_{\mathbf{k,k'}}=\sum_{mn} \mathcal{D}^0_{m,n}(\mathbf{q}) e^{im\phi_{\mathbf{k}}} e^{-in\phi_{\mathbf{k'}}},
\end{gather}
where $\mathbf{q}=q\hat{x}$, and the angles $\phi_{\mathbf{k}}$, $\phi_{\mathbf{k'}}$ are measured relative to the x-axis. We define a kernel for the diffusons
\begin{gather}
\mathcal{K}_{\mathbf{p,k'}}=\mathcal{F}_{\mathbf{p,k'}}\mathcal{F}_{\mathbf{k'+q,p+q}}\frac{2\pi n_{\text{imp}}u_0^2\tau \rho(0)}{1-i\omega\tau-i\tau \mathbf{v}_{\mathbf{p}}\cdot \mathbf{q}}.\label{eq_supp:kernel_diffusion}
\end{gather}
In the diffusive limit of $\omega\tau\ll1$, and $qv_f\tau\ll1$, it can be expanded as
\begin{gather}
\mathcal{K}_{\mathbf{p,k'}}\simeq \mathcal{F}_{\mathbf{p,k'}}
\mathcal{F}_{\mathbf{k'+q,p+q}}2\pi n_{\text{imp}}u_0^2\tau \rho(0)\left(1+i\omega\tau-\omega^2\tau^2+(i\tau-2\omega\tau^2) \mathbf{v}_{\mathbf{p}}\cdot \mathbf{q}-\tau^2(\mathbf{v}_{\mathbf{p}}\cdot \mathbf{q})^2\right)\label{eq_supp_expansion_diffusion_kernel}.
\end{gather}
Further, the kernel in Eq.~\eqref{eq_supp_expansion_diffusion_kernel} can also be expanded in angular harmonics on the Fermi surface
\begin{gather}
    \mathcal{K}_{\mathbf{p,k'}}=\sum_{mn} \mathcal{K}_{m,n}(\mathbf{q}) e^{im\phi_{\mathbf{p}}} e^{-in\phi_{\mathbf{k'}}}.
\end{gather}
Using the decomposition above, we find that the Bethe-Salpeter equation in Eq.~\eqref{eq_supp:BS_diffusion_final} can be written as a matrix equation in the angular harmonic basis as
\begin{align}
    \mathcal{D}_{m,n}= \,\,\mathcal{D}^0_{m,n}+\mathcal{D}_{m,l}\mathcal{K}_{l,n}
    \implies \mathcal{D}_{m,l}(1-\mathcal{K})_{l,n}= \,\,\mathcal{D}^0_{m,n}\label{eq_supp:BS_matrix form_diff}.
\end{align}
Since we focus on the $\mathbf{q}\to 0$ regime, we expand the diffuson kernel $\mathcal{K}_{m,n}(\mathbf{q})$ in powers of $q=|\mathbf{q}|$ as
\begin{gather}
    \mathcal{K}_{m,n}(\mathbf{q})=\sum_{\ell=0}^{\infty}\mathcal{K}^{(\ell)}_{m,n}(q/k_F)^\ell.
\end{gather}
In the weak-disorder limit, i.e., $k_F l_0 =k_F v_F \tau \gg 1$, we retain only the leading-order contributions to Eq.~\eqref{eq_supp_expansion_diffusion_kernel}, which  yields the following 
\begin{align}
\mathcal{K}^{(0)}_{m,n}&=\int\frac{d\phi_{\mathbf{p}}}{2\pi}\int\frac{d\phi_{\mathbf{k}}}{2\pi} e^{-im\phi_{\mathbf{p}}+in\phi_{\mathbf{k}}}
    \left[(1+i\omega\tau)\gamma\,\, |\mathcal{F}_{\mathbf{p,k}}|^2\right],\\
    \mathcal{K}^{(1)}_{m,n}&=\int\frac{d\phi_{\mathbf{p}}}{2\pi}\int\frac{d\phi_{\mathbf{k}}}{2\pi}e^{-im\phi_{\mathbf{p}}+in\phi_{\mathbf{k}}} \left[ i(\tau v_F\,k_F) \,\gamma\,\, |\mathcal{F}_{\mathbf{p,k}}|^2 \cos(\phi_{\mathbf{p}})\right],\\
    \mathcal{K}^{(2)}_{m,n}&=\int\frac{d\phi_{\mathbf{p}}}{2\pi}\int\frac{d\phi_{\mathbf{k}}}{2\pi}e^{-im\phi_{\mathbf{p}}+in\phi_{\mathbf{k}}} \left[ -(\tau v_F\,k_F)^2 \,\gamma\,\, |\mathcal{F}_{\mathbf{p,k}}|^2 \cos^2(\phi_{\mathbf{p}})\right].\label{eq_supp:K_mat_defn_1}
\end{align}
In the above the matrices $\mathcal{K}_{m,n}^{(\ell)}$ are expressed in the basis: $\{e^{im\phi_{\mathbf{p}}}\}\otimes \{e^{-in\phi_{\mathbf{k}}}\}$. Computing them explicitly, we find that these matrices have a banded structure. The matrix $\mathcal{K}^{(0)}$ is diagonal, and $\mathcal{K}^{(1)}$ has non-zero entries only on the upper and lower first-off diagonal elements. $\mathcal{K}^{(2)}$ is non-zero only on the diagonal and the upper and lower second off-diagonals. The central element of these matrices is located at $(m,n)=(0,0)$, and we define all other indices with respect to this central element. Since the integrand in matrix  $\mathcal{K}^{(\ell)}_{m,n}$ contains $\cos^\ell(\phi_\mathbf{p}) = 2^{-\ell}\sum_{j=0}^\ell {}^{\ell}C_j e^{i(\ell-2j)\phi_\mathbf{p}}$, the nonzero elements of $\mathcal{K}^{(\ell)}_{m,n}$ occur only when 
\begin{gather}
     m-n = -\ell, -\ell+2, \cdots, \ell-2, \ell.
\end{gather}
Below we show the matrices $\mathcal{K}^{(0)}$, $\mathcal{K}^{(1)}$, $\mathcal{K}^{(2)}$ for the $\lambda_2$-model (or $\lambda$-jellium), where the central $(0,0)$ element is boxed 
\begin{gather}
    \mathcal{K}^{(0)}
    = \frac{1+i\omega\tau}{1+\beta^2}
    \begin{pmatrix}
        \ddots\\
        &\beta\\
        &&\boxed{1+\beta^2}\\
        &&&\beta\\
        &&&&\ddots
    \end{pmatrix},
    \qquad
    \mathcal{K}^{(1)}
    = \frac{i\tau v_F k_F}{2(1+\beta^2)}
    \begin{pmatrix}
        \ddots & & & & & &\\
        & &\beta & & & &\\
        & &0 &1+\beta^2 & & &\\
        & &\beta &\boxed{0} &\beta & &\\
        & & &1+\beta^2 &0 & &\\
        & & & &\beta & &\\
        & & & & & &\ddots
    \end{pmatrix},
    \\
    \mathcal{K}^{(2)}
    = \frac{-\tau^2v_F^2 k_F^2}{4(1+\beta^2)}
    \begin{pmatrix}
        \ddots & & & & & & & &\\
        & & &\beta & & & & &\\
        & & &0 &1+\beta^2 & & & &\\
        & & &2\beta &0 &\beta & & &\\
        & & &0 &\boxed{2(1+\beta^2)} &0 & & &\\
        & & &\beta &0 &2\beta & & &\\
        & & & &1+\beta^2 &0 & & &\\
        & & & & &\beta & & &\\
        & & & & & & & &\ddots
    \end{pmatrix}.
\end{gather}
Therefore, we find that $\mathcal{K}^{(0)}$ couples only terms with $m-n=0$, $\mathcal{K}^{(1)}$ couples $m-n=\pm1$, and so on. Using this structure, Eq.~\eqref{eq_supp:BS_matrix form_diff} can be written in a block structure as follows, for a fixed index $s$, and indices $a,b\neq s$,
\begin{gather}
    \begin{pmatrix}
        \mathcal{D}_{m,s} & \mathcal{D}_{m,a}
    \end{pmatrix}
    \begin{pmatrix}
        1-\mathcal{K}_{s,s} & -\mathcal{K}_{s,b}\\
        -\mathcal{K}_{a,s} & 1-\mathcal{K}_{a,b}
    \end{pmatrix}
    =
    \begin{pmatrix}
        \mathcal{D}^0_{m,s} & \mathcal{D}^0_{m,b}
    \end{pmatrix}\label{eq_supp:BS_diffuson_block_form}
\end{gather}
Using the two components of Eq.~\eqref{eq_supp:BS_diffuson_block_form}, the diffuson entry $(m,s)$ can be written as follows
\begin{gather}
\mathcal{D}_{m,s}=\frac{\mathcal{D}^0_{m,s}+\mathcal{D}^0_{m,a}[(1-\mathcal{K}_{\overline s})^{-1}]_{a,b}\mathcal{K}_{b,s}}{1-\mathcal{K}_{s,s}-\mathcal{K}_{s,a}[(1-\mathcal{K}_{\overline s})^{-1}]_{a,b}\mathcal{K}_{b,s}}
\simeq
\frac{\mathcal{D}^0_{m,s}}{1-\mathcal{K}^{(0)}_{s,s}-q^2/k_F^2\left(\mathcal{K}^{(2)}_{s,s}+\mathcal{K}^{(1)}_{s,a}[(1-\mathcal{K}^{(0)}_{\overline s})^{-1}]_{a,b} \mathcal{K}^{(1)}_{b,s}\right)}\\
\implies \mathcal{D}_{m,s}=\frac{\mathcal{D}^{0}_{s,s}\delta_{m,s}}{{\Phi}_s+\mathfrak{D}_{s}q^2\tau},\label{eq_supp:Diffusion_comps}
\end{gather}
where the subscript $\overline s$ denotes the submatrix of $\mathcal{K}$ without row and column $s$. Above, we utilized the rotational symmetry of $|\mathcal{F}_{\mathbf{k,p}}|^2$ which depends only on the relative angle between the momenta, implying $\mathcal{D}^0_{m,s}\propto \delta_{m,s}$. We also retain terms of order $\mathcal{O}(q^0)$ in the numerator, and terms up to $\mathcal{O}(q^2)$ in the denominator. The resulting $q^2$ dependence of the denominator is characteristic of diffusive propagation. The diffusion stiffness $\mathfrak{D}_s$ for the diffuson in the channel $s$ contains elements of $\mathcal{K}^{(1)}$, which are non-zero for $\Delta\ell=\pm 1$. Therefore, only the following components contribute
\begin{gather}
    \mathfrak{D_s}=-\left(\mathcal{K}^{(2)}_{s,s}+\mathcal{K}^{(1)}_{s,s+1} [(1-\mathcal{K}_{\overline s}^{(0)})^{-1}]_{s+1,s+1}\mathcal{K}^{(1)}_{s+1,s}+\mathcal{K}^{(1)}_{s,s-1} [(1-\mathcal{K}_{\overline s}^{(0)})^{-1}]_{s-1,s-1}\mathcal{K}^{(1)}_{s-1,s}\right)/(k_F^2 \tau).\label{eq_supp:diffuson_D_n}
\end{gather}
We also have 
$
    {\Phi}_s=1-\mathcal{K}^{(0)}_{s,s}
$,
which for $s=0$ gives $\Phi_0=-i\omega\tau$, characteristic of the diffusion pole,
\begin{gather}
\mathcal{D}_{0,0}=\frac{\mathcal{D}^0_{0,0}}{-i\omega\tau+\mathfrak{D}_0q^2\tau}.
\end{gather}
%In the $q\to0$ limit, the density vertex couples to the $s=0$ diffuson channel. 
The long-wavelength density correlations have contributions only from $\mathcal D_{0,0}$. Using Eqs.~\eqref{eq_supp:K_mat_defn_1}, \eqref{eq_supp:diffuson_D_n}, we can simplify $\mathfrak{D}_0$ as
\begin{gather}
    \mathfrak{D}_0=\frac{v_F^2\tau}{2}\frac{1}{1-\gamma \sum_{m=0}^{N-1}f_mf_{m+1}}=D_{\text{bare}}\gamma\eta=D_{\text{tr}}
\end{gather}
where we have used the definitions: $\gamma^{-1}=\sum_{m=0}^{N-1}f_m^2$, and $\eta=\sum_{m=0}^{N-1}f_m^2/\sum_{m=0}^{N-1}(f_m^2-f_mf_{m+1})$, expressed in terms of general orbital weight introduced in Eq.~\eqref{eq:bloch_states_fourier_expansion}. We find that the diffusion stiffness $\mathfrak{D}_0$ for the diffuson in channel $n=0$ is identical to the transport diffusion coefficient $D_{\text{tr}}$ that enters the dc conductivity. We show this explicitly in the Appendices.~\ref{sec_supp:vertex_correction}, \ref{sec_supp:drude_conductivity}, where we find the same $D_{\text{tr}}$ using the ladder vertex corrections to the velocity vertex entering the Drude conductivity.

\section{Vertex correction to bare current vertex}\label{sec_supp:vertex_correction}
The uniform part of the bare current vertex is related to the velocity vertex by, 
$j^i_{\mathbf{k}}=-ev^i_{\mathbf{k}}$. 
By construction the bare velocity vertex in orbital basis is
\begin{gather}
\Tilde{v}^i_{\mathbf{k}}=\frac{\partial \xi_{\mathbf{k}}}{\partial k_i}\mathds{1}_{\text{orb}}.
\end{gather}
Projecting into the ``active'' band, the velocity vertex in the band-basis is
\begin{gather}
{v}^i_{\mathbf{k}}=\Tilde{v}^i_{\mathbf{k}}\mathcal{F}_{\mathbf{k,k}}=\frac{\partial \xi_{\mathbf{k}}}{\partial k_i},
\end{gather}
where $\mathcal{F}_{\mathbf{k,k}}=1$. The renormalized velocity vertex $\overline{v}_{\mathbf{k}}^i$ is calculated using the impurity-ladder vertex correction shown in Fig.~\ref{fig:fig_2_diagrams}~(c) as follows
\begin{align}
    \overline{v}_{\mathbf{k}}^i=&v_{\mathbf{k}}^i+\frac{n_{\text{imp}}u_0^2}{\mathcal{V}}\sum_{\mathbf{p}}|\mathcal{F}_{\mathbf{k,p}}|^2G^{R}_{\mathbf{p}}G^{A}_{\mathbf{p}}\overline{v}_{\mathbf{p}}^i\\
    \underset{(|\mathbf{k}|\approx|\mathbf{p}|\approx k_F)}{\simeq}&v_{\mathbf{k}}^i+n_{\text{imp}}u_0^2\rho(0) \int \frac{d \phi_{\mathbf{p}}}{2\pi} \int d\xi_{\mathbf{p}}\frac{1}{\omega+i/2\tau-\xi_{\mathbf{p}}}\frac{1}{\omega-i/2\tau-\xi_{\mathbf{p}}}|\mathcal{F}_{\mathbf{k,p}}|^2\overline{v}_{\mathbf{p}}^i\\
    =&v_{\mathbf{k}}^i+2\pi n_{\text{imp}}u_0^2\rho(0) \tau \int \frac{d\phi_{\mathbf{p}}}{2\pi}|\mathcal{F}_{\mathbf{k,p}}|^2\overline{v}_{\mathbf{p}}^i\label{eq_supp:renormlized_v}
\end{align}
As all momenta are on the isotropic Fermi surface, we assume $\overline{v}_{\mathbf{k}}^i=\eta {v}_{\mathbf{k}}^i$. Eqs.~\eqref{eq_supp:def_delta}, \eqref{eq_supp:renormlized_v} then imply that
\begin{align}
   \eta v_F \hat{\mathbf{k}}^i =& v_F \hat{\mathbf{k}}^i+\gamma \int \frac{d\phi_{\mathbf{p}}}{2\pi}|\mathcal{F}_{\mathbf{k,p}}|^2\eta v_F \hat{\mathbf{p}}^i\label{eq_supp:ev_eta_current_correction}.
\end{align}
Utilizing the rotational symmetry of the form factors we find
\begin{gather}
    \eta=1+\eta\gamma \int \frac{d\theta}{2\pi} |\mathcal{F}(k_F,\theta)|^2\cos\theta\\
    \implies\eta=\left[1-{\sum_{n=0}^{N-2} \frac{\beta^{2n+1}}{n!(n+1)!}/\sum_{n=0}^{N-1} \frac{\beta^{2n}}{(n!)^2}} \right]^{-1}\label{eq_supp:defn_eta}
\end{gather}
Therefore, the velocity vertex is renormalized by $\eta$, i.e., $\overline{v}_{\mathbf{k}}^i=\eta v_{\mathbf{k}}^i$.

\section{Drude conductivity} \label{sec_supp:drude_conductivity}
The leading order term to the dc conductivity takes the usual Drude form
\begin{align}
\sigma_i^{\text{Drude}} (\mathbf{q=0}, \omega\rightarrow 0)=&\frac{e^2}{2\pi \mathcal{V}}\sum_{\mathbf{k}}\overline{v}^i_{\mathbf{k}}{v}^i_{\mathbf{k}}G^R_{\mathbf{k}}G^A_{\mathbf{k}}\\
=&\frac{e^2}{2\pi} \int \frac{d\phi_{\mathbf{k}}}{2\pi}\cos^2(\phi_{\mathbf{k}})\int d\xi_{\mathbf{k}} \rho(\xi_{\mathbf{k}}) \eta v_F^2 \frac{1}{\omega+i/2\tau-\xi_{\mathbf{k}}}\frac{1}{\omega-i/2\tau-\xi_{\mathbf{k}}}\\
    =&\frac{e^2}{2} \rho(0)\eta v_F^2  \tau=e^2\rho(0)D_{\text{tr}}.
\end{align}
Here, the transport diffusion coefficient is $D_{\text{tr}}=\eta v_F^2  \tau/2$, and is the same as the diffusion stiffness $\mathfrak{D}_0$ obtained for the diffuson. For the $\lambda_N$-model, it can be expressed as
\begin{gather}
    D_{\text{tr}}=D_{\mathrm{bare}}\eta\gamma=D_{\mathrm{bare}}
    \left[\sum_{n=0}^{N-1} \frac{\beta^n}{n!} \right]^2/\left[\sum_{n=0}^{N-1}\frac{\beta^{2n}}{(n!)^2}-\sum_{n=0}^{N-2}\frac{\beta^{2n+1}}{n!(n+1)!} \right],
\label{eq_supp:transport_diff_coeff}
\end{gather}
where $D_{\mathrm{bare}}=\frac{v_F^2 \tau_0}{2}$ is the bare diffusion coefficient in the absence of form factors endowing the quantum geometry.

\begin{figure}
\includegraphics[scale=1]{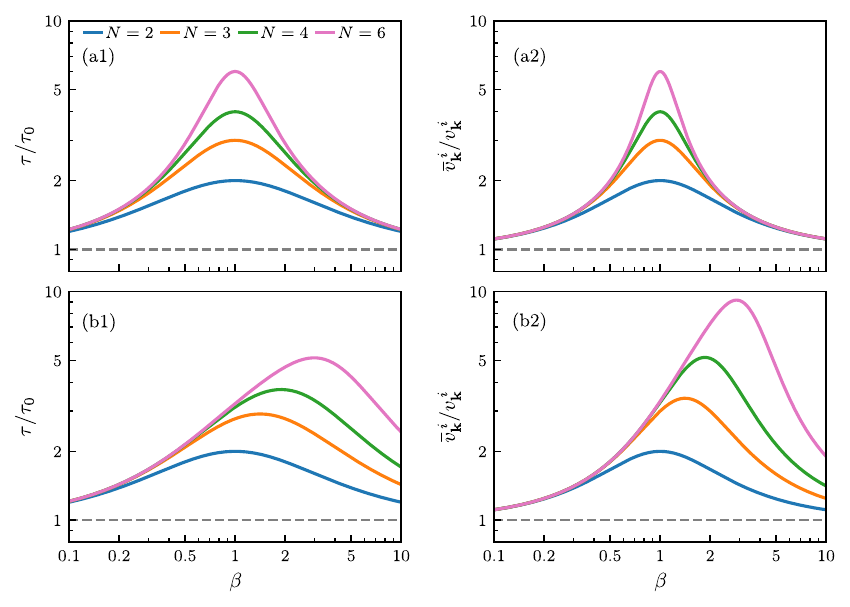}
    \caption{Column (1): the quantum geometric modification factor of the elastic scattering rate $\gamma=\tau/\tau_0$. Second column (2): the vertex renormalization $\eta=\overline v_{\mathbf{k}}^{\,i}/v_{\mathbf{k}}^{\,i}$. Both quantities are shown as functions of $\beta$ for several values of $N$ for (a) the ideal RNG model, and (b) the $\lambda_N$-model.}
    \label{fig:gamma_D_s}
\end{figure}
\section{Cooperons}\label{sec_supp:Cooperon}
The bare Cooperon, as shown in Fig.~\ref{fig:fig_2_diagrams}~(b), can be written as
\begin{gather}
    C^0_{\mathbf{k},\mathbf{k'}}=\frac{1}{\mathcal{V}}\left\langle U_{\mathbf{k,k'}} U_{\mathbf{Q-k,Q-k'}} \right\rangle_{\text{imp}}=n_{\text{imp}}u_0^2\mathcal{F}_{\mathbf{k,k'}}\mathcal{F}_{\mathbf{Q-k,Q-k'}}
\end{gather}
where $\mathbf{Q}$ denotes the total momentum of the particle-particle pair. Since the bare Cooperon is nonsingular in $\mathbf{Q}$, its momentum dependence may be evaluated at $\mathbf{Q}=0$ when extracting the leading contribution near the Cooperon pole
\begin{gather}
C^0_{\mathbf{k},\mathbf{k'}}\simeq n_{\text{imp}}u_0^2\mathcal{F}_{\mathbf{k,k'}}^2.
\end{gather}
The Bethe-Salpeter equation for the dressed Cooperon $C_{\mathbf{k,k'}}$ (shown in Fig.~\ref{fig:fig_2_diagrams}~(b)) can be written as
\begin{align}
C_{\mathbf{k,k'}}=&C^0_{\mathbf{k,k'}}+\frac{1}{\mathcal{V}}\sum_{\mathbf{p}}C_{\mathbf{k,p}} G^R_{\mathbf{p}}G^A_{\mathbf{Q-p}} C^0_{\mathbf{p,k'}}\\
=&n_{\text{imp}}u_0^2 \mathcal{F}_{\mathbf{k,k'}}^2+n_{\text{imp}}u_0^2\rho(0) \int \frac{d\phi_{\mathbf{p}}}{2\pi}C_{\mathbf{k,p}}\mathcal{F}_{\mathbf{p,k'}}\mathcal{F}_{\mathbf{Q-p,Q-k'}}\frac{2\pi \tau}{1-i\omega\tau+i\tau \mathbf{v}_{\mathbf{p}}\cdot \mathbf{Q}}\label{eq_supp:BS_Cooperon_final}.
\end{align}
With all momenta restricted to the Fermi surface, i.e. $\mathbf{|k|=|\mathbf{p}|=|k'|}=k_F$, the Cooperons are also expanded in angular harmonics on the Fermi surface as
\begin{gather}
    C_{\mathbf{k,k'}}=\sum_{mn} C_{m,n}(\mathbf{Q}) e^{im\phi_{\mathbf{k}}} e^{-in\phi_{\mathbf{k'}}},\\
    C^0_{\mathbf{k,k'}}=\sum_{mn} C^0_{m,n}(\mathbf{Q}) e^{im\phi_{\mathbf{k}}} e^{-in\phi_{\mathbf{k'}}},
\end{gather}
where $\mathbf{Q}=Q\hat{x}$, and the angles $\phi_{\mathbf{k}}$, $\phi_{\mathbf{k'}}$ are measured relative to the x-axis. The kernel from Eq.~\eqref{eq_supp:BS_Cooperon_final} is defined as
\begin{gather}
K_{\mathbf{p,k'}}=\mathcal{F}_{\mathbf{p,k'}}\mathcal{F}_{\mathbf{Q-p,Q-k'}}\frac{2\pi n_{\text{imp}}u_0^2\tau \rho(0)}{1-i\omega\tau+i\tau \mathbf{v}_{\mathbf{p}}\cdot \mathbf{Q}}\label{eq_supp:kernel_Cooperon}.
\end{gather}
In the diffusive limit of $\omega\tau\ll1$, and $Qv_f\tau\ll1$, it can be expanded as
\begin{gather}
K_{\mathbf{p,k'}}\simeq \mathcal{F}_{\mathbf{p,k'}}\mathcal{F}_{\mathbf{Q-p,Q-k'}}2\pi n_{\text{imp}}u_0^2\tau \rho(0)\left(1+i\omega\tau-\omega^2\tau^2+(2\omega\tau^2-i\tau) \mathbf{v}_{\mathbf{p}}\cdot \mathbf{Q}-\tau^2(\mathbf{v}_{\mathbf{p}}\cdot \mathbf{Q})^2\right)\label{eq_supp_expansion_Cooper_kernel}.
\end{gather}
The kernel in Eq.~\eqref{eq_supp:kernel_Cooperon} is also expanded in the angular harmonics
\begin{gather}
    K_{\mathbf{p,k'}}=\sum_{mn} K_{m,n}(\mathbf{Q}) e^{im\phi_{\mathbf{p}}} e^{-in\phi_{\mathbf{k'}}}.
\end{gather}
The Bethe-Salpeter equation in Eq.~\eqref{eq_supp:BS_Cooperon_final} then becomes a matrix equation in the angular harmonic basis
\begin{align}
    C_{m,n}= \,\,C^0_{m,n}+C_{m,l}K_{l,n}
    \implies C_{m,l}(1-K)_{l,n}= \,\,C^0_{m,n}\label{eq_supp:BS_matrix form}.
\end{align}
Since we are interested in the limit $Q\to 0$, the Cooperon kernel $K_{m,n}(\mathbf{Q})$ can be expanded in powers of $Q$ as
\begin{gather}
    K_{m,n}(\mathbf{Q})=\sum_{\ell=0}^{\infty}K^{(\ell)}_{m,n}(Q/k_F)^\ell.
\end{gather}
In the weak-disorder limit, $k_F l_0=k_F v_F \tau \gg 1$, retaining only the leading-order contribution to Eq.~\eqref{eq_supp_expansion_Cooper_kernel} at $\omega = 0$, we find the following
\begin{align}
    K^{(0)}_{m,n}&=\int\frac{d\phi_{\mathbf{p}}}{2\pi}\int\frac{d\phi_{\mathbf{k}}}{2\pi} e^{-im\phi_{\mathbf{p}}+in\phi_{\mathbf{k}}}
    \left[\gamma\,\, \mathcal{F}_{\mathbf{p,k}}^2\right],\\
    K^{(1)}_{m,n}&=\int\frac{d\phi_{\mathbf{p}}}{2\pi}\int\frac{d\phi_{\mathbf{k}}}{2\pi}e^{-im\phi_{\mathbf{p}}+in\phi_{\mathbf{k}}} \left[ -i(\tau v_F\,k_F) \,\gamma\,\, \mathcal{F}_{\mathbf{p,k}}^2 \cos(\phi_{\mathbf{p}})\right],\\
    K^{(2)}_{m,n}&=\int\frac{d\phi_{\mathbf{p}}}{2\pi}\int\frac{d\phi_{\mathbf{k}}}{2\pi}e^{-im\phi_{\mathbf{p}}+in\phi_{\mathbf{k}}} \left[ -(\tau v_F\,k_F)^2 \,\gamma\,\, \mathcal{F}_{\mathbf{p,k}}^2 \cos^2(\phi_{\mathbf{p}})\right]
\end{align}
Similar to the diffusons discussed previously, we find that the matrix $K^{(0)}$ is diagonal, and $K^{(1)}$ has non-zero entries only on the upper and lower first-off diagonals. $K^{(2)}$ is non-zero only on the diagonal, and the upper and lower second off-diagonals. The central element of these matrices is located at $(m,n)=(N-1,N-1)$. We define the indices relative to this central element as
\begin{gather}
    (\Delta m,\Delta n)=\bigl(m-(N-1),\,n-(N-1)\bigr),
\end{gather}
similar to the diffuson kernel $\mathcal{K}$, the nonzero elements of $K^{(\ell)}_{m,n}$ occur only when
\begin{gather}
    \Delta m-\Delta n = -\ell, -\ell+2, \cdots, \ell-2, \ell.
\end{gather} 
Below we show the matrices $K^{(0)}$, $K^{(1)}$, $K^{(2)}$ for the $\lambda_2$-model ($\lambda$-jellium), where the central $(1,1)$ element is boxed
\begin{gather}
    K^{(0)} = \frac{1}{1+\beta^2}
    \begin{pmatrix}
        \ddots\\
        &1\\
        &&\boxed{2\beta}\\
        &&&\beta^2\\
        &&&&\ddots
    \end{pmatrix},
    \qquad
    K^{(1)} = \frac{-i\tau v_F k_F}{2(1+\beta^2)}
    \begin{pmatrix}
        \ddots & & & & & &\\
        &0 &1 & & & &\\
        & &0 &2\beta & & &\\
        & &1 &\boxed{0} &\beta^2 & &\\
        & & &2\beta &0 & &\\
        & & & &\beta^2 &0 &\\
        & & & & & &\ddots
    \end{pmatrix},
    \\
    K^{(2)} = \frac{-\tau^2v_F^2 k_F^2}{4(1+\beta^2)}
    \begin{pmatrix}
        \ddots & & & & & & & &\\
        &0 & &1 & & & & &\\
        & &0 &0 &2\beta & & & &\\
        & & &2 &0 &\beta^2 & & &\\
        & & &0 &\boxed{4\beta} &0 & & &\\
        & & &1 &0 &2\beta^2 & & &\\
        & & & &2\beta &0 &0 & &\\
        & & & & &\beta^2 & &0 &\\
        & & & & & & & &\ddots
    \end{pmatrix}.
\end{gather}
Proceeding similarly to the diffuson calculation, Eq.~\eqref{eq_supp:BS_matrix form} can be written in a matrix-block structure as follows, for a fixed index $s$, and with indices $a,b\neq s$
\begin{gather}
    \begin{pmatrix}
        C_{m,s} & C_{m,a}
    \end{pmatrix}
    \begin{pmatrix}
        1-K_{s,s} & -K_{s,b}\\
        -K_{a,s} & 1-K_{a,b}
    \end{pmatrix}
    =
    \begin{pmatrix}
        C^0_{m,s} & C^0_{m,b}
    \end{pmatrix}\label{eq_supp:BS_Cooperon_block_form}
\end{gather}
Using the two scalar equations derived from Eq.~\eqref{eq_supp:BS_Cooperon_block_form}, the Cooperon entry $(m,s)$ can be written as follows
\begin{gather}
    C_{m,s}=\frac{C^0_{m,s}+C^0_{m,a}[(1-K_{\overline s})^{-1}]_{a,b}K_{b,s}}{1-K_{s,s}-K_{s,a}[(1-K_{\overline s})^{-1}]_{a,b}K_{b,s}}
    \simeq
    \frac{C^0_{m,s}}{1-K^{(0)}_{s,s}-Q^2/k_F^2\left(K^{(2)}_{s,s}+K^{(1)}_{s,a}[(1-K_{\overline s}^{(0)})^{-1}]_{a,b} K^{(1)}_{b,s}\right)}\\
    \implies C_{m,s}=\frac{C^{0}_{m,s}}{\Delta_s+D_{s}Q^2\tau},\label{eq_supp:Cooperon_comps}
\end{gather}
where we make the same approximations as the diffusons discussed earlier. We find that a non-zero gap can develop for the Cooperon in channel $s$, which is given by
\begin{gather}
\Delta_s=1-K^{(0)}_{s,s}.\label{eq_supp:Cooperon_gap}
\end{gather}
Additionally, the diffusion stiffness $D_s$ for the Cooperon channel $s$ can be written as
\begin{gather}
    D_s=-\left(K^{(2)}_{s,s}+K^{(1)}_{s,s+1} [(1-K^{(0)}_{\overline s})^{-1}]_{s+1,s+1}K^{(1)}_{s+1,s}+K^{(1)}_{s,s-1} [(1-K^{(0)}_{\overline s})^{-1}]_{s-1,s-1}K^{(1)}_{s-1,s}\right)/(k_F^2 \tau).\label{eq_supp:Cooperon_D_n}
\end{gather}
As shown later, the Cooperon entering the expression of the conductivity is of the form $C_{\mathbf{Q-k,k}}$, which can be expressed using Eq.~\eqref{eq_supp:Cooperon_comps} as follows
\begin{align}
    C_{\mathbf{Q-k,k}}\simeq\sum_{m,n}\frac{C^0_{m,n}e^{im(\phi_{\mathbf{k}}+\pi)}e^{-in\phi_{\mathbf{k}}}}{\Delta_n+D_nQ^2\tau}=\sum_{l}e^{il\phi_{\mathbf{k}}}\left(
    \sum_n \frac{C^0_{n+l,n}(-1)^{l+n}}{\Delta_n+D_nQ^2\tau}\right)
    =\sum_n \frac{C^0_{n,n}(-1)^n}{\Delta_n+D_nQ^2\tau}
\end{align}
The sign of the contribution from the $n$-th channel alternates depending on even or odd values of $n$. The total number of angular channels is set by the bare Cooperon $C^0_{n,n}$. For the $\lambda_N$-model this is equal to $2N - 1$.

When the gap of the Cooperon channel $n=N-1$ closes or $\Delta_{n=N-1}\to 0$, we can express the diffusion stiffness $D_{N-1}$ for the Cooperon, using Eqs.~\eqref{eq:bloch_states_fourier_expansion}, \eqref{eq:cooperon_gap_condition}, \eqref{eq_supp:diffusion_stiffness_simp}, as follows
\begin{gather}
    \frac{D_{N-1}}{D_{\text{bare}}}=\frac{\gamma}{\Delta_N}=\gamma\eta \Rightarrow D_{N-1}=D_{\text{tr}}.
\end{gather}
Therefore, when the Cooperon gap closes in the presence of the emergent antiunitary symmetry $\mathcal{T}_N$, the diffusion stiffness of the gapless Cooperon mode becomes equal the transport diffusion coefficient $D_{\mathrm{tr}}$.

\section{Quantum-interference correction to the Conductivity}\label{sec_supp:conductivity_correction}
\subsection{Bare Hikami box}
The quantum-interference correction to the conductivity from the bare Hikami box in Fig. \ref{fig:fig_2_diagrams}~(d) is given by
\begin{align}
\delta\sigma_{1i}
=&\frac{e^2}{2\pi\mathcal{V}^2}\sum_{\mathbf{k,Q}}\overline{v}_{\mathbf{k}}^i\overline{v}_{\mathbf{Q-k}}^i G^R_{\mathbf{k}} G^A_{\mathbf{k}} G^R_{\mathbf{Q-k}} G^A_{\mathbf{Q-k}} C_{\mathbf{Q-k,k}}\label{eq_supp:bare_Hikami_box}\\
=&\frac{e^2}{2\pi\mathcal{V}^2}\sum_{\mathbf{Q}}C_{\mathbf{Q-k,k}} \sum_{\mathbf{k}}\overline{v}_{\mathbf{k}}^i\overline{v}_{\mathbf{-k}}^i G^R_{\mathbf{k}} G^A_{\mathbf{k}} G^R_{\mathbf{-k}} G^A_{\mathbf{-k}} 
\end{align}
In the last step, it is assumed that the Cooperon $C_{\mathbf{Q-k},\mathbf{k}}$ diverges near $\mathbf{Q}=0$, and $C_{\mathbf{Q-k},\mathbf{k}}$ is independent of $\mathbf{k}$. Evaluating the second summation we find
\begin{align}
\frac{1}{\mathcal{V}}\sum_{\mathbf{k}}\overline{v}_{\mathbf{k}}^i\overline{v}_{\mathbf{-k}}^i G^R_{\mathbf{k}} G^A_{\mathbf{k}} G^R_{\mathbf{-k}} G^A_{\mathbf{-k}} =&\eta^2 \rho(0)\int \frac{d\phi_{\mathbf{k}}}{2\pi}\int d\xi_{\mathbf{k}}{v}_{\mathbf{k}}^i {v}_{\mathbf{-k}}^i G^R_{\mathbf{k}} G^A_{\mathbf{k}} G^R_{\mathbf{-k}} G^A_{\mathbf{-k}}\\
=&-\eta^2v_F^2\rho(0) \int \frac{d\phi_{\mathbf{k}}}{2\pi}\cos^2(\phi_{\mathbf{k}})\int d\xi_{\mathbf{k}}\frac{1}{(0-\xi_{\mathbf{k}}+i/2\tau)^2}\frac{1}{(0-\xi_{\mathbf{k}}-i/2\tau)^2}\\
=& -2\pi\eta^2v_F^2\rho(0)\tau^3.
\end{align}
Therefore, the conductivity correction is
\begin{align}
\delta\sigma_{1i}=&-\frac{e^2 2\pi \eta^2 v_F^2 \rho(0)\tau^3}{2\pi} \int \frac{d^2Q}{(2\pi)^2}
\sum_{n}\frac{C^0_{n,n}(-1)^n}{\Delta_n+D_nQ^2\tau}\\
=&-e^2\eta^2v_F^2\rho(0)\tau^3 \frac{1}{4\pi} \sum_{n} \int d(Q^2) \frac{C^0_{n,n}(-1)^n}{\Delta_n+D_nQ^2\tau}\\
=&-\frac{e^2\eta^2v_F^2\rho(0)\tau^2}{4\pi} \sum_n \frac{C^0_{n,n}(-1)^n}{D_n}\log\left[\frac{\Delta_n+D_nQ_{\text{max}}^2\tau}{\Delta_n+D_nQ_{\text{min}}^2\tau} \right]\\
=&-\frac{e^2\eta^2\gamma^2}{4\pi^2 } \sum_{n} \frac{(-1)^n[\mathcal{F}^2]_{n,n}}{D_n/D_{\mathrm{bare}}}\log \left[\frac{\Delta_n+D_n\tau/l_0^2}{\Delta_n+D_n\tau/L_{\text{max}}^2} \right]\label{eq_supp:sigma_1}.
\end{align}
In the above we have assumed that the $Q$ integral is bounded from above by the inverse mean-free path $l_0=v_F\tau$, and from below by inverse of the coherence length $l_{\phi}$ or the sample size $L$, i.e., $L_{\text{max}}=\min\{L,l_{\phi} \}$. Assuming $L_{\text{max}}=\Lambda l_0$, the total conductivity correction in Eq.~\eqref{eq_supp:sigma_1} can be written as follows
\begin{gather}
    \delta\sigma_{1i}=-\frac{e^2\eta^2\gamma^2}{4\pi^2 } \sum_{n} \frac{(-1)^n[\mathcal{F}^2]_{n,n}}{D_n/D_{\mathrm{bare}}}\log \left[\frac{\Delta_n+D_n/(2\gamma\, D_{\mathrm{bare}})}{\Delta_n+D_n/(2\Lambda^2\gamma\, D_{\mathrm{bare}})} \right].
\end{gather}

\subsection{Dressed Hikami box}

The two dressed Hikami boxes shown in Fig.~\ref{fig:fig_2_diagrams}~(d) contribute equally.
Here, the retarded Hikami box is
\begin{align}
    \delta\sigma_{2i}=&\frac{e^2}{2\pi\mathcal{V}^3}\sum_{\mathbf{k,Q,p}}\overline{v}^i_{\mathbf{k}}\overline{v}^i_{\mathbf{Q-p}} G^A_{\mathbf{k}} G^A_{\mathbf{Q-p}} C_{\mathbf{Q-k,p}} G^R_{\mathbf{k}} G^R_{\mathbf{p}} C^0_{\mathbf{p,k}} G^R_{\mathbf{Q-k}} G^R_{\mathbf{Q-p}}\\
    =& \frac{e^2\eta^2}{2\pi \mathcal{V}^3}
    \sum_{\mathbf{Q}}
    \sum_{\mathbf{k,p}}  
    {v}^i_{\mathbf{k}}{v}^i_{\mathbf{-p}}
    G^A_{\mathbf{k}} G^A_{\mathbf{-p}}  G^R_{\mathbf{k}} G^R_{\mathbf{p}} G^R_{\mathbf{-k}} G^R_{\mathbf{-p}}
    C^0_{\mathbf{p,k}}
    C_{\mathbf{Q-k,p}}.
\end{align}
Using Eq.~\eqref{eq_supp:comp_F2_nn} and the rotational symmetry of the form factors, we write the bare and dressed Cooperons as
\begin{align}
    C^0_{\mathbf{p},\mathbf{k}}=& n_{\mathrm{imp}}u_0^2\sum_n[\mathcal{F}^2]_{n,n}e^{-in(\phi_{\mathbf{k}}-\phi_{\mathbf{p}})},  \label{eq_supp:dressed_C0_expansion}\\
    C_{\mathbf{Q-k},\mathbf{p}}=&n_{\mathrm{imp}}u_0^2\sum_n\frac{(-1)^n[\mathcal{F}^2]_{n,n}e^{in(\phi_{\mathbf{k}}-\phi_{\mathbf{p}})}}{\Delta_n+D_nQ^2\tau}.\label{eq_supp:dressed_C_expansion}
\end{align}

The inner summation is evaluated first,
\begin{align}
    &\frac{1}{\mathcal{V}^2}\sum_{\mathbf{k,p}}  
    {v}^i_{\mathbf{k}}{v}^i_{\mathbf{-p}}
    G^A_{\mathbf{k}} G^A_{\mathbf{-p}}  G^R_{\mathbf{k}} G^R_{\mathbf{p}} G^R_{\mathbf{-k}} G^R_{\mathbf{-p}}
    C^0_{\mathbf{p,k}}
    C_{\mathbf{Q-k,p}} \\
    =&-v_F^2
    \rho(0)^2\left(\int d\xi_{\mathbf{k}} G^A_{\mathbf{k}} G^R_{\mathbf{k}} G^R_{\mathbf{-k}}\right)^2
    \int \frac{d\phi_{\mathbf{k}}}{2\pi} 
    \int \frac{d\phi_{\mathbf{p}}}{2\pi}
    C^0_{\mathbf{p,k}}
    C_{\mathbf{Q-k,p}}
    \cos(\phi_{\mathbf{k}})\cos(\phi_{\mathbf{p}})\\
    =&4\pi^2v_F^2(n_{\text{imp}}u_0^2)^2\rho(0)^2\tau^4
    \sum_n
    \frac{(-1)^n[\mathcal{F}^2]_{n,n}}{\Delta_n+D_nQ^2\tau}
    \left(\frac{[\mathcal{F}^2]_{n+1,n+1}+[\mathcal{F}^2]_{n-1,n-1}}{4}\right)
\end{align}
The contribution from the dressed Hikami box is therefore
\begin{gather}
    \delta\sigma_{2i}=\frac{e^2\eta^2\gamma^3}{4\pi^2 } \sum_{n} \frac{(-1)^n[\mathcal{F}^2]_{n,n}}{D_n/D_{\mathrm{bare}}} \left(\frac{[\mathcal{F}^2]_{n+1,n+1}+[\mathcal{F}^2]_{n-1,n-1}}{4}\right)  \log \left[\frac{\Delta_n+D_n/(2\gamma\, D_{\mathrm{bare}})}{\Delta_n+D_n/(2\Lambda^2\gamma\, D_{\mathrm{bare}})} \right]
\end{gather}
Therefore, the total quantum correction to the conductivity from the bare and dressed Hikami boxes is
\begin{gather}
    \delta \sigma_i=-\frac{e^2\eta^2\gamma^2}{4\pi^2 } \sum_{n} \frac{(-1)^n[\mathcal{F}^2]_{n,n}}{D_n/D_{\mathrm{bare}}}\left[1-\frac{\gamma}{2} ([\mathcal{F}^2]_{n+1,n+1}+[\mathcal{F}^2]_{n-1,n-1})\right]\log \left[\frac{\Delta_n+D_n/(2\gamma\, D_{\mathrm{bare}})}{\Delta_n+D_n/(2\Lambda^2\gamma\, D_{\mathrm{bare}})} \right] \label{eq_supp:delta_sigma_supp}
\end{gather}
The expressions for $\gamma$, $\eta$, $\Delta_n$, $D_n$, $[\mathcal{F}^2]_{n,n}$ are expressed in Eqs.~\eqref{eq_supp:def_delta}, \eqref{eq_supp:defn_eta}, \eqref{eq_supp:Cooperon_gap}, \eqref{eq_supp:Cooperon_D_n}, and \eqref{eq_supp:comp_F2_nn} respectively.

Using the expressions of the Cooperon gap and diffusion stiffness given by Eq.~\eqref{eq_supp:Cooperon_gap} and ~\eqref{eq_supp:Cooperon_D_n}, 
\begin{align}
    \Delta_n &= 1 - \gamma[F^2]_{n,n}, \label{eq_supp:cooperon_gap_simp}\\
    D_n &= \frac{\gamma v_F^2\tau}{4}[\mathcal{F}^2]_{n,n}\left(\Delta_{n+1,n+1}^{-1}+\Delta_{n-1,n-1}^{-1}\right) \nonumber\\
    &= \frac{\gamma^2[\mathcal{F}^2]_{n,n}\left(\Delta_{n+1,n+1}+\Delta_{n-1,n-1}\right)}{2\Delta_{n+1,n+1}\Delta_{n-1,n-1}}D_\mathrm{bare}, \label{eq_supp:diffusion_stiffness_simp}
\end{align}
we further simplify Eq.~\eqref{eq_supp:delta_sigma_supp} to the following
\begin{gather}
    \delta \sigma_i=-\frac{e^2\eta^2}{4\pi^2 } \sum_{n} (-1)^n\Delta_{n+1,n+1}\Delta_{n-1,n-1}\log \left[\frac{\Delta_n+D_n/(2\gamma\, D_{\mathrm{bare}})}{\Delta_n+D_n/(2\Lambda^2\gamma\, D_{\mathrm{bare}})} \right] \label{eq_supp:delta_sigma_supp_simplified}
\end{gather}

\section{Emergent antiunitary symmetry $\mathcal{T}_N$}
\label{sec_supp:effective_TRS}
In this section, we explicitly establish the antiunitarity of $\mathcal{T}_N$ and show that it is a symmetry of the projected Hamiltonian on the Fermi surface for the ideal RNG model at $\beta=1$ for all $N$, and additionally the $\lambda_3$-model at $\beta=\sqrt{2}$.

As defined in the main text Eq.~\eqref{eq:emergent_TRS_def}, 
\begin{gather}
    \mathcal{T}_N = \mathcal{R}_N\mathcal{K}, \quad \mathcal{R}_N\ket{m} = (-1)^m\ket{N-1-m}, \quad m=0,1,\cdots,N-1,
\end{gather}
where $\ket{m}$ is an orthonormal basis. For any arbitrary states
\begin{gather}
    \ket{u} = \sum_{m=0}^{N-1} u_m \ket{m}, \quad \ket{v} = \sum_{m=0}^{N-1} v_m \ket{m},
\end{gather}
the action of $\mathcal{R}_N$ yields
\begin{align}
    \braket{\mathcal{R}_Nu|\mathcal{R}_Nv} &= \sum_{n,m=0}^{N-1}(-1)^{n+m}u_n^*v_m\braket{N-1-n|N-1-m} \\
    &= \sum_{n,m=0}^{N-1} u_n^*v_m \delta_{nm} = \braket{u|v}.
\end{align}
Thus, $\mathcal{R}_N$ is unitary. Meanwhile, since the action of $\mathcal{T}_N$ yields
\begin{align}
    \braket{\mathcal{T}_Nu|\mathcal{T}_Nv} &= \sum_{n,m=0}^{N-1}(-1)^{n+m}u_nv_m^*\braket{N-1-n|N-1-m} \\
    &= \sum_{n,m=0}^{N-1} u_nv_m^* \delta_{nm} = \braket{u|v}^*.
\end{align}
Therefore, $\mathcal{T}_N$ is antiunitary.

Next, we explicitly show that $\mathcal{T}_N$ is a symmetry for the ideal RNG model at $\beta_\mathbf{k}=1$ for all $N$ and the $\lambda_N$-model at $\beta_\mathbf{k}=\sqrt{2}$ and $N=3$ only. As discussed in Sec.~\ref{sec:antiunitary_symm} of the main text, the symmetry of the projected Hamiltonian is encoded in the active-band projector $P_\mathbf{k}$. In the ideal RNG model, it is 
\begin{gather}
    P_\mathbf{k}^\mathrm{ideal\,RNG} =\frac{\sum_{n,m=0}^{N-1}\beta_{\mathbf{k}}^{(n+m)/2}e^{-i(n-m)\phi_{\mathbf{k}}}\ket{n}\bra{m}}{\sum_{n=0}^{N-1}\beta_{\mathbf{k}}^n}.
\end{gather}
Applying $\mathcal{T}_N$ and relabeling the indices yields
\begin{align}
    \mathcal{T}_N P_{\mathbf{k}}^{\mathrm{ideal\,RNG}}\mathcal{T}_N^{-1}&=\frac{\sum_{n,m=0}^{N-1}(-1)^{n+m}\beta_{\mathbf{k}}^{(n+m)/2}e^{i(n-m)\phi_{\mathbf{k}}}\ket{N-1-n}\bra{N-1-m}}{\sum_{n=0}^{N-1}\beta_{\mathbf{k}}^n}\\
    &=\frac{\sum_{n,m=0}^{N-1}\beta_{\mathbf{k}}^{N-1-(n+m)/2}e^{i(n-m)(\phi_{\mathbf{k}}+\pi)}\ket{n}\bra{m}}{\sum_{n=0}^{N-1}\beta_{\mathbf{k}}^n} = \begin{cases}
        P_{\mathbf{-k}}^{\mathrm{ideal\,RNG}}, &\beta_\mathbf{k} = 1\\
        \neq P_{\mathbf{-k}}^{\mathrm{ideal\,RNG}}, &\beta_\mathbf{k} \neq 1
    \end{cases}.
\end{align}
This also shows that $\mathcal{T} = \mathcal{T}_2$ is a symmetry of the $\lambda$-jellium model, projected onto the active-band.

Similarly, in the $\lambda_N$ model, the active-band projector is
\begin{gather}
    P_\mathbf{k}^{\lambda_N} =  \frac{\sum_{n,m=0}^{N-1}\frac{\beta_{\mathbf{k}}^{(n+m)/2}}{\sqrt{n!m!}}e^{-i(n-m)\phi_{\mathbf{k}}}\ket{n}\bra{m}}{\sum_{n=0}^{N-1}\frac{\beta_\mathbf{k}^n}{n!}}.
\end{gather}
Applying $\mathcal{T}_N$ and relabeling the indices gives
\begin{gather}
    \mathcal{T}_NP_{\mathbf{k}}^{\lambda_N}\mathcal{T}_N^{-1}=\frac{\sum_{n,m=0}^{N-1}\frac{\beta_{\mathbf{k}}^{N-1-(n+m)/2}}{\sqrt{(N-1-n)!(N-1-m)!}}e^{-i(n-m)(\phi_{\mathbf{k}}+\pi)}\ket{n}\bra{m}}{\sum_{n=0}^{N-1}\frac{\beta_\mathbf{k}^n}{n!}}.
\end{gather}
Notice that for $N>2$, only at $N=3$ and $\beta_{\mathbf{k}}=\sqrt{2}$, the orbital amplitudes
satisfy
\begin{gather}
    \frac{\beta_{\mathbf{k}}^{(N-1-n)/2}}{\sqrt{(N-1-n)!}}=\frac{\beta_{\mathbf{k}}^{n/2}}{\sqrt{n!}},\qquad n=0,1,\cdots,N-1.
\end{gather}
Hence,
\begin{gather}
    \mathcal{T}_3P_{\mathbf{k}}^{\lambda_3}\mathcal{T}_3^{-1} = \begin{cases}
        P_{\mathbf{-k}}^{\lambda_3}, &\beta_\mathbf{k} =\sqrt{2}\\
        \neq P_{\mathbf{-k}}^{\lambda_3}, &\beta_\mathbf{k} \neq \sqrt{2}
    \end{cases}.
\end{gather}

\section{Uniform quantum geometry and Cooperon weights}\label{sec_app:oscillations}
In this section, we demonstrate that in the limit of uniform quantum geometry in the $\lambda_N$-model, i.e., $N\gg2\beta\gg1$, the distribution of Cooperon weights becomes symmetric at $\beta = n/2$. 

The Fermi surface form factor of $\lambda_N$-model is 
\begin{gather}
    \mathcal{F}_N(\mathbf{k}_F,\mathbf{k}'_F)\equiv\sum_{n=0}^{N-1}f_ne^{in\theta}, \quad f_n\equiv\frac{\frac{\beta^n}{n!}}{\sum_{m=0}^{N-1}\frac{\beta^m}{m!}}.
\end{gather}
For $N\gg\beta$, where the truncation at $n=N-1$ is insignificant, we obtain
\begin{gather}
    f_n\approx e^{-\beta}\frac{\beta^n}{n!}.\label{eq_supp:oscillation_poisson_approx}
\end{gather}
Hence, the bare Cooperon weight $C^0_n\propto[F^2]_{n,n}$, where $[F^2]_{n,n}$ is given by the convolution
\begin{gather}
    [\mathcal{F}^2]_{n,n}=\sum_l f_lf_{n-l}\approx e^{-2\beta} \sum_{l=0}^{n} \frac{\beta^l}{l!} \frac{\beta^{n-l}}{(n-l)!} =e^{-2\beta}\frac{(2\beta)^n}{n!}.
\end{gather}
Thus, $[F^2]_{n,n}$ follows a Poisson distribution with mean $2\beta$. The ratio between successive bare Cooperon weights is 
\begin{gather}
    \frac{C^0_{n,n}}{C^0_{n-1,n-1}} = \frac{[\mathcal{F}^2]_{n,n}}{[\mathcal{F}^2]_{n-1,n-1}}=\frac{2\beta}{n}, \label{eq_supp:succ_bare_Cooperon_ratio}
\end{gather}
Therefore at $\beta=n/2$, the two neighboring bare Cooperon weights $C^0_{n-1,n-1}$ and $C^0_{n,n}$ are equal. More generally, in the regime $N\gg n=2\beta\gg1$, the weights are approximately reflection symmetric about the index $n-1/2$. To show this symmetry, consider pairs of channels with indices $n-1-m$ and $n+m$. Within the Poisson approximation,
\begin{align}
    \frac{[\mathcal{F}^2]_{n-1-m,n-1-m}}
         {[\mathcal{F}^2]_{n+m,n+m}}
    &=\frac{(n+m)!}{n^{2m+1}(n-1-m)!}
    =\prod_{j=1}^{m}\left(1-\frac{j^2}{n^2}\right),
\end{align}
where $0\leq m\leq n-1$. As the coefficient $[F^2]_{n,n}$, where the dominant contribution comes from a window of width $\sim \sqrt n$, has a Poisson distribution, we can take $m\sim\sqrt n \ll n$. This gives
\begin{gather}
    \ln\left(
    \frac{[\mathcal{F}^2]_{n-1-m,n-1-m}}
         {[\mathcal{F}^2]_{n+m,n+m}}
    \right)
    =-\frac{m(m+1)(2m+1)}{6n^2} = O(n^{-1/2}).
\end{gather}
Hence,
\begin{gather}
    \frac{C^0_{n-1-m,n-1-m}}{C^0_{n+m,n+m}} = \frac{[\mathcal{F}^2]_{n-1-m,n-1-m}}
         {[\mathcal{F}^2]_{n+m,n+m}} \approx 1+O(n^{-1/2}),
\end{gather}
and so the dominant bare Cooperons are symmetric about a reflection center lying between $n-1$ and $n$. Further, from Eq.~\eqref{eq_supp:cooperon_gap_simp},~\eqref{eq_supp:diffusion_stiffness_simp}, the Cooperon gap and diffusion stiffness both depend on $[F^2]_{n,n}$, their dominant contributions are also symmetrically distributed. Therefore, the dominant Cooperon weight and conductivity contributions have equal magnitudes in the paired channels
\begin{gather}
    C_{n+m,n+m}\approx C_{n-1-m,n-1-m}, \quad m=O(\sqrt n),\, n\gg1.
\end{gather}

\twocolumngrid


\begin{thebibliography}{56}%
\makeatletter
\providecommand \@ifxundefined [1]{%
 \@ifx{#1\undefined}
}%
\providecommand \@ifnum [1]{%
 \ifnum #1\expandafter \@firstoftwo
 \else \expandafter \@secondoftwo
 \fi
}%
\providecommand \@ifx [1]{%
 \ifx #1\expandafter \@firstoftwo
 \else \expandafter \@secondoftwo
 \fi
}%
\providecommand \natexlab [1]{#1}%
\providecommand \enquote  [1]{``#1''}%
\providecommand \bibnamefont  [1]{#1}%
\providecommand \bibfnamefont [1]{#1}%
\providecommand \citenamefont [1]{#1}%
\providecommand \href@noop [0]{\@secondoftwo}%
\providecommand \href [0]{\begingroup \@sanitize@url \@href}%
\providecommand \@href[1]{\@@startlink{#1}\@@href}%
\providecommand \@@href[1]{\endgroup#1\@@endlink}%
\providecommand \@sanitize@url [0]{\catcode `\\12\catcode `\$12\catcode `\&12\catcode `\#12\catcode `\^12\catcode `\_12\catcode `\%12\relax}%
\providecommand \@@startlink[1]{}%
\providecommand \@@endlink[0]{}%
\providecommand \url  [0]{\begingroup\@sanitize@url \@url }%
\providecommand \@url [1]{\endgroup\@href {#1}{\urlprefix }}%
\providecommand \urlprefix  [0]{URL }%
\providecommand \Eprint [0]{\href }%
\providecommand \doibase [0]{https://doi.org/}%
\providecommand \selectlanguage [0]{\@gobble}%
\providecommand \bibinfo  [0]{\@secondoftwo}%
\providecommand \bibfield  [0]{\@secondoftwo}%
\providecommand \translation [1]{[#1]}%
\providecommand \BibitemOpen [0]{}%
\providecommand \bibitemStop [0]{}%
\providecommand \bibitemNoStop [0]{.\EOS\space}%
\providecommand \EOS [0]{\spacefactor3000\relax}%
\providecommand \BibitemShut  [1]{\csname bibitem#1\endcsname}%
\let\auto@bib@innerbib\@empty
%</preamble>
\bibitem [{\citenamefont {{Van den Dries}}\ \emph {et~al.}(1981)\citenamefont {{Van den Dries}}, \citenamefont {{Van Haesendonck}}, \citenamefont {Bruynseraede},\ and\ \citenamefont {Deutscher}}]{VanDenDries1981}%
  \BibitemOpen
  \bibfield  {author} {\bibinfo {author} {\bibfnamefont {L.}~\bibnamefont {{Van den Dries}}}, \bibinfo {author} {\bibfnamefont {C.}~\bibnamefont {{Van Haesendonck}}}, \bibinfo {author} {\bibfnamefont {Y.}~\bibnamefont {Bruynseraede}},\ and\ \bibinfo {author} {\bibfnamefont {G.}~\bibnamefont {Deutscher}},\ }\bibfield  {title} {\bibinfo {title} {Two-dimensional localization in thin copper films},\ }\href {https://doi.org/10.1103/PhysRevLett.46.565} {\bibfield  {journal} {\bibinfo  {journal} {Phys. Rev. Lett.}\ }\textbf {\bibinfo {volume} {46}},\ \bibinfo {pages} {565} (\bibinfo {year} {1981})}\BibitemShut {NoStop}%
\bibitem [{\citenamefont {Lin}\ \emph {et~al.}(1984)\citenamefont {Lin}, \citenamefont {Paalanen}, \citenamefont {Gossard},\ and\ \citenamefont {Tsui}}]{Lin1984}%
  \BibitemOpen
  \bibfield  {author} {\bibinfo {author} {\bibfnamefont {B.~J.~F.}\ \bibnamefont {Lin}}, \bibinfo {author} {\bibfnamefont {M.~A.}\ \bibnamefont {Paalanen}}, \bibinfo {author} {\bibfnamefont {A.~C.}\ \bibnamefont {Gossard}},\ and\ \bibinfo {author} {\bibfnamefont {D.~C.}\ \bibnamefont {Tsui}},\ }\bibfield  {title} {\bibinfo {title} {Weak localization of two-dimensional electrons in heterostructures},\ }\href {https://doi.org/10.1103/PhysRevB.29.927} {\bibfield  {journal} {\bibinfo  {journal} {Phys. Rev. B}\ }\textbf {\bibinfo {volume} {29}},\ \bibinfo {pages} {927} (\bibinfo {year} {1984})}\BibitemShut {NoStop}%
\bibitem [{\citenamefont {Hikami}\ \emph {et~al.}(1980)\citenamefont {Hikami}, \citenamefont {Larkin},\ and\ \citenamefont {Nagaoka}}]{Hikami_1980}%
  \BibitemOpen
  \bibfield  {author} {\bibinfo {author} {\bibfnamefont {S.}~\bibnamefont {Hikami}}, \bibinfo {author} {\bibfnamefont {A.~I.}\ \bibnamefont {Larkin}},\ and\ \bibinfo {author} {\bibfnamefont {Y.}~\bibnamefont {Nagaoka}},\ }\bibfield  {title} {\bibinfo {title} {Spin-orbit interaction and magnetoresistance in the two dimensional random system},\ }\href {https://doi.org/10.1143/PTP.63.707} {\bibfield  {journal} {\bibinfo  {journal} {Prog. Theor. Phys.}\ }\textbf {\bibinfo {volume} {63}},\ \bibinfo {pages} {707} (\bibinfo {year} {1980})}\BibitemShut {NoStop}%
\bibitem [{\citenamefont {{Van Haesendonck}}\ \emph {et~al.}(1982)\citenamefont {{Van Haesendonck}}, \citenamefont {{Van den Dries}}, \citenamefont {Bruynseraede},\ and\ \citenamefont {Deutscher}}]{VanHaesendonck1982}%
  \BibitemOpen
  \bibfield  {author} {\bibinfo {author} {\bibfnamefont {C.}~\bibnamefont {{Van Haesendonck}}}, \bibinfo {author} {\bibfnamefont {L.}~\bibnamefont {{Van den Dries}}}, \bibinfo {author} {\bibfnamefont {Y.}~\bibnamefont {Bruynseraede}},\ and\ \bibinfo {author} {\bibfnamefont {G.}~\bibnamefont {Deutscher}},\ }\bibfield  {title} {\bibinfo {title} {Localization and negative magnetoresistance in thin copper films},\ }\href {https://doi.org/10.1103/PhysRevB.25.5090} {\bibfield  {journal} {\bibinfo  {journal} {Phys. Rev. B}\ }\textbf {\bibinfo {volume} {25}},\ \bibinfo {pages} {5090} (\bibinfo {year} {1982})}\BibitemShut {NoStop}%
\bibitem [{\citenamefont {Ando}\ and\ \citenamefont {Nakanishi}(1998)}]{Ando1998}%
  \BibitemOpen
  \bibfield  {author} {\bibinfo {author} {\bibfnamefont {T.}~\bibnamefont {Ando}}\ and\ \bibinfo {author} {\bibfnamefont {T.}~\bibnamefont {Nakanishi}},\ }\bibfield  {title} {\bibinfo {title} {Impurity scattering in carbon nanotubes – absence of back scattering –},\ }\href {https://doi.org/10.1143/JPSJ.67.1704} {\bibfield  {journal} {\bibinfo  {journal} {J. Phys. Soc. Jpn.}\ }\textbf {\bibinfo {volume} {67}},\ \bibinfo {pages} {1704} (\bibinfo {year} {1998})}\BibitemShut {NoStop}%
\bibitem [{\citenamefont {Oh}\ \emph {et~al.}(2024)\citenamefont {Oh}, \citenamefont {Kim},\ and\ \citenamefont {Rhim}}]{Oh2024}%
  \BibitemOpen
  \bibfield  {author} {\bibinfo {author} {\bibfnamefont {C.-g.}\ \bibnamefont {Oh}}, \bibinfo {author} {\bibfnamefont {K.~W.}\ \bibnamefont {Kim}},\ and\ \bibinfo {author} {\bibfnamefont {J.-W.}\ \bibnamefont {Rhim}},\ }\bibfield  {title} {\bibinfo {title} {Thermoelectric transport driven by the {Hilbert--Schmidt} distance},\ }\href {https://doi.org/10.1002/advs.202411313} {\bibfield  {journal} {\bibinfo  {journal} {Adv. Science}\ }\textbf {\bibinfo {volume} {11}},\ \bibinfo {pages} {2411313} (\bibinfo {year} {2024})}\BibitemShut {NoStop}%
\bibitem [{\citenamefont {McCann}\ \emph {et~al.}(2006)\citenamefont {McCann}, \citenamefont {Kechedzhi}, \citenamefont {Fal'ko}, \citenamefont {Suzuura}, \citenamefont {Ando},\ and\ \citenamefont {Altshuler}}]{mccannWeakLocalizationMagnetoresistanceValley2006}%
  \BibitemOpen
  \bibfield  {author} {\bibinfo {author} {\bibfnamefont {E.}~\bibnamefont {McCann}}, \bibinfo {author} {\bibfnamefont {K.}~\bibnamefont {Kechedzhi}}, \bibinfo {author} {\bibfnamefont {V.~I.}\ \bibnamefont {Fal'ko}}, \bibinfo {author} {\bibfnamefont {H.}~\bibnamefont {Suzuura}}, \bibinfo {author} {\bibfnamefont {T.}~\bibnamefont {Ando}},\ and\ \bibinfo {author} {\bibfnamefont {B.~L.}\ \bibnamefont {Altshuler}},\ }\bibfield  {title} {\bibinfo {title} {Weak-localization magnetoresistance and valley symmetry in graphene},\ }\href {https://doi.org/10.1103/PhysRevLett.97.146805} {\bibfield  {journal} {\bibinfo  {journal} {Phys. Rev. Lett.}\ }\textbf {\bibinfo {volume} {97}},\ \bibinfo {pages} {146805} (\bibinfo {year} {2006})}\BibitemShut {NoStop}%
\bibitem [{\citenamefont {Abrahams}\ \emph {et~al.}(1979)\citenamefont {Abrahams}, \citenamefont {Anderson}, \citenamefont {Licciardello},\ and\ \citenamefont {Ramakrishnan}}]{Abrahams_1979}%
  \BibitemOpen
  \bibfield  {author} {\bibinfo {author} {\bibfnamefont {E.}~\bibnamefont {Abrahams}}, \bibinfo {author} {\bibfnamefont {P.~W.}\ \bibnamefont {Anderson}}, \bibinfo {author} {\bibfnamefont {D.~C.}\ \bibnamefont {Licciardello}},\ and\ \bibinfo {author} {\bibfnamefont {T.~V.}\ \bibnamefont {Ramakrishnan}},\ }\bibfield  {title} {\bibinfo {title} {Scaling theory of localization: Absence of quantum diffusion in two dimensions},\ }\href {https://doi.org/10.1103/PhysRevLett.42.673} {\bibfield  {journal} {\bibinfo  {journal} {Phys. Rev. Lett.}\ }\textbf {\bibinfo {volume} {42}},\ \bibinfo {pages} {673} (\bibinfo {year} {1979})}\BibitemShut {NoStop}%
\bibitem [{\citenamefont {Suzuura}\ and\ \citenamefont {Ando}(2002)}]{suzuuraCrossoverSymplecticOrthogonal2002}%
  \BibitemOpen
  \bibfield  {author} {\bibinfo {author} {\bibfnamefont {H.}~\bibnamefont {Suzuura}}\ and\ \bibinfo {author} {\bibfnamefont {T.}~\bibnamefont {Ando}},\ }\bibfield  {title} {\bibinfo {title} {Crossover from symplectic to orthogonal class in a two-dimensional honeycomb lattice},\ }\href {https://doi.org/10.1103/PhysRevLett.89.266603} {\bibfield  {journal} {\bibinfo  {journal} {Phys. Rev. Lett.}\ }\textbf {\bibinfo {volume} {89}},\ \bibinfo {pages} {266603} (\bibinfo {year} {2002})}\BibitemShut {NoStop}%
\bibitem [{\citenamefont {Haldane}(2004)}]{HaldaneBerryCurvature2004}%
  \BibitemOpen
  \bibfield  {author} {\bibinfo {author} {\bibfnamefont {F.~D.~M.}\ \bibnamefont {Haldane}},\ }\bibfield  {title} {\bibinfo {title} {Berry curvature on the fermi surface: Anomalous hall effect as a topological fermi-liquid property},\ }\href {https://doi.org/10.1103/PhysRevLett.93.206602} {\bibfield  {journal} {\bibinfo  {journal} {Phys. Rev. Lett.}\ }\textbf {\bibinfo {volume} {93}},\ \bibinfo {pages} {206602} (\bibinfo {year} {2004})}\BibitemShut {NoStop}%
\bibitem [{\citenamefont {Nagaosa}\ \emph {et~al.}(2010)\citenamefont {Nagaosa}, \citenamefont {Sinova}, \citenamefont {Onoda}, \citenamefont {MacDonald},\ and\ \citenamefont {Ong}}]{NagaosaAnomalousHall2010}%
  \BibitemOpen
  \bibfield  {author} {\bibinfo {author} {\bibfnamefont {N.}~\bibnamefont {Nagaosa}}, \bibinfo {author} {\bibfnamefont {J.}~\bibnamefont {Sinova}}, \bibinfo {author} {\bibfnamefont {S.}~\bibnamefont {Onoda}}, \bibinfo {author} {\bibfnamefont {A.~H.}\ \bibnamefont {MacDonald}},\ and\ \bibinfo {author} {\bibfnamefont {N.~P.}\ \bibnamefont {Ong}},\ }\bibfield  {title} {\bibinfo {title} {Anomalous hall effect},\ }\href {https://doi.org/10.1103/RevModPhys.82.1539} {\bibfield  {journal} {\bibinfo  {journal} {Rev. Mod. Phys.}\ }\textbf {\bibinfo {volume} {82}},\ \bibinfo {pages} {1539} (\bibinfo {year} {2010})}\BibitemShut {NoStop}%
\bibitem [{\citenamefont {Xiao}\ \emph {et~al.}(2010)\citenamefont {Xiao}, \citenamefont {Chang},\ and\ \citenamefont {Niu}}]{XiaoBerryPhase2010}%
  \BibitemOpen
  \bibfield  {author} {\bibinfo {author} {\bibfnamefont {D.}~\bibnamefont {Xiao}}, \bibinfo {author} {\bibfnamefont {M.-C.}\ \bibnamefont {Chang}},\ and\ \bibinfo {author} {\bibfnamefont {Q.}~\bibnamefont {Niu}},\ }\bibfield  {title} {\bibinfo {title} {Berry phase effects on electronic properties},\ }\href {https://doi.org/10.1103/RevModPhys.82.1959} {\bibfield  {journal} {\bibinfo  {journal} {Rev. Mod. Phys.}\ }\textbf {\bibinfo {volume} {82}},\ \bibinfo {pages} {1959} (\bibinfo {year} {2010})}\BibitemShut {NoStop}%
\bibitem [{\citenamefont {Roy}(2014)}]{royBandGeometryFractional2014}%
  \BibitemOpen
  \bibfield  {author} {\bibinfo {author} {\bibfnamefont {R.}~\bibnamefont {Roy}},\ }\bibfield  {title} {\bibinfo {title} {Band geometry of fractional topological insulators},\ }\href {https://doi.org/10.1103/PhysRevB.90.165139} {\bibfield  {journal} {\bibinfo  {journal} {Phys. Rev. B}\ }\textbf {\bibinfo {volume} {90}},\ \bibinfo {pages} {165139} (\bibinfo {year} {2014})}\BibitemShut {NoStop}%
\bibitem [{\citenamefont {Peotta}\ and\ \citenamefont {T{\"o}rm{\"a}}(2015)}]{Peotta2015}%
  \BibitemOpen
  \bibfield  {author} {\bibinfo {author} {\bibfnamefont {S.}~\bibnamefont {Peotta}}\ and\ \bibinfo {author} {\bibfnamefont {P.}~\bibnamefont {T{\"o}rm{\"a}}},\ }\bibfield  {title} {\bibinfo {title} {Superfluidity in topologically nontrivial flat bands},\ }\href {https://doi.org/10.1038/ncomms9944} {\bibfield  {journal} {\bibinfo  {journal} {Nat. Comms.}\ }\textbf {\bibinfo {volume} {6}},\ \bibinfo {pages} {8944} (\bibinfo {year} {2015})}\BibitemShut {NoStop}%
\bibitem [{\citenamefont {Sodemann}\ and\ \citenamefont {Fu}(2015)}]{Sodemann2015}%
  \BibitemOpen
  \bibfield  {author} {\bibinfo {author} {\bibfnamefont {I.}~\bibnamefont {Sodemann}}\ and\ \bibinfo {author} {\bibfnamefont {L.}~\bibnamefont {Fu}},\ }\bibfield  {title} {\bibinfo {title} {Quantum nonlinear hall effect induced by berry curvature dipole in time-reversal invariant materials},\ }\href {https://doi.org/10.1103/PhysRevLett.115.216806} {\bibfield  {journal} {\bibinfo  {journal} {Phys. Rev. Letters}\ }\textbf {\bibinfo {volume} {115}},\ \bibinfo {pages} {216806} (\bibinfo {year} {2015})}\BibitemShut {NoStop}%
\bibitem [{\citenamefont {Gao}\ \emph {et~al.}(2023)\citenamefont {Gao} \emph {et~al.}}]{Gao2023}%
  \BibitemOpen
  \bibfield  {author} {\bibinfo {author} {\bibfnamefont {A.}~\bibnamefont {Gao}} \emph {et~al.},\ }\bibfield  {title} {\bibinfo {title} {Quantum metric nonlinear hall effect in a topological antiferromagnetic heterostructure},\ }\href {https://doi.org/10.1126/science.adf1506} {\bibfield  {journal} {\bibinfo  {journal} {Science}\ }\textbf {\bibinfo {volume} {381}},\ \bibinfo {pages} {181} (\bibinfo {year} {2023})}\BibitemShut {NoStop}%
\bibitem [{\citenamefont {Li}\ \emph {et~al.}(2024)\citenamefont {Li}, \citenamefont {Su}, \citenamefont {Kim}, \citenamefont {Kee}, \citenamefont {Sun},\ and\ \citenamefont {Lin}}]{Li2024}%
  \BibitemOpen
  \bibfield  {author} {\bibinfo {author} {\bibfnamefont {H.}~\bibnamefont {Li}}, \bibinfo {author} {\bibfnamefont {Y.}~\bibnamefont {Su}}, \bibinfo {author} {\bibfnamefont {Y.~B.}\ \bibnamefont {Kim}}, \bibinfo {author} {\bibfnamefont {H.-Y.}\ \bibnamefont {Kee}}, \bibinfo {author} {\bibfnamefont {K.}~\bibnamefont {Sun}},\ and\ \bibinfo {author} {\bibfnamefont {S.-Z.}\ \bibnamefont {Lin}},\ }\bibfield  {title} {\bibinfo {title} {Contrasting twisted bilayer graphene and transition metal dichalcogenides for fractional chern insulators: An emergent gauge picture},\ }\href {https://doi.org/10.1103/PhysRevB.109.245131} {\bibfield  {journal} {\bibinfo  {journal} {Phys. Rev. B}\ }\textbf {\bibinfo {volume} {109}},\ \bibinfo {pages} {245131} (\bibinfo {year} {2024})}\BibitemShut {NoStop}%
\bibitem [{\citenamefont {Kim}\ \emph {et~al.}(2025)\citenamefont {Kim}, \citenamefont {Chung}, \citenamefont {Qian}, \citenamefont {Park}, \citenamefont {Jozwiak}, \citenamefont {Rotenberg}, \citenamefont {Bostwick}, \citenamefont {Kim},\ and\ \citenamefont {Yang}}]{Kim2025}%
  \BibitemOpen
  \bibfield  {author} {\bibinfo {author} {\bibfnamefont {S.}~\bibnamefont {Kim}}, \bibinfo {author} {\bibfnamefont {Y.}~\bibnamefont {Chung}}, \bibinfo {author} {\bibfnamefont {Y.}~\bibnamefont {Qian}}, \bibinfo {author} {\bibfnamefont {S.}~\bibnamefont {Park}}, \bibinfo {author} {\bibfnamefont {C.}~\bibnamefont {Jozwiak}}, \bibinfo {author} {\bibfnamefont {E.}~\bibnamefont {Rotenberg}}, \bibinfo {author} {\bibfnamefont {A.}~\bibnamefont {Bostwick}}, \bibinfo {author} {\bibfnamefont {K.~S.}\ \bibnamefont {Kim}},\ and\ \bibinfo {author} {\bibfnamefont {B.-J.}\ \bibnamefont {Yang}},\ }\bibfield  {title} {\bibinfo {title} {Direct measurement of the quantum metric tensor in solids},\ }\href {https://doi.org/10.1126/science.ado6049} {\bibfield  {journal} {\bibinfo  {journal} {Science}\ }\textbf {\bibinfo {volume} {388}},\ \bibinfo {pages} {1050} (\bibinfo {year} {2025})}\BibitemShut {NoStop}%
\bibitem [{\citenamefont {Le~Nir}\ \emph {et~al.}(2026)\citenamefont {Le~Nir}, \citenamefont {Mitra},\ and\ \citenamefont {Kim}}]{LeNir2026}%
  \BibitemOpen
  \bibfield  {author} {\bibinfo {author} {\bibfnamefont {L.~D.}\ \bibnamefont {Le~Nir}}, \bibinfo {author} {\bibfnamefont {A.}~\bibnamefont {Mitra}},\ and\ \bibinfo {author} {\bibfnamefont {Y.~B.}\ \bibnamefont {Kim}},\ }\bibfield  {title} {\bibinfo {title} {Chiral superconductors from parent states with nonuniform berry curvature: Momentum-space vortices, bogoliubov--de gennes topology, and thermal hall conductivity},\ }\href {https://doi.org/10.1103/tx56-fym2} {\bibfield  {journal} {\bibinfo  {journal} {Phys. Rev. B}\ }\textbf {\bibinfo {volume} {114}},\ \bibinfo {pages} {154504} (\bibinfo {year} {2026})}\BibitemShut {NoStop}%
\bibitem [{\citenamefont {Hirsbrunner}\ \emph {et~al.}(2026)\citenamefont {Hirsbrunner}, \citenamefont {Desrochers}, \citenamefont {Huxford},\ and\ \citenamefont {Kim}}]{Hirsbrunner_2026}%
  \BibitemOpen
  \bibfield  {author} {\bibinfo {author} {\bibfnamefont {M.~R.}\ \bibnamefont {Hirsbrunner}}, \bibinfo {author} {\bibfnamefont {F.}~\bibnamefont {Desrochers}}, \bibinfo {author} {\bibfnamefont {J.}~\bibnamefont {Huxford}},\ and\ \bibinfo {author} {\bibfnamefont {Y.~B.}\ \bibnamefont {Kim}},\ }\bibfield  {title} {\bibinfo {title} {Topological phonons in anomalous hall crystals},\ }\href {https://doi.org/10.1103/9hlx-thl1} {\bibfield  {journal} {\bibinfo  {journal} {Phys. Rev. B}\ }\textbf {\bibinfo {volume} {114}},\ \bibinfo {pages} {045119} (\bibinfo {year} {2026})}\BibitemShut {NoStop}%
\bibitem [{\citenamefont {Froese}\ \emph {et~al.}(2026)\citenamefont {Froese}, \citenamefont {Hirsbrunner},\ and\ \citenamefont {Kim}}]{froese2026}%
  \BibitemOpen
  \bibfield  {author} {\bibinfo {author} {\bibfnamefont {P.}~\bibnamefont {Froese}}, \bibinfo {author} {\bibfnamefont {M.~R.}\ \bibnamefont {Hirsbrunner}},\ and\ \bibinfo {author} {\bibfnamefont {Y.~B.}\ \bibnamefont {Kim}},\ }\href {https://arxiv.org/abs/2609.04330} {\bibinfo {title} {The influence of quantum geometry on the phase boundary and collective excitations of electron liquids and crystals}} (\bibinfo {year} {2026}),\ \Eprint {https://arxiv.org/abs/2609.04330} {arXiv:2609.04330} \BibitemShut {NoStop}%
\bibitem [{\citenamefont {Desrochers}\ \emph {et~al.}(2026)\citenamefont {Desrochers}, \citenamefont {Huxford}, \citenamefont {Hirsbrunner},\ and\ \citenamefont {Kim}}]{desrochersElectronicCrystalPhases2026}%
  \BibitemOpen
  \bibfield  {author} {\bibinfo {author} {\bibfnamefont {F.}~\bibnamefont {Desrochers}}, \bibinfo {author} {\bibfnamefont {J.}~\bibnamefont {Huxford}}, \bibinfo {author} {\bibfnamefont {M.~R.}\ \bibnamefont {Hirsbrunner}},\ and\ \bibinfo {author} {\bibfnamefont {Y.~B.}\ \bibnamefont {Kim}},\ }\bibfield  {title} {\bibinfo {title} {Electronic crystal phases in the presence of nonuniform berry curvature and tunable berry flux: The ${\ensuremath{\lambda}}_{N}$-jellium model},\ }\href {https://doi.org/10.1103/5w7g-j4cf} {\bibfield  {journal} {\bibinfo  {journal} {Phys. Rev. B}\ }\textbf {\bibinfo {volume} {113}},\ \bibinfo {pages} {045148} (\bibinfo {year} {2026})}\BibitemShut {NoStop}%
\bibitem [{\citenamefont {Zhou}\ \emph {et~al.}(2021{\natexlab{a}})\citenamefont {Zhou}, \citenamefont {Xie}, \citenamefont {Ghazaryan}, \citenamefont {Holder}, \citenamefont {Ehrets}, \citenamefont {Spanton}, \citenamefont {Taniguchi}, \citenamefont {Watanabe}, \citenamefont {Berg}, \citenamefont {Serbyn},\ and\ \citenamefont {Young}}]{zhouHalfQuartermetalsRhombohedral2021}%
  \BibitemOpen
  \bibfield  {author} {\bibinfo {author} {\bibfnamefont {H.}~\bibnamefont {Zhou}}, \bibinfo {author} {\bibfnamefont {T.}~\bibnamefont {Xie}}, \bibinfo {author} {\bibfnamefont {A.}~\bibnamefont {Ghazaryan}}, \bibinfo {author} {\bibfnamefont {T.}~\bibnamefont {Holder}}, \bibinfo {author} {\bibfnamefont {J.~R.}\ \bibnamefont {Ehrets}}, \bibinfo {author} {\bibfnamefont {E.~M.}\ \bibnamefont {Spanton}}, \bibinfo {author} {\bibfnamefont {T.}~\bibnamefont {Taniguchi}}, \bibinfo {author} {\bibfnamefont {K.}~\bibnamefont {Watanabe}}, \bibinfo {author} {\bibfnamefont {E.}~\bibnamefont {Berg}}, \bibinfo {author} {\bibfnamefont {M.}~\bibnamefont {Serbyn}},\ and\ \bibinfo {author} {\bibfnamefont {A.~F.}\ \bibnamefont {Young}},\ }\bibfield  {title} {\bibinfo {title} {Half- and quarter-metals in rhombohedral trilayer graphene},\ }\href {https://doi.org/10.1038/s41586-021-03938-w} {\bibfield  {journal} {\bibinfo  {journal} {Nature}\ }\textbf {\bibinfo {volume} {598}},\ \bibinfo {pages} {429} (\bibinfo {year}
  {2021}{\natexlab{a}})}\BibitemShut {NoStop}%
\bibitem [{\citenamefont {Lu}\ \emph {et~al.}(2024)\citenamefont {Lu}, \citenamefont {Han}, \citenamefont {Yao}, \citenamefont {Reddy}, \citenamefont {Yang}, \citenamefont {Seo}, \citenamefont {Watanabe}, \citenamefont {Taniguchi}, \citenamefont {Fu},\ and\ \citenamefont {Ju}}]{Lu2024FQAH}%
  \BibitemOpen
  \bibfield  {author} {\bibinfo {author} {\bibfnamefont {Z.}~\bibnamefont {Lu}}, \bibinfo {author} {\bibfnamefont {T.}~\bibnamefont {Han}}, \bibinfo {author} {\bibfnamefont {Y.}~\bibnamefont {Yao}}, \bibinfo {author} {\bibfnamefont {A.~P.}\ \bibnamefont {Reddy}}, \bibinfo {author} {\bibfnamefont {J.}~\bibnamefont {Yang}}, \bibinfo {author} {\bibfnamefont {J.}~\bibnamefont {Seo}}, \bibinfo {author} {\bibfnamefont {K.}~\bibnamefont {Watanabe}}, \bibinfo {author} {\bibfnamefont {T.}~\bibnamefont {Taniguchi}}, \bibinfo {author} {\bibfnamefont {L.}~\bibnamefont {Fu}},\ and\ \bibinfo {author} {\bibfnamefont {L.}~\bibnamefont {Ju}},\ }\bibfield  {title} {\bibinfo {title} {Fractional quantum anomalous hall effect in multilayer graphene},\ }\href {https://doi.org/10.1038/s41586-023-07010-7} {\bibfield  {journal} {\bibinfo  {journal} {Nature}\ }\textbf {\bibinfo {volume} {626}},\ \bibinfo {pages} {759} (\bibinfo {year} {2024})}\BibitemShut {NoStop}%
\bibitem [{\citenamefont {Seo}\ \emph {et~al.}(2026)\citenamefont {Seo}, \citenamefont {Cotten}, \citenamefont {Ye}, \citenamefont {Xu}, \citenamefont {Sedeh}, \citenamefont {Weldeyesus}, \citenamefont {Han}, \citenamefont {Lu}, \citenamefont {Wu}, \citenamefont {Xu}, \citenamefont {Yang}, \citenamefont {Aitken}, \citenamefont {Liong}, \citenamefont {Pattanakanvijit}, \citenamefont {Hadjri}, \citenamefont {Gazizulin}, \citenamefont {Watanabe}, \citenamefont {Taniguchi}, \citenamefont {Li}, \citenamefont {Zumb{\"u}hl},\ and\ \citenamefont {Ju}}]{seo2026family}%
  \BibitemOpen
  \bibfield  {author} {\bibinfo {author} {\bibfnamefont {J.}~\bibnamefont {Seo}}, \bibinfo {author} {\bibfnamefont {A.~A.}\ \bibnamefont {Cotten}}, \bibinfo {author} {\bibfnamefont {S.}~\bibnamefont {Ye}}, \bibinfo {author} {\bibfnamefont {M.}~\bibnamefont {Xu}}, \bibinfo {author} {\bibfnamefont {O.~S.}\ \bibnamefont {Sedeh}}, \bibinfo {author} {\bibfnamefont {H.}~\bibnamefont {Weldeyesus}}, \bibinfo {author} {\bibfnamefont {T.}~\bibnamefont {Han}}, \bibinfo {author} {\bibfnamefont {Z.}~\bibnamefont {Lu}}, \bibinfo {author} {\bibfnamefont {Z.}~\bibnamefont {Wu}}, \bibinfo {author} {\bibfnamefont {W.}~\bibnamefont {Xu}}, \bibinfo {author} {\bibfnamefont {J.}~\bibnamefont {Yang}}, \bibinfo {author} {\bibfnamefont {E.}~\bibnamefont {Aitken}}, \bibinfo {author} {\bibfnamefont {P.~P.}\ \bibnamefont {Liong}}, \bibinfo {author} {\bibfnamefont {P.}~\bibnamefont {Pattanakanvijit}}, \bibinfo {author} {\bibfnamefont {Z.}~\bibnamefont {Hadjri}}, \bibinfo {author} {\bibfnamefont {R.}~\bibnamefont {Gazizulin}}, \bibinfo
  {author} {\bibfnamefont {K.}~\bibnamefont {Watanabe}}, \bibinfo {author} {\bibfnamefont {T.}~\bibnamefont {Taniguchi}}, \bibinfo {author} {\bibfnamefont {M.}~\bibnamefont {Li}}, \bibinfo {author} {\bibfnamefont {D.~M.}\ \bibnamefont {Zumb{\"u}hl}},\ and\ \bibinfo {author} {\bibfnamefont {L.}~\bibnamefont {Ju}},\ }\bibfield  {title} {\bibinfo {title} {Family of magnetic field-boosted superconductors in rhombohedral graphene},\ }\href {https://doi.org/10.1038/s41586-026-10815-x} {\bibfield  {journal} {\bibinfo  {journal} {Nature}\ }\textbf {\bibinfo {volume} {656}},\ \bibinfo {pages} {60} (\bibinfo {year} {2026})}\BibitemShut {NoStop}%
\bibitem [{\citenamefont {Zhou}\ \emph {et~al.}(2021{\natexlab{b}})\citenamefont {Zhou}, \citenamefont {Xie}, \citenamefont {Taniguchi}, \citenamefont {Watanabe},\ and\ \citenamefont {Young}}]{zhou2021}%
  \BibitemOpen
  \bibfield  {author} {\bibinfo {author} {\bibfnamefont {H.}~\bibnamefont {Zhou}}, \bibinfo {author} {\bibfnamefont {T.}~\bibnamefont {Xie}}, \bibinfo {author} {\bibfnamefont {T.}~\bibnamefont {Taniguchi}}, \bibinfo {author} {\bibfnamefont {K.}~\bibnamefont {Watanabe}},\ and\ \bibinfo {author} {\bibfnamefont {A.~F.}\ \bibnamefont {Young}},\ }\bibfield  {title} {\bibinfo {title} {Superconductivity in rhombohedral trilayer graphene},\ }\href {https://doi.org/10.1038/s41586-021-03926-0} {\bibfield  {journal} {\bibinfo  {journal} {Nature}\ }\textbf {\bibinfo {volume} {598}},\ \bibinfo {pages} {434} (\bibinfo {year} {2021}{\natexlab{b}})}\BibitemShut {NoStop}%
\bibitem [{\citenamefont {Han}\ \emph {et~al.}(2025)\citenamefont {Han}, \citenamefont {Lu}, \citenamefont {Hadjri}, \citenamefont {Shi}, \citenamefont {Wu}, \citenamefont {Xu}, \citenamefont {Yao}, \citenamefont {Cotten}, \citenamefont {Sharifi~Sedeh}, \citenamefont {Weldeyesus}, \citenamefont {Yang}, \citenamefont {Seo}, \citenamefont {Ye}, \citenamefont {Zhou}, \citenamefont {Liu}, \citenamefont {Shi}, \citenamefont {Hua}, \citenamefont {Watanabe}, \citenamefont {Taniguchi}, \citenamefont {Xiong}, \citenamefont {Zumb{\"u}hl}, \citenamefont {Fu},\ and\ \citenamefont {Ju}}]{HanChiralSuperconductivity2025}%
  \BibitemOpen
  \bibfield  {author} {\bibinfo {author} {\bibfnamefont {T.}~\bibnamefont {Han}}, \bibinfo {author} {\bibfnamefont {Z.}~\bibnamefont {Lu}}, \bibinfo {author} {\bibfnamefont {Z.}~\bibnamefont {Hadjri}}, \bibinfo {author} {\bibfnamefont {L.}~\bibnamefont {Shi}}, \bibinfo {author} {\bibfnamefont {Z.}~\bibnamefont {Wu}}, \bibinfo {author} {\bibfnamefont {W.}~\bibnamefont {Xu}}, \bibinfo {author} {\bibfnamefont {Y.}~\bibnamefont {Yao}}, \bibinfo {author} {\bibfnamefont {A.~A.}\ \bibnamefont {Cotten}}, \bibinfo {author} {\bibfnamefont {O.}~\bibnamefont {Sharifi~Sedeh}}, \bibinfo {author} {\bibfnamefont {H.}~\bibnamefont {Weldeyesus}}, \bibinfo {author} {\bibfnamefont {J.}~\bibnamefont {Yang}}, \bibinfo {author} {\bibfnamefont {J.}~\bibnamefont {Seo}}, \bibinfo {author} {\bibfnamefont {S.}~\bibnamefont {Ye}}, \bibinfo {author} {\bibfnamefont {M.}~\bibnamefont {Zhou}}, \bibinfo {author} {\bibfnamefont {H.}~\bibnamefont {Liu}}, \bibinfo {author} {\bibfnamefont {G.}~\bibnamefont {Shi}}, \bibinfo {author} {\bibfnamefont
  {Z.}~\bibnamefont {Hua}}, \bibinfo {author} {\bibfnamefont {K.}~\bibnamefont {Watanabe}}, \bibinfo {author} {\bibfnamefont {T.}~\bibnamefont {Taniguchi}}, \bibinfo {author} {\bibfnamefont {P.}~\bibnamefont {Xiong}}, \bibinfo {author} {\bibfnamefont {D.~M.}\ \bibnamefont {Zumb{\"u}hl}}, \bibinfo {author} {\bibfnamefont {L.}~\bibnamefont {Fu}},\ and\ \bibinfo {author} {\bibfnamefont {L.}~\bibnamefont {Ju}},\ }\bibfield  {title} {\bibinfo {title} {Signatures of chiral superconductivity in rhombohedral graphene},\ }\href {https://doi.org/10.1038/s41586-025-09169-7} {\bibfield  {journal} {\bibinfo  {journal} {Nature}\ }\textbf {\bibinfo {volume} {643}},\ \bibinfo {pages} {654} (\bibinfo {year} {2025})}\BibitemShut {NoStop}%
\bibitem [{\citenamefont {Kumar}\ \emph {et~al.}(2026)\citenamefont {Kumar}, \citenamefont {Waleffe}, \citenamefont {Okounkova}, \citenamefont {Tejani}, \citenamefont {Watanabe}, \citenamefont {Taniguchi}, \citenamefont {Étienne Lantagne-Hurtubise}, \citenamefont {Folk},\ and\ \citenamefont {Yankowitz}}]{KumarPervasiveSpinTriplet2026}%
  \BibitemOpen
  \bibfield  {author} {\bibinfo {author} {\bibfnamefont {M.}~\bibnamefont {Kumar}}, \bibinfo {author} {\bibfnamefont {D.}~\bibnamefont {Waleffe}}, \bibinfo {author} {\bibfnamefont {A.}~\bibnamefont {Okounkova}}, \bibinfo {author} {\bibfnamefont {R.}~\bibnamefont {Tejani}}, \bibinfo {author} {\bibfnamefont {K.}~\bibnamefont {Watanabe}}, \bibinfo {author} {\bibfnamefont {T.}~\bibnamefont {Taniguchi}}, \bibinfo {author} {\bibnamefont {Étienne Lantagne-Hurtubise}}, \bibinfo {author} {\bibfnamefont {J.}~\bibnamefont {Folk}},\ and\ \bibinfo {author} {\bibfnamefont {M.}~\bibnamefont {Yankowitz}},\ }\href {https://arxiv.org/abs/2511.16578} {\bibinfo {title} {Pervasive spin-triplet superconductivity in rhombohedral graphene}} (\bibinfo {year} {2026}),\ \Eprint {https://arxiv.org/abs/2511.16578} {arXiv:2511.16578} \BibitemShut {NoStop}%
\bibitem [{\citenamefont {Qin}\ \emph {et~al.}(2026)\citenamefont {Qin}, \citenamefont {Wu}, \citenamefont {Nguyen}, \citenamefont {Morissette}, \citenamefont {Zhang}, \citenamefont {Watanabe}, \citenamefont {Taniguchi},\ and\ \citenamefont {Li}}]{qin2026stripeordermetallicsuperconducting}%
  \BibitemOpen
  \bibfield  {author} {\bibinfo {author} {\bibfnamefont {P.}~\bibnamefont {Qin}}, \bibinfo {author} {\bibfnamefont {H.-T.}\ \bibnamefont {Wu}}, \bibinfo {author} {\bibfnamefont {R.~Q.}\ \bibnamefont {Nguyen}}, \bibinfo {author} {\bibfnamefont {E.}~\bibnamefont {Morissette}}, \bibinfo {author} {\bibfnamefont {N.~J.}\ \bibnamefont {Zhang}}, \bibinfo {author} {\bibfnamefont {K.}~\bibnamefont {Watanabe}}, \bibinfo {author} {\bibfnamefont {T.}~\bibnamefont {Taniguchi}},\ and\ \bibinfo {author} {\bibfnamefont {J.~I.~A.}\ \bibnamefont {Li}},\ }\href {https://arxiv.org/abs/2504.05129} {\bibinfo {title} {Stripe order in the metallic and superconducting phases of rhombohedral hexalayer graphene}} (\bibinfo {year} {2026}),\ \Eprint {https://arxiv.org/abs/2504.05129} {arXiv:2504.05129} \BibitemShut {NoStop}%
\bibitem [{\citenamefont {Cao}\ \emph {et~al.}(2018)\citenamefont {Cao}, \citenamefont {Fatemi}, \citenamefont {Fang}, \citenamefont {Watanabe}, \citenamefont {Taniguchi}, \citenamefont {Kaxiras},\ and\ \citenamefont {{Jarillo-Herrero}}}]{caoUnconventionalSuperconductivityMagicangle2018b}%
  \BibitemOpen
  \bibfield  {author} {\bibinfo {author} {\bibfnamefont {Y.}~\bibnamefont {Cao}}, \bibinfo {author} {\bibfnamefont {V.}~\bibnamefont {Fatemi}}, \bibinfo {author} {\bibfnamefont {S.}~\bibnamefont {Fang}}, \bibinfo {author} {\bibfnamefont {K.}~\bibnamefont {Watanabe}}, \bibinfo {author} {\bibfnamefont {T.}~\bibnamefont {Taniguchi}}, \bibinfo {author} {\bibfnamefont {E.}~\bibnamefont {Kaxiras}},\ and\ \bibinfo {author} {\bibfnamefont {P.}~\bibnamefont {{Jarillo-Herrero}}},\ }\bibfield  {title} {\bibinfo {title} {Unconventional superconductivity in magic-angle graphene superlattices},\ }\href {https://doi.org/10.1038/nature26160} {\bibfield  {journal} {\bibinfo  {journal} {Nature}\ }\textbf {\bibinfo {volume} {556}},\ \bibinfo {pages} {43} (\bibinfo {year} {2018})}\BibitemShut {NoStop}%
\bibitem [{\citenamefont {Park}\ \emph {et~al.}(2023)\citenamefont {Park}, \citenamefont {Cai}, \citenamefont {Anderson}, \citenamefont {Zhang}, \citenamefont {Zhu}, \citenamefont {Liu}, \citenamefont {Wang}, \citenamefont {Holtzmann}, \citenamefont {Hu}, \citenamefont {Liu}, \citenamefont {Taniguchi}, \citenamefont {Watanabe}, \citenamefont {Chu}, \citenamefont {Cao}, \citenamefont {Fu}, \citenamefont {Yao}, \citenamefont {Chang}, \citenamefont {Cobden}, \citenamefont {Xiao},\ and\ \citenamefont {Xu}}]{Park2023FQAH}%
  \BibitemOpen
  \bibfield  {author} {\bibinfo {author} {\bibfnamefont {H.}~\bibnamefont {Park}}, \bibinfo {author} {\bibfnamefont {J.}~\bibnamefont {Cai}}, \bibinfo {author} {\bibfnamefont {E.}~\bibnamefont {Anderson}}, \bibinfo {author} {\bibfnamefont {Y.}~\bibnamefont {Zhang}}, \bibinfo {author} {\bibfnamefont {J.}~\bibnamefont {Zhu}}, \bibinfo {author} {\bibfnamefont {X.}~\bibnamefont {Liu}}, \bibinfo {author} {\bibfnamefont {C.}~\bibnamefont {Wang}}, \bibinfo {author} {\bibfnamefont {W.}~\bibnamefont {Holtzmann}}, \bibinfo {author} {\bibfnamefont {C.}~\bibnamefont {Hu}}, \bibinfo {author} {\bibfnamefont {Z.}~\bibnamefont {Liu}}, \bibinfo {author} {\bibfnamefont {T.}~\bibnamefont {Taniguchi}}, \bibinfo {author} {\bibfnamefont {K.}~\bibnamefont {Watanabe}}, \bibinfo {author} {\bibfnamefont {J.-H.}\ \bibnamefont {Chu}}, \bibinfo {author} {\bibfnamefont {T.}~\bibnamefont {Cao}}, \bibinfo {author} {\bibfnamefont {L.}~\bibnamefont {Fu}}, \bibinfo {author} {\bibfnamefont {W.}~\bibnamefont {Yao}}, \bibinfo {author}
  {\bibfnamefont {C.-Z.}\ \bibnamefont {Chang}}, \bibinfo {author} {\bibfnamefont {D.}~\bibnamefont {Cobden}}, \bibinfo {author} {\bibfnamefont {D.}~\bibnamefont {Xiao}},\ and\ \bibinfo {author} {\bibfnamefont {X.}~\bibnamefont {Xu}},\ }\bibfield  {title} {\bibinfo {title} {Observation of fractionally quantized anomalous hall effect},\ }\href {https://doi.org/10.1038/s41586-023-06536-0} {\bibfield  {journal} {\bibinfo  {journal} {Nature}\ }\textbf {\bibinfo {volume} {622}},\ \bibinfo {pages} {74} (\bibinfo {year} {2023})}\BibitemShut {NoStop}%
\bibitem [{\citenamefont {Yu}\ \emph {et~al.}(2025)\citenamefont {Yu}, \citenamefont {Bernevig}, \citenamefont {Queiroz}, \citenamefont {Rossi}, \citenamefont {T{\"o}rm{\"a}},\ and\ \citenamefont {Yang}}]{yuQuantumGeometryQuantum2025}%
  \BibitemOpen
  \bibfield  {author} {\bibinfo {author} {\bibfnamefont {J.}~\bibnamefont {Yu}}, \bibinfo {author} {\bibfnamefont {B.~A.}\ \bibnamefont {Bernevig}}, \bibinfo {author} {\bibfnamefont {R.}~\bibnamefont {Queiroz}}, \bibinfo {author} {\bibfnamefont {E.}~\bibnamefont {Rossi}}, \bibinfo {author} {\bibfnamefont {P.}~\bibnamefont {T{\"o}rm{\"a}}},\ and\ \bibinfo {author} {\bibfnamefont {B.-J.}\ \bibnamefont {Yang}},\ }\bibfield  {title} {\bibinfo {title} {Quantum geometry in quantum materials},\ }\href {https://doi.org/10.1038/s41535-025-00801-3} {\bibfield  {journal} {\bibinfo  {journal} {npj Quant. Mater.}\ }\textbf {\bibinfo {volume} {10}},\ \bibinfo {pages} {101} (\bibinfo {year} {2025})}\BibitemShut {NoStop}%
\bibitem [{\citenamefont {Tanaka}\ \emph {et~al.}(2025)\citenamefont {Tanaka}, \citenamefont {Wang}, \citenamefont {Dinh}, \citenamefont {Rodan-Legrain}, \citenamefont {Zaman}, \citenamefont {Hays}, \citenamefont {Almanakly}, \citenamefont {Kannan}, \citenamefont {Kim}, \citenamefont {Niedzielski}, \citenamefont {Serniak}, \citenamefont {Schwartz}, \citenamefont {Watanabe}, \citenamefont {Taniguchi}, \citenamefont {Orlando}, \citenamefont {Gustavsson}, \citenamefont {Grover}, \citenamefont {Jarillo-Herrero},\ and\ \citenamefont {Oliver}}]{tanakaSuperfluidStiffnessMagicangle2025}%
  \BibitemOpen
  \bibfield  {author} {\bibinfo {author} {\bibfnamefont {M.}~\bibnamefont {Tanaka}}, \bibinfo {author} {\bibfnamefont {J.~I.-J.}\ \bibnamefont {Wang}}, \bibinfo {author} {\bibfnamefont {T.~H.}\ \bibnamefont {Dinh}}, \bibinfo {author} {\bibfnamefont {D.}~\bibnamefont {Rodan-Legrain}}, \bibinfo {author} {\bibfnamefont {S.}~\bibnamefont {Zaman}}, \bibinfo {author} {\bibfnamefont {M.}~\bibnamefont {Hays}}, \bibinfo {author} {\bibfnamefont {A.}~\bibnamefont {Almanakly}}, \bibinfo {author} {\bibfnamefont {B.}~\bibnamefont {Kannan}}, \bibinfo {author} {\bibfnamefont {D.~K.}\ \bibnamefont {Kim}}, \bibinfo {author} {\bibfnamefont {B.~M.}\ \bibnamefont {Niedzielski}}, \bibinfo {author} {\bibfnamefont {K.}~\bibnamefont {Serniak}}, \bibinfo {author} {\bibfnamefont {M.~E.}\ \bibnamefont {Schwartz}}, \bibinfo {author} {\bibfnamefont {K.}~\bibnamefont {Watanabe}}, \bibinfo {author} {\bibfnamefont {T.}~\bibnamefont {Taniguchi}}, \bibinfo {author} {\bibfnamefont {T.~P.}\ \bibnamefont {Orlando}}, \bibinfo {author}
  {\bibfnamefont {S.}~\bibnamefont {Gustavsson}}, \bibinfo {author} {\bibfnamefont {J.~A.}\ \bibnamefont {Grover}}, \bibinfo {author} {\bibfnamefont {P.}~\bibnamefont {Jarillo-Herrero}},\ and\ \bibinfo {author} {\bibfnamefont {W.~D.}\ \bibnamefont {Oliver}},\ }\bibfield  {title} {\bibinfo {title} {Superfluid stiffness of magic-angle twisted bilayer graphene},\ }\href {https://doi.org/10.1038/s41586-024-08494-7} {\bibfield  {journal} {\bibinfo  {journal} {Nature}\ }\textbf {\bibinfo {volume} {638}},\ \bibinfo {pages} {99} (\bibinfo {year} {2025})}\BibitemShut {NoStop}%
\bibitem [{\citenamefont {Tan}\ \emph {et~al.}(2025)\citenamefont {Tan}, \citenamefont {Ledwith},\ and\ \citenamefont {Devakul}}]{tan2025ideallimitrhombohedralgraphene}%
  \BibitemOpen
  \bibfield  {author} {\bibinfo {author} {\bibfnamefont {T.}~\bibnamefont {Tan}}, \bibinfo {author} {\bibfnamefont {P.~J.}\ \bibnamefont {Ledwith}},\ and\ \bibinfo {author} {\bibfnamefont {T.}~\bibnamefont {Devakul}},\ }\href {https://arxiv.org/abs/2511.07402} {\bibinfo {title} {The ideal limit of rhombohedral graphene: Interaction-induced layer-skyrmion lattices and their collective excitations}} (\bibinfo {year} {2025}),\ \Eprint {https://arxiv.org/abs/2511.07402} {arXiv:2511.07402} \BibitemShut {NoStop}%
\bibitem [{\citenamefont {Soejima}\ \emph {et~al.}(2025)\citenamefont {Soejima}, \citenamefont {Dong}, \citenamefont {Vishwanath},\ and\ \citenamefont {Parker}}]{soejima$l$JelliumModelAnomalous2025}%
  \BibitemOpen
  \bibfield  {author} {\bibinfo {author} {\bibfnamefont {T.}~\bibnamefont {Soejima}}, \bibinfo {author} {\bibfnamefont {J.}~\bibnamefont {Dong}}, \bibinfo {author} {\bibfnamefont {A.}~\bibnamefont {Vishwanath}},\ and\ \bibinfo {author} {\bibfnamefont {D.~E.}\ \bibnamefont {Parker}},\ }\bibfield  {title} {\bibinfo {title} {$\ensuremath{\lambda}$-jellium model for the anomalous hall crystal},\ }\href {https://doi.org/10.1103/x53d-12s6} {\bibfield  {journal} {\bibinfo  {journal} {Phys. Rev. Lett.}\ }\textbf {\bibinfo {volume} {135}},\ \bibinfo {pages} {186505} (\bibinfo {year} {2025})}\BibitemShut {NoStop}%
\bibitem [{\citenamefont {Vollhardt}\ and\ \citenamefont {W{\"o}lfle}(1980)}]{vollhardtAndersonLocalization<1980}%
  \BibitemOpen
  \bibfield  {author} {\bibinfo {author} {\bibfnamefont {D.}~\bibnamefont {Vollhardt}}\ and\ \bibinfo {author} {\bibfnamefont {P.}~\bibnamefont {W{\"o}lfle}},\ }\bibfield  {title} {\bibinfo {title} {Anderson localization in {$d<\sim2$} dimensions: A self-consistent diagrammatic theory},\ }\href {https://doi.org/10.1103/PhysRevLett.45.842} {\bibfield  {journal} {\bibinfo  {journal} {Phys. Rev. Lett.}\ }\textbf {\bibinfo {volume} {45}},\ \bibinfo {pages} {842} (\bibinfo {year} {1980})}\BibitemShut {NoStop}%
\bibitem [{\citenamefont {Altland}\ and\ \citenamefont {Zirnbauer}(1997)}]{Altland1997}%
  \BibitemOpen
  \bibfield  {author} {\bibinfo {author} {\bibfnamefont {A.}~\bibnamefont {Altland}}\ and\ \bibinfo {author} {\bibfnamefont {M.~R.}\ \bibnamefont {Zirnbauer}},\ }\bibfield  {title} {\bibinfo {title} {Nonstandard symmetry classes in mesoscopic normal-superconducting hybrid structures},\ }\href {https://doi.org/10.1103/PhysRevB.55.1142} {\bibfield  {journal} {\bibinfo  {journal} {Phys. Rev. B}\ }\textbf {\bibinfo {volume} {55}},\ \bibinfo {pages} {1142} (\bibinfo {year} {1997})}\BibitemShut {NoStop}%
\bibitem [{\citenamefont {Ledwith}\ \emph {et~al.}(2023)\citenamefont {Ledwith}, \citenamefont {Vishwanath},\ and\ \citenamefont {Parker}}]{ledwithVortexabilityUnifyingCriterion2023}%
  \BibitemOpen
  \bibfield  {author} {\bibinfo {author} {\bibfnamefont {P.~J.}\ \bibnamefont {Ledwith}}, \bibinfo {author} {\bibfnamefont {A.}~\bibnamefont {Vishwanath}},\ and\ \bibinfo {author} {\bibfnamefont {D.~E.}\ \bibnamefont {Parker}},\ }\bibfield  {title} {\bibinfo {title} {Vortexability: A unifying criterion for ideal fractional chern insulators},\ }\href {https://doi.org/10.1103/PhysRevB.108.205144} {\bibfield  {journal} {\bibinfo  {journal} {Phys. Rev. B}\ }\textbf {\bibinfo {volume} {108}},\ \bibinfo {pages} {205144} (\bibinfo {year} {2023})}\BibitemShut {NoStop}%
\bibitem [{\citenamefont {Wang}\ \emph {et~al.}(2021)\citenamefont {Wang}, \citenamefont {Cano}, \citenamefont {Millis}, \citenamefont {Liu},\ and\ \citenamefont {Yang}}]{wangExactLandauLevel2021}%
  \BibitemOpen
  \bibfield  {author} {\bibinfo {author} {\bibfnamefont {J.}~\bibnamefont {Wang}}, \bibinfo {author} {\bibfnamefont {J.}~\bibnamefont {Cano}}, \bibinfo {author} {\bibfnamefont {A.~J.}\ \bibnamefont {Millis}}, \bibinfo {author} {\bibfnamefont {Z.}~\bibnamefont {Liu}},\ and\ \bibinfo {author} {\bibfnamefont {B.}~\bibnamefont {Yang}},\ }\bibfield  {title} {\bibinfo {title} {Exact landau level description of geometry and interaction in a flatband},\ }\href {https://doi.org/10.1103/PhysRevLett.127.246403} {\bibfield  {journal} {\bibinfo  {journal} {Phys. Rev. Lett.}\ }\textbf {\bibinfo {volume} {127}},\ \bibinfo {pages} {246403} (\bibinfo {year} {2021})}\BibitemShut {NoStop}%
\bibitem [{\citenamefont {Akkermans}\ and\ \citenamefont {Montambaux}(2007)}]{Akkermans_Montambaux_2007}%
  \BibitemOpen
  \bibfield  {author} {\bibinfo {author} {\bibfnamefont {E.}~\bibnamefont {Akkermans}}\ and\ \bibinfo {author} {\bibfnamefont {G.}~\bibnamefont {Montambaux}},\ }\bibinfo {title} {Electronic transport},\ in\ \href@noop {} {\emph {\bibinfo {booktitle} {Mesoscopic Physics of Electrons and Photons}}}\ (\bibinfo  {publisher} {Cambridge University Press},\ \bibinfo {year} {2007})\ p.\ \bibinfo {pages} {270–319}\BibitemShut {NoStop}%
\bibitem [{\citenamefont {Lee}\ and\ \citenamefont {Ramakrishnan}(1985)}]{Lee_1985}%
  \BibitemOpen
  \bibfield  {author} {\bibinfo {author} {\bibfnamefont {P.~A.}\ \bibnamefont {Lee}}\ and\ \bibinfo {author} {\bibfnamefont {T.~V.}\ \bibnamefont {Ramakrishnan}},\ }\bibfield  {title} {\bibinfo {title} {Disordered electronic systems},\ }\href {https://doi.org/10.1103/RevModPhys.57.287} {\bibfield  {journal} {\bibinfo  {journal} {Rev. Mod. Phys.}\ }\textbf {\bibinfo {volume} {57}},\ \bibinfo {pages} {287} (\bibinfo {year} {1985})}\BibitemShut {NoStop}%
\bibitem [{\citenamefont {Chen}\ \emph {et~al.}(2019)\citenamefont {Chen}, \citenamefont {Lu},\ and\ \citenamefont {Zilberberg}}]{chenWeakLocalizationAntilocalization2019}%
  \BibitemOpen
  \bibfield  {author} {\bibinfo {author} {\bibfnamefont {W.}~\bibnamefont {Chen}}, \bibinfo {author} {\bibfnamefont {H.-Z.}\ \bibnamefont {Lu}},\ and\ \bibinfo {author} {\bibfnamefont {O.}~\bibnamefont {Zilberberg}},\ }\bibfield  {title} {\bibinfo {title} {Weak localization and antilocalization in nodal-line semimetals: Dimensionality and topological effects},\ }\href {https://doi.org/10.1103/PhysRevLett.122.196603} {\bibfield  {journal} {\bibinfo  {journal} {Phys. Rev. Lett.}\ }\textbf {\bibinfo {volume} {122}},\ \bibinfo {pages} {196603} (\bibinfo {year} {2019})}\BibitemShut {NoStop}%
\bibitem [{\citenamefont {Hikami}(1981)}]{hikamiAndersonLocalizationNonlinearsigmamodel1981}%
  \BibitemOpen
  \bibfield  {author} {\bibinfo {author} {\bibfnamefont {S.}~\bibnamefont {Hikami}},\ }\bibfield  {title} {\bibinfo {title} {Anderson localization in a nonlinear-$\ensuremath{\sigma}$-model representation},\ }\href {https://doi.org/10.1103/PhysRevB.24.2671} {\bibfield  {journal} {\bibinfo  {journal} {Phys. Rev. B}\ }\textbf {\bibinfo {volume} {24}},\ \bibinfo {pages} {2671} (\bibinfo {year} {1981})}\BibitemShut {NoStop}%
\bibitem [{\citenamefont {Tkachov}\ and\ \citenamefont {Hankiewicz}(2011)}]{Tkachov_2011}%
  \BibitemOpen
  \bibfield  {author} {\bibinfo {author} {\bibfnamefont {G.}~\bibnamefont {Tkachov}}\ and\ \bibinfo {author} {\bibfnamefont {E.~M.}\ \bibnamefont {Hankiewicz}},\ }\bibfield  {title} {\bibinfo {title} {Weak antilocalization in hgte quantum wells and topological surface states: Massive versus massless dirac fermions},\ }\href {https://doi.org/10.1103/PhysRevB.84.035444} {\bibfield  {journal} {\bibinfo  {journal} {Phys. Rev. B}\ }\textbf {\bibinfo {volume} {84}},\ \bibinfo {pages} {035444} (\bibinfo {year} {2011})}\BibitemShut {NoStop}%
\bibitem [{\citenamefont {Lu}\ \emph {et~al.}(2011)\citenamefont {Lu}, \citenamefont {Shi},\ and\ \citenamefont {Shen}}]{Lu_2011}%
  \BibitemOpen
  \bibfield  {author} {\bibinfo {author} {\bibfnamefont {H.-Z.}\ \bibnamefont {Lu}}, \bibinfo {author} {\bibfnamefont {J.}~\bibnamefont {Shi}},\ and\ \bibinfo {author} {\bibfnamefont {S.-Q.}\ \bibnamefont {Shen}},\ }\bibfield  {title} {\bibinfo {title} {Competition between weak localization and antilocalization in topological surface states},\ }\href {https://doi.org/10.1103/PhysRevLett.107.076801} {\bibfield  {journal} {\bibinfo  {journal} {Phys. Rev. Lett.}\ }\textbf {\bibinfo {volume} {107}},\ \bibinfo {pages} {076801} (\bibinfo {year} {2011})}\BibitemShut {NoStop}%
\bibitem [{\citenamefont {Chen}\ \emph {et~al.}(2010)\citenamefont {Chen}, \citenamefont {Qin}, \citenamefont {Yang}, \citenamefont {Liu}, \citenamefont {Guan}, \citenamefont {Qu}, \citenamefont {Zhang}, \citenamefont {Shi}, \citenamefont {Xie}, \citenamefont {Yang}, \citenamefont {Wu}, \citenamefont {Li},\ and\ \citenamefont {Lu}}]{Chen_2010}%
  \BibitemOpen
  \bibfield  {author} {\bibinfo {author} {\bibfnamefont {J.}~\bibnamefont {Chen}}, \bibinfo {author} {\bibfnamefont {H.~J.}\ \bibnamefont {Qin}}, \bibinfo {author} {\bibfnamefont {F.}~\bibnamefont {Yang}}, \bibinfo {author} {\bibfnamefont {J.}~\bibnamefont {Liu}}, \bibinfo {author} {\bibfnamefont {T.}~\bibnamefont {Guan}}, \bibinfo {author} {\bibfnamefont {F.~M.}\ \bibnamefont {Qu}}, \bibinfo {author} {\bibfnamefont {G.~H.}\ \bibnamefont {Zhang}}, \bibinfo {author} {\bibfnamefont {J.~R.}\ \bibnamefont {Shi}}, \bibinfo {author} {\bibfnamefont {X.~C.}\ \bibnamefont {Xie}}, \bibinfo {author} {\bibfnamefont {C.~L.}\ \bibnamefont {Yang}}, \bibinfo {author} {\bibfnamefont {K.~H.}\ \bibnamefont {Wu}}, \bibinfo {author} {\bibfnamefont {Y.~Q.}\ \bibnamefont {Li}},\ and\ \bibinfo {author} {\bibfnamefont {L.}~\bibnamefont {Lu}},\ }\bibfield  {title} {\bibinfo {title} {Gate-voltage control of chemical potential and weak antilocalization in ${\mathrm{bi}}_{2}{\mathrm{se}}_{3}$},\ }\href
  {https://doi.org/10.1103/PhysRevLett.105.176602} {\bibfield  {journal} {\bibinfo  {journal} {Phys. Rev. Lett.}\ }\textbf {\bibinfo {volume} {105}},\ \bibinfo {pages} {176602} (\bibinfo {year} {2010})}\BibitemShut {NoStop}%
\bibitem [{\citenamefont {He}\ \emph {et~al.}(2011)\citenamefont {He}, \citenamefont {Wang}, \citenamefont {Zhang}, \citenamefont {Sou}, \citenamefont {Wong}, \citenamefont {Wang}, \citenamefont {Lu}, \citenamefont {Shen},\ and\ \citenamefont {Zhang}}]{He_2011}%
  \BibitemOpen
  \bibfield  {author} {\bibinfo {author} {\bibfnamefont {H.-T.}\ \bibnamefont {He}}, \bibinfo {author} {\bibfnamefont {G.}~\bibnamefont {Wang}}, \bibinfo {author} {\bibfnamefont {T.}~\bibnamefont {Zhang}}, \bibinfo {author} {\bibfnamefont {I.-K.}\ \bibnamefont {Sou}}, \bibinfo {author} {\bibfnamefont {G.~K.~L.}\ \bibnamefont {Wong}}, \bibinfo {author} {\bibfnamefont {J.-N.}\ \bibnamefont {Wang}}, \bibinfo {author} {\bibfnamefont {H.-Z.}\ \bibnamefont {Lu}}, \bibinfo {author} {\bibfnamefont {S.-Q.}\ \bibnamefont {Shen}},\ and\ \bibinfo {author} {\bibfnamefont {F.-C.}\ \bibnamefont {Zhang}},\ }\bibfield  {title} {\bibinfo {title} {Impurity effect on weak antilocalization in the topological insulator ${\mathrm{bi}}_{2}{\mathrm{te}}_{3}$},\ }\href {https://doi.org/10.1103/PhysRevLett.106.166805} {\bibfield  {journal} {\bibinfo  {journal} {Phys. Rev. Lett.}\ }\textbf {\bibinfo {volume} {106}},\ \bibinfo {pages} {166805} (\bibinfo {year} {2011})}\BibitemShut {NoStop}%
\bibitem [{\citenamefont {Kim}\ \emph {et~al.}(2026)\citenamefont {Kim}, \citenamefont {Lee}, \citenamefont {Furusaki},\ and\ \citenamefont {Yang}}]{kim2026weaklocalizationmagneticeuler}%
  \BibitemOpen
  \bibfield  {author} {\bibinfo {author} {\bibfnamefont {D.~Y.}\ \bibnamefont {Kim}}, \bibinfo {author} {\bibfnamefont {S.~H.}\ \bibnamefont {Lee}}, \bibinfo {author} {\bibfnamefont {A.}~\bibnamefont {Furusaki}},\ and\ \bibinfo {author} {\bibfnamefont {B.-J.}\ \bibnamefont {Yang}},\ }\href {https://arxiv.org/abs/2608.15529} {\bibinfo {title} {Weak localization in magnetic euler bands}} (\bibinfo {year} {2026}),\ \Eprint {https://arxiv.org/abs/2608.15529} {arXiv:2608.15529} \BibitemShut {NoStop}%
\bibitem [{\citenamefont {Zhang}\ \emph {et~al.}(2010)\citenamefont {Zhang}, \citenamefont {Sahu}, \citenamefont {Min},\ and\ \citenamefont {MacDonald}}]{zhangBandStructureStacked2010}%
  \BibitemOpen
  \bibfield  {author} {\bibinfo {author} {\bibfnamefont {F.}~\bibnamefont {Zhang}}, \bibinfo {author} {\bibfnamefont {B.}~\bibnamefont {Sahu}}, \bibinfo {author} {\bibfnamefont {H.}~\bibnamefont {Min}},\ and\ \bibinfo {author} {\bibfnamefont {A.~H.}\ \bibnamefont {MacDonald}},\ }\bibfield  {title} {\bibinfo {title} {Band structure of $abc$-stacked graphene trilayers},\ }\href {https://doi.org/10.1103/PhysRevB.82.035409} {\bibfield  {journal} {\bibinfo  {journal} {Phys. Rev. B}\ }\textbf {\bibinfo {volume} {82}},\ \bibinfo {pages} {035409} (\bibinfo {year} {2010})}\BibitemShut {NoStop}%
\bibitem [{\citenamefont {Koshino}\ and\ \citenamefont {McCann}(2009)}]{koshinoTrigonalWarpingBerrys2009}%
  \BibitemOpen
  \bibfield  {author} {\bibinfo {author} {\bibfnamefont {M.}~\bibnamefont {Koshino}}\ and\ \bibinfo {author} {\bibfnamefont {E.}~\bibnamefont {McCann}},\ }\bibfield  {title} {\bibinfo {title} {Trigonal warping and berry's phase $n\ensuremath{\pi}$ in abc-stacked multilayer graphene},\ }\href {https://doi.org/10.1103/PhysRevB.80.165409} {\bibfield  {journal} {\bibinfo  {journal} {Phys. Rev. B}\ }\textbf {\bibinfo {volume} {80}},\ \bibinfo {pages} {165409} (\bibinfo {year} {2009})}\BibitemShut {NoStop}%
\bibitem [{\citenamefont {Holleis}\ \emph {et~al.}(2026)\citenamefont {Holleis}, \citenamefont {Choi}, \citenamefont {Zhang}, \citenamefont {Farrell}, \citenamefont {Bargas}, \citenamefont {Hsu}, \citenamefont {Chen}, \citenamefont {Sackin}, \citenamefont {Zhou}, \citenamefont {Guo}, \citenamefont {Charpentier}, \citenamefont {Jiang}, \citenamefont {Foutty}, \citenamefont {Keough}, \citenamefont {Huber}, \citenamefont {Taniguchi}, \citenamefont {Watanabe}, \citenamefont {Lucas},\ and\ \citenamefont {Young}}]{holleis2026cryogenicshockexfoliationultrahigh}%
  \BibitemOpen
  \bibfield  {author} {\bibinfo {author} {\bibfnamefont {L.}~\bibnamefont {Holleis}}, \bibinfo {author} {\bibfnamefont {Y.}~\bibnamefont {Choi}}, \bibinfo {author} {\bibfnamefont {C.}~\bibnamefont {Zhang}}, \bibinfo {author} {\bibfnamefont {J.~H.}\ \bibnamefont {Farrell}}, \bibinfo {author} {\bibfnamefont {G.}~\bibnamefont {Bargas}}, \bibinfo {author} {\bibfnamefont {A.}~\bibnamefont {Hsu}}, \bibinfo {author} {\bibfnamefont {Z.}~\bibnamefont {Chen}}, \bibinfo {author} {\bibfnamefont {I.}~\bibnamefont {Sackin}}, \bibinfo {author} {\bibfnamefont {W.}~\bibnamefont {Zhou}}, \bibinfo {author} {\bibfnamefont {Y.}~\bibnamefont {Guo}}, \bibinfo {author} {\bibfnamefont {T.}~\bibnamefont {Charpentier}}, \bibinfo {author} {\bibfnamefont {Y.}~\bibnamefont {Jiang}}, \bibinfo {author} {\bibfnamefont {B.~A.}\ \bibnamefont {Foutty}}, \bibinfo {author} {\bibfnamefont {A.}~\bibnamefont {Keough}}, \bibinfo {author} {\bibfnamefont {M.~E.}\ \bibnamefont {Huber}}, \bibinfo {author} {\bibfnamefont {T.}~\bibnamefont {Taniguchi}},
  \bibinfo {author} {\bibfnamefont {K.}~\bibnamefont {Watanabe}}, \bibinfo {author} {\bibfnamefont {A.}~\bibnamefont {Lucas}},\ and\ \bibinfo {author} {\bibfnamefont {A.~F.}\ \bibnamefont {Young}},\ }\href {https://arxiv.org/abs/2604.21912} {\bibinfo {title} {Cryogenic shock exfoliation for ultrahigh mobility rhombohedral graphite nanoelectronics}} (\bibinfo {year} {2026}),\ \Eprint {https://arxiv.org/abs/2604.21912} {arXiv:2604.21912} \BibitemShut {NoStop}%
\bibitem [{\citenamefont {Kechedzhi}\ \emph {et~al.}(2007)\citenamefont {Kechedzhi}, \citenamefont {Fal'ko}, \citenamefont {McCann},\ and\ \citenamefont {Altshuler}}]{Kechedzhi_2007_PRL}%
  \BibitemOpen
  \bibfield  {author} {\bibinfo {author} {\bibfnamefont {K.}~\bibnamefont {Kechedzhi}}, \bibinfo {author} {\bibfnamefont {V.~I.}\ \bibnamefont {Fal'ko}}, \bibinfo {author} {\bibfnamefont {E.}~\bibnamefont {McCann}},\ and\ \bibinfo {author} {\bibfnamefont {B.~L.}\ \bibnamefont {Altshuler}},\ }\bibfield  {title} {\bibinfo {title} {Influence of trigonal warping on interference effects in bilayer graphene},\ }\href {https://doi.org/10.1103/PhysRevLett.98.176806} {\bibfield  {journal} {\bibinfo  {journal} {Phys. Rev. Lett.}\ }\textbf {\bibinfo {volume} {98}},\ \bibinfo {pages} {176806} (\bibinfo {year} {2007})}\BibitemShut {NoStop}%
\bibitem [{\citenamefont {Dai}\ \emph {et~al.}(2026)\citenamefont {Dai}, \citenamefont {Zhao}, \citenamefont {Chen},\ and\ \citenamefont {Law}}]{Dai_2026}%
  \BibitemOpen
  \bibfield  {author} {\bibinfo {author} {\bibfnamefont {W.-B.}\ \bibnamefont {Dai}}, \bibinfo {author} {\bibfnamefont {J.}~\bibnamefont {Zhao}}, \bibinfo {author} {\bibfnamefont {S.~A.}\ \bibnamefont {Chen}},\ and\ \bibinfo {author} {\bibfnamefont {K.~T.}\ \bibnamefont {Law}},\ }\href {https://arxiv.org/abs/2605.03987} {\bibinfo {title} {Quantum metric localization and quantum metric protection}} (\bibinfo {year} {2026}),\ \Eprint {https://arxiv.org/abs/2605.03987} {arXiv:2605.03987} \BibitemShut {NoStop}%
\bibitem [{\citenamefont {Chau}\ \emph {et~al.}(2026)\citenamefont {Chau}, \citenamefont {Xiang}, \citenamefont {Chen},\ and\ \citenamefont {Law}}]{Chau_2026}%
  \BibitemOpen
  \bibfield  {author} {\bibinfo {author} {\bibfnamefont {C.~W.}\ \bibnamefont {Chau}}, \bibinfo {author} {\bibfnamefont {T.}~\bibnamefont {Xiang}}, \bibinfo {author} {\bibfnamefont {S.~A.}\ \bibnamefont {Chen}},\ and\ \bibinfo {author} {\bibfnamefont {K.~T.}\ \bibnamefont {Law}},\ }\href {https://arxiv.org/abs/2602.01354} {\bibinfo {title} {Quantum metric length as a fundamental length scale in disordered flat band materials}} (\bibinfo {year} {2026}),\ \Eprint {https://arxiv.org/abs/2602.01354} {arXiv:2602.01354} \BibitemShut {NoStop}%
\bibitem [{\citenamefont {Wang}\ and\ \citenamefont {Qi}(2026)}]{Wang_2026}%
  \BibitemOpen
  \bibfield  {author} {\bibinfo {author} {\bibfnamefont {X.-C.}\ \bibnamefont {Wang}}\ and\ \bibinfo {author} {\bibfnamefont {Y.}~\bibnamefont {Qi}},\ }\href {https://arxiv.org/abs/2608.08042} {\bibinfo {title} {Quantum geometric localization length and localization criticality in an ideally flat chern band}} (\bibinfo {year} {2026}),\ \Eprint {https://arxiv.org/abs/2608.08042} {arXiv:2608.08042} \BibitemShut {NoStop}%
\bibitem [{\citenamefont {Burkov}(2026)}]{Burkov_2026}%
  \BibitemOpen
  \bibfield  {author} {\bibinfo {author} {\bibfnamefont {A.~A.}\ \bibnamefont {Burkov}},\ }\bibfield  {title} {\bibinfo {title} {Quantum geometric contribution to the diffusion constant in metals and semimetals},\ }\href {https://doi.org/10.1103/8lkl-x1nq} {\bibfield  {journal} {\bibinfo  {journal} {Phys. Rev. B}\ }\textbf {\bibinfo {volume} {113}},\ \bibinfo {pages} {155145} (\bibinfo {year} {2026})}\BibitemShut {NoStop}%
\end{thebibliography}
\end{document}